\pdfoutput=1
\documentclass[11pt,twoside,a4paper,cmspaper,final,collab]{cms-tdr}
\def\svnVersion{915d973-D}\def\svnDate{2021/03/23}\def\cmsCernNoTag{CERN-EP-2020-121}\def\cmsCernDate{\today}\def\cmsMessage{"Published in Physical Review D as \href{http://dx.doi.org/10.1103/PhysRevD.103.052008}{\doi{10.1103/PhysRevD.103.052008}.}"}
\begin{document}\cmsNoteHeader{TOP-18-013}

\RCS$Revision$
\RCS$HeadURL$
\RCS$Id$

\newlength\cmsFigWidth
\ifthenelse{\boolean{cms@external}}{\setlength\cmsFigWidth{0.49\textwidth}}{\setlength\cmsFigWidth{0.47\textwidth}}
\ifthenelse{\boolean{cms@external}}{\providecommand{\cmsLeft}{upper\xspace}}{\providecommand{\cmsLeft}{left\xspace}}
\ifthenelse{\boolean{cms@external}}{\providecommand{\cmsRight}{lower\xspace}}{\providecommand{\cmsRight}{right\xspace}}

\renewcommand{\theenumi}{\roman{enumi}}
\newcommand{\mSD}{\ensuremath{m_{\mathrm{SD}}}\xspace}
\newlength\cmsTabSkip\setlength{\cmsTabSkip}{0.5ex}

\cmsNoteHeader{TOP-18-013} 

\title{Measurement of differential \texorpdfstring{\ttbar}{ttbar} production cross sections using top quarks at large transverse momenta in \texorpdfstring{$\Pp\Pp$}{pp} collisions at \texorpdfstring{$\sqrt{s} = 13\TeV$}{sqrt(s) = 13 TeV}}

\date{\today}

\abstract{    
A measurement is reported of differential top quark pair (\ttbar) production cross sections, where top quarks are produced at large transverse momenta. 
The data collected with the CMS detector at the LHC are from $\Pp\Pp$ collisions at a center-of-mass energy of 13\TeV corresponding to an integrated luminosity of 35.9\fbinv. 
The measurement uses events where at least one top quark decays as $\PQt \to \PW \PQb \to \PQq \PAQq' \PQb$ and is reconstructed as a large-radius jet with transverse momentum in excess of 400\GeV. The second top quark is required to decay either in a similar way, or leptonically, as inferred from a reconstructed electron or muon, a bottom quark jet, and missing transverse momentum due to the undetected neutrino.
The cross section is extracted as a function of kinematic variables of individual top quarks or of the \ttbar system. The results are presented at the particle level, within a region of phase space close to that of the experimental acceptance, and at the parton level, and are compared to various theoretical models. 
In both decay channels, the observed absolute cross sections are significantly lower than the predictions from theory, while the normalized differential measurements are well described.
}

\hypersetup{
pdfauthor={CMS Collaboration},
pdftitle={Measurement of differential ttbar production cross sections using top quarks at large transverse momenta in pp collisions at sqrt(s) = 13 TeV},
pdfsubject={CMS},
pdfkeywords={CMS, physics, top, ttbar, jets, boosted}}

\maketitle

\section{Introduction}\label{sec:Intro}

The top quark completes the third generation of quarks in the standard model (SM), and a precise understanding of its properties is critical for the overall consistency of the theory. Measurements of the top quark-antiquark pair (\ttbar) production cross section confront the expectations from quantum chromodynamics (QCD), but could also be sensitive to effects of physics beyond the SM. In particular, \ttbar production constitutes a dominant SM background to many direct searches for beyond-the-SM phenomena, and its detailed characterization is therefore important for confirming possible discoveries. 

The large \ttbar yield expected in $\Pp\Pp$ collisions at the CERN LHC enables measurements of the \ttbar production rate as functions of kinematic variables of individual top quarks and the \ttbar system. Such measurements have been performed at the ATLAS~\cite{ATLAShadronicResolved13TeV,Aad:2019ntk,Aad:2019hzw,Aaboud:2017fha,Aaboud:2016syx,Aaboud:2017ujq,Aaboud:2016iot,Aad:2015eia,Aad:2012hg} and CMS~\cite{Sirunyan:2017mzl,CMS-TOP-18-004,Sirunyan:2018ucr,Khachatryan:2016mnb,Sirunyan:2018ptc,Sirunyan:2017azo,Khachatryan:2015oqa,Khachatryan:2016oou,Chatrchyan:2012saa,TOP17002} experiments at 7, 8, and 13\TeV center-of-mass energies, assuming a resolved final state where the decay products of the \ttbar system can be reconstructed individually. Resolved top quark reconstruction is possible for top quark transverse momenta (\pt) up to about 500\GeV. At higher \pt, the top quark decay products are highly collimated (``Lorentz boosted") and they can no longer be reconstructed separately. To explore the highly boosted phase space, top quark decays are reconstructed as large-radius ($R$) jets in this analysis. Previous efforts in this domain by ATLAS~\cite{Aaboud:2018eqg,Aad:2015hna} and CMS~\cite{Khachatryan:2016gxp} confirm that it is feasible to perform precise differential measurements of high-{\pt} \ttbar production and have also indicated possibly interesting deviations from theory.
 
This paper reports a measurement of the differential \ttbar production cross section in the boosted regime in the all-jet and lepton+jets final states. The results are based on $\Pp\Pp$ collisions at $\sqrt{s}=13\TeV$ recorded by the CMS detector, corresponding to a total integrated luminosity of 35.9\fbinv. 
In the all-jet decay channel, each \PW boson arising from the $\PQt \to \PW \PQb$ transition decays into a quark (\PQq) and antiquark ($\PAQq'$). As a result, the final state consists of at least six quarks, two of which are bottom quarks. Additional partons, gluons or quarks, can arise from initial-state radiation (ISR) and final-state radiation (FSR). The sizable boost of the top quarks in this measurement ($\pt>400\GeV$) provides two top quarks reconstructed as large-$R$ jets and the final state therefore consists of at least two such jets. 
In the lepton+jets channel, one top quark decays according to $\PQt \to \PW \PQb \to \PQq \PAQq' \PQb$ and is reconstructed as a single large-$R$ jet, while the second top quark decays to a bottom quark and a \PW boson that in turn decays to a charged lepton ($\ell$), either an electron (\Pe) or a muon (\Pgm), and a neutrino ($\PQt \to \PW \PQb \to \ell \PGn \PQb$). 
Decays of \PW bosons via \PGt leptons to electrons or muons are treated as signal.
The measurements were performed using larger integrated luminosity and higher center-of-mass energy compared to previous CMS results~\cite{Khachatryan:2016gxp}. 
This provides a sharper confrontation with theory over data in a wider region of phase space. 

The paper is organized as follows: Section~\ref{sec:CMS} describes the main features of the CMS detector and the triggering system. Section~\ref{sec:MC} gives the details of the Monte Carlo (MC) simulations. Event reconstruction and selection are outlined in Sections~\ref{sec:ObjectReco} and~\ref{sec:Reco}, respectively. In Section~\ref{sec:Bkg}, we discuss the estimation of the background contributions, followed by a description of signal extraction in Section~\ref{sec:SignalExtraction}. Systematic uncertainties are discussed in Section~\ref{sec:Sys}. The unfolding procedure used to obtain the particle- and parton-level cross sections and the resulting measurements are presented in Section~\ref{sec:Meas}. Finally, Section~\ref{sec:Sum} provides a brief summary of the paper.

\section{The CMS detector}\label{sec:CMS}

The central feature of the CMS apparatus is a superconducting solenoid of 6\unit{m} internal diameter, providing a magnetic field of 3.8\unit{T}. A silicon pixel and strip tracker, a lead tungstate crystal electromagnetic calorimeter (ECAL), and a brass and scintillator hadron calorimeter (HCAL), each composed of a barrel and two endcap sections, reside within the magnetic volume. Forward calorimeters extend the pseudorapidity ($\eta$) coverage provided by the barrel and endcap detectors. Muons are detected in gas-ionization chambers embedded in the steel flux-return yoke outside the solenoid. A more detailed description of the CMS detector, together with a definition of the coordinate system and kinematic variables can be found in Ref.~\cite{Chatrchyan:2008zzk}. 

Events of interest are selected using a two-tiered trigger system~\cite{Khachatryan:2016bia}. The first level (L1), composed of specialized hardware processors, uses information from the calorimeters and muon detectors to select events at a rate of about 100\unit{kHz} within a fixed time interval of 4\mus. The second level, known as the high-level trigger (HLT), consists of a farm of processors that run the full event reconstruction software in a configuration for fast processing, and reduces the event rate to about 1\unit{kHz} before data storage. 

\section{Event simulation}\label{sec:MC}

We use MC simulation to generate event samples for the \ttbar signal and also to model the contributions from some of the background processes. 
The \ttbar events are generated at next-to-leading order (NLO) in QCD using \POWHEG~(v2)~\cite{Nason:2004rx,Frixione:2007nw,Frixione:2007vw,Alioli:2010xd,Powheg_tt}, assuming a top quark mass $m_{\PQt} = 172.5\GeV$. Single top quark production in the $t$ channel and in association with a \PW boson is simulated at NLO with \POWHEG~\cite{Powheg_st}, while $s$ channel production is negligible in this analysis. The production of \PW and \PZ bosons in association with jets ($\PV$+jets), as well as multijet events, are simulated using the \MGvATNLO~\cite{Alwall:2014hca} (v2.2.2) generator at leading order (LO), with the MLM matching algorithm~\cite{Alwall:2007fs} to avoid double-counting of partons.
Samples of diboson ($\PW\PW$, $\PW\PZ$, or $\PZ\PZ$) events are simulated at LO using \PYTHIA~(v8.212)~\cite{Sjostrand:2006za,Sjostrand:2007gs}. 

All simulated events are processed using \PYTHIA to model parton showering, hadronization, and the underlying event (UE). The NNPDF~3.0~\cite{Ball:2014uwa} parton distribution functions (PDFs) are used to generate the events, and the CUETP8M1 UE tune~\cite{Khachatryan:2110213} is used for all but the \ttbar and single top quark processes. For these, the CUETP8M2T4 tune with an adjusted value of the strong coupling \alpS is used, yielding an improved modeling of \ttbar event properties~\cite{CMS-PAS-TOP-16-021}. 
The simulation of the response of the CMS detector is based on \GEANTfour~\cite{Agostinelli2003250}. Additional $\Pp\Pp$ interactions in the same or neighboring bunch crossings (pileup) are simulated through \PYTHIA and overlaid with events generated according to the pileup distribution measured in data. An average of 27 pileup interactions was observed for the collected data. 

The simulated processes are normalized to their best known theoretical cross sections. Specifically, the \ttbar, $\PV$+jets, and single top quark event samples are normalized to next-to-NLO (NNLO) precision in QCD~\cite{Czakon:2011xx,Li:2012wna,Kidonakis:2013zqa}.

{\tolerance=800 The measured differential cross sections for \ttbar production are compared with state-of-the-art theoretical expectations provided by the NLO \POWHEG generator, combined with \PYTHIA for parton showering, as described above, or combined with NLO \HERWIGpp~\cite{Bahr:2008pv} and the corresponding EE5C UE tune~\cite{Gieseke:2012ft}. In addition, a comparison is performed with \MGvATNLO~\cite{Alwall:2014hca} using \PYTHIA for the parton showering. \par}

\section{Event reconstruction}\label{sec:ObjectReco}

Global event reconstruction, also called particle-flow (PF) event reconstruction~\cite{CMS-PRF-14-001}, aims to reconstruct and identify each individual particle in an event through an optimized combination of information from all subdetectors. In this process, the particle type (photon, electron, muon, and charged or neutral hadron) plays an important role in the determination of particle direction and energy. Photons are identified as ECAL energy clusters not linked to the extrapolation of any charged-particle trajectory to the ECAL. 
Electrons are identified as primary charged particle tracks and potentially multiple ECAL energy clusters corresponding to extrapolation of these tracks to the ECAL and to possible bremsstrahlung photons emitted along the way through the tracker material. 
Muons are identified as tracks in the central tracker consistent with either a track or several hits in the muon system associated with calorimeter deposition compatible with the muon hypothesis. Charged hadrons are identified as charged-particle tracks that are neither identified as electrons nor as muons. Finally, neutral hadrons are identified as HCAL energy clusters not linked to any charged-hadron trajectory, or as a combined ECAL and HCAL energy excess relative to the expected deposit of the charged-hadron energy.

The energy of photons is obtained from the ECAL measurement. The energy of electrons is determined from a combination of the track momentum at the main interaction vertex, the energy of the corresponding ECAL cluster, and the energy sum of all bremsstrahlung photons spatially compatible with originating from the electron track. The momentum of muons is obtained from the curvature of the corresponding track. The energy of charged hadrons is determined from a combination of their momentum measured in the tracker and the matching ECAL and HCAL energy deposits, corrected for the response function of the calorimeters to hadronic showers. Finally, the energy of neutral hadrons is obtained from the corresponding corrected ECAL and HCAL energies. 
 
Leptons and charged hadrons are required to be compatible with originating from the primary interaction vertex. 
The candidate vertex with the largest value of summed physics-object $\pt^2$ is taken to be the primary $\Pp\Pp$ interaction vertex. For this purpose the physics objects are the jets, clustered using the jet finding algorithm~\cite{Cacciari:2008gp, Cacciari:2011ma} with the tracks assigned to candidate vertices as inputs, and the negative vector \pt sum of those jets.
Charged hadrons that are associated with a pileup vertex are classified as pileup candidates and are ignored in the subsequent event reconstruction. Electron and muon objects are first identified from corresponding electron or muon PF candidates. Next, jet clustering is performed on all PF candidates that are not classified as pileup candidates. The jet clustering does not exclude the electron and muon PF candidates, even if these have already been assigned to electron/muon objects. A dedicated removal of overlapping physics objects is therefore used at the analysis level to avoid double counting. 

Electrons and muons selected in the $\ell$+jets channel must have $\pt > 50\GeV$ and $\abs{\eta} < 2.1$. 
For vetoing leptons in the all-jet channel, they are instead required to have $\pt > 20\GeV$ and $\abs{\eta} < 2.1$. Leptons are also required to be isolated according to the "mini-isolation" ($I_{\text{mini}}$) algorithm, which requires the scalar \pt sum of tracks in a cone around the electron or muon to be less than a given fraction of the lepton \pt ($\pt^{\ell}$)~\cite{Rehermann_2011}. The width of the cone ($\Delta R$) depends on the lepton \pt, being defined as $\Delta R = (10\GeV)/\pt^{\ell}$ for $\pt^{\ell} < 200\GeV$ and $\Delta R = 0.05$ for $\pt^{\ell}>200\GeV$. This algorithm retains high isolation efficiency for leptons originating from decays of highly-boosted top quarks. A value of $I_{\text{mini}}< 0.1$ is chosen, corresponding to approximately a 95\% efficiency
For vetoing additional leptons in the $\ell$+jets channel, the same lepton selection is used with the isolation requirement removed.
Correction factors are applied to account for differences between data and simulation in the modeling of lepton identification, isolation, and trigger efficiencies, determined as functions of $\abs{\eta}$ and \pt of the electron or muon using a ``tag-and-probe'' method~\cite{Khachatryan_2011}. 

In each event, jets are clustered using the reconstructed PF candidates through the infrared- and collinear-safe anti-\kt algorithm~\cite{Cacciari:2008gp, Cacciari:2011ma}.
Two jet collections are considered to identify \PQb and \PQt jet candidates. Small-$R$ jets are clustered using a distance parameter of 0.4 in the $\ell$+jets channel and large-$R$ jets using a distance parameter of 0.8 in the all-jet and $\ell$+jets channels. 
The jet momenta are determined through the vector sum of all particle momenta in the jet, and found from simulation to be typically within 5--10\% of the true momentum over the entire spectrum and detector acceptance. Additional $\Pp\Pp$  interactions can contribute more tracks and calorimetric energy depositions to the jet momentum. To mitigate this effect, the pileup candidates are discarded before the clustering and an offset correction is applied to correct for the remaining contributions from neutral particles~\cite{Cacciari_2008}. 

Jet energy corrections are obtained from simulation to bring the average measured response of jets to that of particle-level jets. In situ measurements of the momentum balance in dijet, photon+jet, $\PZ$+jet, and multijet events are used to account for any residual differences in the jet energy scale (JES) between data and simulation~\cite{Khachatryan:2016kdb}. The jet energy resolution (JER) amounts typically to 15--20\% at 30\GeV, 10\% at 100\GeV, and 5\% at 1\TeV. Additional criteria are applied to remove jets that are due to anomalous signals in the subdetectors or due to reconstruction failures~\cite{CMS-PAS-JME-16-003}. 

A grooming technique is used to remove soft, wide-angle radiation from the large-$R$ jets and to thereby improve the mass resolution. The algorithm employed is the ``modified mass drop tagger''~\cite{Dasgupta:2013ihk,Butterworth:2008iy}, also known as the ``soft-drop" (SD) algorithm~\cite{Larkoski:2014wba}, with angular exponent $\beta = 0$, soft cutoff threshold $z_{\mathrm{cut}} < 0.1$, and characteristic radius $R_{0} = 0.8$~\cite{Larkoski:2014wba}. 
The corresponding SD jet mass is referred to as \mSD.
The subjets within large-$R$ jets are identified through a reclustering of their constituents using the Cambridge--Aachen algorithm~\cite{Dokshitzer:1997in,Wobisch:1998wt} and then reversing the last step of the clustering history.
 
To identify jets originating from top quarks that decay according to $\PQt \to \PW \PQb \to \PQq \PAQq' \PQb$ ($\PQt$ tagging), we use the
$N$-subjettiness variables~\cite{Thaler:2010tr} $\tau_3$, $\tau_2$, and $\tau_1$ computed using the jet constituents according to
\begin{equation}\label{eq:tau}
\tau_N=\frac{1}{\sum_{j}{p_{\mathrm{T},j}}R}\sum_k{p_{\mathrm{T},k}\min\{\Delta R_{1,k},\Delta R_{2,k},\dots \Delta R_{N,k}\}}, 
\end{equation}
where $N$ denotes the number of reconstructed candidate subjets and $k$ runs over the constituent particles in the jet~\cite{JME-18-002}. The term min refers to the minimum value of the items within the curly parentheses, and the variable $\Delta R_{i,k}=\sqrt{\smash[b]{(\Delta \eta_{i,k})^2+(\Delta \phi_{i,k})^2}}$, where $\phi$ is the azimuthal angle, is the angular distance between the candidate subjet $i$ axis and the jet constituent $k$. The variable $R$ corresponds to the characteristic jet distance parameter ($R=0.8$ in our case). 
The directions of enhanced energy flow in jets are found by applying the exclusive \kt algorithm~\cite{Catani:1993hr,Thaler:2011gf} to the jet constituents before proceeding with jet grooming techniques. 

Small-$R$ jets and subjets of large-$R$ jets are identified as bottom quark candidates (\PQb-tagged) using the combined secondary vertex (CSV) algorithm~\cite{BTV-16-002}. Data-to-simulation correction factors are used to match the \PQb tagging efficiency observed in simulation to that measured in data. The typical efficiencies of the \PQb tagging algorithm for small-$R$ jets and subjets of large-$R$ jets are, respectively, 63 and 58\% for genuine \PQb~(sub)jets, while the misidentification probability for light-flavor (sub)jets is 1\%. For the subjets of large-$R$ jets, the efficiency for tagging genuine \PQb~subjets drops from 65 to 40\% as the \pt increases from 20\GeV to 1\TeV.

The missing transverse momentum vector \ptvecmiss is defined as the projection onto the plane perpendicular to the beam axis of the negative momentum vector sum of all PF candidates in an event. Its magnitude is referred to as \ptmiss, which is calculated after applying the aforementioned jet energy corrections. 

\section{Event selection}\label{sec:Reco}

\subsection{Trigger}
 
Different triggers were employed to collect signal events in the all-jet and $\ell$+jets channels, according to each event topology. 
The trigger used in the all-jet channel required the presence of a jet with $\pt>180\GeV$ at L1. At the HLT, large-$R$ jets were reconstructed from PF candidates using the anti-\kt algorithm with a distance parameter of 0.8. The mass of the jets at the HLT, after removal of soft particles, was required to be greater than 30\GeV. Selected events had to contain at least two such jets with $\pt>280$ and 200\GeV for the leading and trailing jets, respectively. Finally, at least one of these jets had to be $\PQb$-tagged using the CSV algorithm suitably adjusted for the HLT at an average identification efficiency of 90\% for \PQb jets. The aforementioned trigger ran for the entire 2016 data run, collecting an integrated luminosity of 35.9\fbinv. A second trigger with identical kinematic criteria but without any \PQb tagging requirement was employed and ran on average every 21 bunch crossings, collecting an integrated luminosity of 1.67\fbinv. The events collected with the latter trigger were intended for use as a control data sample to estimate the multijet background in the all-jet channel, as described below.
For the $\ell$+jets channel, the data were selected using triggers requiring a single lepton without imposing any isolation criteria, either an electron with $\pt > 45\GeV$ and $\abs{\eta}<2.5$ or a muon with $\pt > 40\GeV$ and $\abs{\eta} < 2.1$, as well as two small-$R$ jets with $\pt > 200$ and 50\GeV. 

\subsection{All-jet channel}

The events considered in the all-jet final state are required to fulfill a common baseline selection. 
This requires the presence of at least two large-$R$ jets in the event with $\pt>400\GeV$, $\abs{\eta}<2.4$, and $50<\mSD<300\GeV$. In addition, events with at least one lepton are vetoed to suppress leptonic final states originating from top quarks. 

Jet substructure variables are used to discriminate between events that originate from \ttbar decays and multijet production. 
These are sensitive to the type of jet, and in particular to whether the jet arises from a single parton, such as those in the case of ordinary quark or gluon evolutions into jets, or from three partons, such as in the $\PQt \to \PW\PQb \to \cPq\cPaq^\prime\PQb$ decay considered here. 
The $\tau_{1,2,3}$ variables of the two large-$R$ jets with highest \pt are combined through a neural network (NN) to form a multivariate discriminant that characterizes each event, with values close to zero indicating dijet production, and values close to one favoring \ttbar production. These variables are chosen such that the correlation with the number of $\PQb$-tagged subjets, which is used to define control regions for the multijet background, is minimal. The NN consists of two hidden layers with 16 and 4 nodes,  implemented in the \textsc{tmva} toolkit~\cite{hoecker2007tmva}. More complex architectures do not improve the discriminating capabilities of the NN. The training of the NN is performed with simulated multijet (background) and \ttbar (signal) events that satisfy the baseline selection, through the back-propagation method and a sigmoid activation function for the nodes. 
Excellent agreement between data and simulation is observed for the input variables in the phase space of the training.   

Besides the baseline selection, sub-regions are defined based on the NN output, the \mSD of the jets, and the number of $\PQb$-tagged subjets in each large-$R$ jet. The signal region (SR) used to extract the differential measurements contains events collected with the signal trigger where both large-$R$ jets contain a $\PQb$-tagged subjet, have masses in the range of 120--220\GeV, and NN output values greater than 0.8. This value is chosen to ensure that the ratio of \ttbar signal to background is large, while keeping a sufficient number of signal events with a top quark $\pt > 1\TeV$. In this region, more than 95\% of the selected \ttbar events originate from all-jet top quark decays according to simulation. The multijet control region (CR) contains events collected via a control trigger that satisfy the same requirements as those in the SR, but with an inverted \PQb tagging requirement. In addition, expanded regions that include both SR and CR events are defined to estimate background contributions. 
Signal region A (SR$_{\mathrm{A}}$) and control region A (CR$_{\mathrm{A}}$) are the same as the SR and CR, but have an extended requirement on the \mSD of large-$R$ jets of 50--300\GeV. 
It should be noted that the events selected in SR$_{\mathrm{A}}$ and CR$_{\mathrm{A}}$ were collected with the signal and control triggers, respectively. Finally, signal region B (SR$_{\mathrm{B}}$) has the same selection criteria as the SR, except without an NN requirement, and is used to constrain some of the signal modeling uncertainties.

\subsection{\texorpdfstring{$\ell$+jets}{ell+jets} channel}

The $\ell$+jets final state is identified through the presence of an electron or a muon, a small-$R$ jet that reflects the bottom quark emitted in the $\PQt \to \PW \PQb \to \ell \PGn \PQb$ decay, and a large-$R$ jet corresponding to the top quark decaying according to $\PQt \to \PW \PQb \to \PQq \PAQq' \PQb$. 
Small-$R$ (large-$R$) jets are required to have $\pt > 50\,(400)\GeV$ and $\abs{\eta} < 2.4$.

All events are required to pass the following preselection criteria, to contain:
\begin{enumerate}
\item Exactly one electron or muon;
\item No additional veto leptons;
\item At least one small-$R$ jet near the lepton, with $0.3 < \Delta R(\ell,\text{jet}) < \pi/2$;
\item At least one large-$R$ jet away from the lepton, with $\Delta R(\ell,\text{jet}) > \pi/2$;
\item $\ptmiss > 50$ or 35\GeV for the electron or muon channel, and;
\item For events in the electron channel, a cutoff to ensure that \ptvecmiss does not point along the transverse direction of the electron or the leading jet: $\abs{\Delta \phi(\ptvec^X,\ptvecmiss)} < 1.5 \ptmiss / 110\GeV$, where $X$ stands for the electron or the leading small-$R$ jet.
\end{enumerate}
The more stringent \ptmiss selection and criterion (vi) in the electron channel are applied to further reduce background from multijet production.

Events that fulfill the preselection criteria are categorized according to whether the jet candidates pass or fail the relevant \PQb or \PQt tagging criteria. 
The \PQb jet candidate is the highest-\pt leptonic-side jet in the event while the \PQt jet candidate is the highest-\pt jet on the non-leptonic side.
The $N$-subjettiness ratio $\tau_3/\tau_2$ (abbreviated as $\tau_{32}$) is used to distinguish a three-pronged top quark decay from background processes by requiring $\tau_{32} < 0.81$. In addition, the \PQt jet candidate must have $105<\mSD<220\GeV$. A data-to-simulation efficiency correction factor is extracted simultaneously with the integrated signal yield, as described in Section~\ref{sec:SignalExtraction}, to correct the \PQt tagging efficiency in simulation to match that in data.  

Events are divided into the following categories:
\begin{enumerate}
\item No \PQt tags (0t): the \PQt jet candidate fails the \PQt tagging requirement;
\item 1 \PQt tag, no \PQb tags (1t0b): the \PQt jet candidate passes the \PQt tagging requirement, but the \PQb jet candidate fails the \PQb tagging requirement, and;
\item 1 \PQt tag, 1 \PQb tag (1t1b): both the \PQt jet candidate and the \PQb jet candidate pass their respective tagging requirement.
\end{enumerate}
These event categories are designed to produce different admixtures of signal and background, with the 0t region having most background and the 1t1b region most signal.

\section{Background estimation}\label{sec:Bkg}
 
The dominant background in the all-jet channel is multijet production, while in the $\ell$+jets channel the dominant sources of background include nonsignal \ttbar, single top quark, $\PW$+jets, and multijet production events. Nonsignal \ttbar events, referred to as ``\ttbar other", comprise dilepton (where one lepton is not identified) and all-jet final states (where a lepton arises from one of the jets), in addition to $\PGt$+jets events where the \PGt lepton decays hadronically. 
 
\begin{figure}[hbt]
\centering
    \includegraphics[width=0.9\cmsFigWidth]{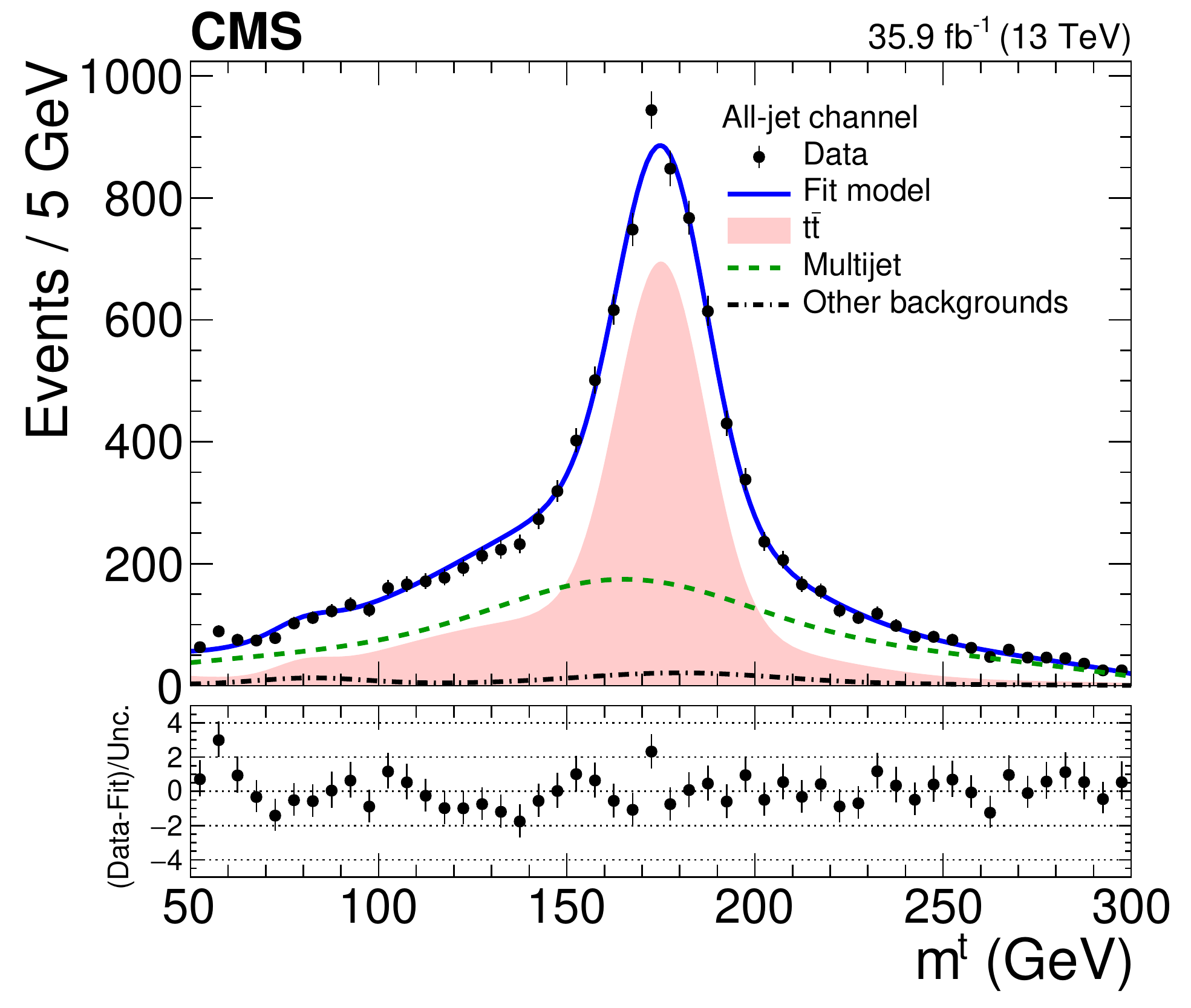}
    \caption{Result of the fit of \mSD of the \PQt jet candidate, $m^{\PQt}$, in the signal region SR$_{\mathrm{A}}$ to data in the all-jet events. The shaded area shows the \ttbar contribution, the dashed line the multijet background, and the dash-dotted line the other subdominant backgrounds. The solid line is the fit to the combined signal+background model, and the data points are represented by the filled circles. The lower panel shows the difference between the data and the fit model, divided by the uncertainty in the fit.}
    \label{fig:fit}
\end{figure}

In the all-jet channel, the background from multijet production is significantly suppressed through a combination of \PQb tagging requirements for the subjets within the large-$R$ jets and the event NN output and it is estimated from a control data sample. The two items determined from data are the shape of the multijet background as a function of an observable of interest $x$, and the absolute normalization $N_\text{multijet}$. The shape is taken from CR$_{\mathrm{A}}$, where the \ttbar signal contamination, based on simulation, is about $1\%$. The value of $N_\text{multijet}$ is extracted through a binned maximum likelihood fit of the data in SR$_{\mathrm{A}}$ of the \mSD of the \PQt jet candidate, $m^{\PQt}$, where the \PQt jet candidate is taken as the large-$R$ jet with highest \pt. The expected number of events $D(m^{\PQt})$ is modeled according to
\begin{linenomath}
\ifthenelse{\boolean{cms@external}}
{
\begin{multline}\label{eq:signal_fit}
D(m^{\PQt})=N_\ttbar T(m^{\PQt};k_\text{scale},k_\text{res})+ \\ N_\text{multijet}(1+k_\text{slope}m^{\PQt})Q(m^{\PQt})+ N_\text{bkg}B(m^{\PQt}), 
\end{multline}
}
{
\begin{equation}\label{eq:signal_fit}
D(m^{\PQt})=N_\ttbar T(m^{\PQt};k_\text{scale},k_\text{res})+ N_\text{multijet}(1+k_\text{slope}m^{\PQt})Q(m^{\PQt})+ N_\text{bkg}B(m^{\PQt}), 
\end{equation}
}
\end{linenomath}
which contains the distributions $T(m^{\PQt})$ and $B(m^{\PQt})$ of the signal and the subdominant backgrounds, respectively, taken from MC simulation, and the distribution $Q(m^{\PQt})$ of the multijet background. To account for a possible difference in the multijet $m^{\PQt}$ dependence in the CR$_{\mathrm{A}}$ and SR$_{\mathrm{A}}$, a multiplicative factor $(1+k_\text{slope}m^{\PQt})$ is introduced, inspired by the simulation, but with the slope parameter $k_\text{slope}$ left free in the fit. Also free in the fit are the normalization factors $N_\ttbar$, $N_\text{multijet}$, and $N_\text{bkg}$. Two additional nuisance parameters are introduced in the analytic parametrization of the $m^{\PQt}$ distribution for simulated \ttbar events, $k_\text{scale}$ and $k_\text{res}$, which account for possible differences between data and simulation in the scale and resolution in the $m^{\PQt}$ parameter. The fit is performed using the \textsc{RooFit} toolkit~\cite{Verkerke:2003ir} and the results are shown in Fig.~\ref{fig:fit} and Table~\ref{tab:fit}. The fitted \ttbar yield of $6238\pm181$ is significantly lower than the 9885 events expected in the SR$_{\mathrm{A}}$ according to \ttbar simulation and the theoretical cross section discussed in Section~\ref{sec:MC}, which implies that the fiducial cross section is smaller than the $\POWHEG$+$\PYTHIA$8 prediction, and corresponds to a fitted
signal strength $r=0.64 \pm 0.03$. This result is consistent with the softer top quark \pt spectrum compared to NLO predictions that has been reported in previous measurements~\cite{Sirunyan:2017mzl,Khachatryan:2016mnb}. The fitted signal strength is used to scale down the expected \ttbar signal yields from the $\POWHEG$+$\PYTHIA$8 simulation in various SRs in the subsequent figures containing comparisons between data and simulations but not in the subsequent derivation of the differential cross sections. The nuisance parameters that control the scale and the resolution of the reconstructed mass are consistent with unity, confirming thereby the good agreement between data and simulation in this variable.

\begin{table}[h]
\setlength{\tabcolsep}{1pt}
\centering
\topcaption{Fitted values of the nuisance parameters for the fit to data in the SR$_{\mathrm{A}}$ in the all-jet channel.}
\begin{scotch}{crl}
Parameter & \multicolumn{2}{c}{Value $\pm$ statistical uncertainty} \\ [\cmsTabSkip] \hline \noalign{\vskip\cmsTabSkip}
$k_\text{res}$     & ~~~~~~$0.960$ & $\pm\,0.026$   \\
$k_\text{scale}$ & $1.002$ & $\pm\,0.002$   \\
$k_\text{slope}$ & $(5.7$   & $\pm\,1.4)\times 10^{-3}$ \\
$N_\text{bkg}$   & $400$   & $\pm\,255$     \\
$N_\text{multijet}$ & $4539$ & $\pm\,247$     \\
$N_{\ttbar}$       & $6238$  & $\pm\,181$     \\
\end{scotch}\label{tab:fit}
\end{table}

The subdominant background processes, namely single top quark production and vector bosons produced in association with jets, have a negligible contribution in the SR (${<}1\%$ in the entire phase space) and are fixed to the predictions from simulation.

\begin{figure}[hbt]
\centering
    \includegraphics[width=\cmsFigWidth]{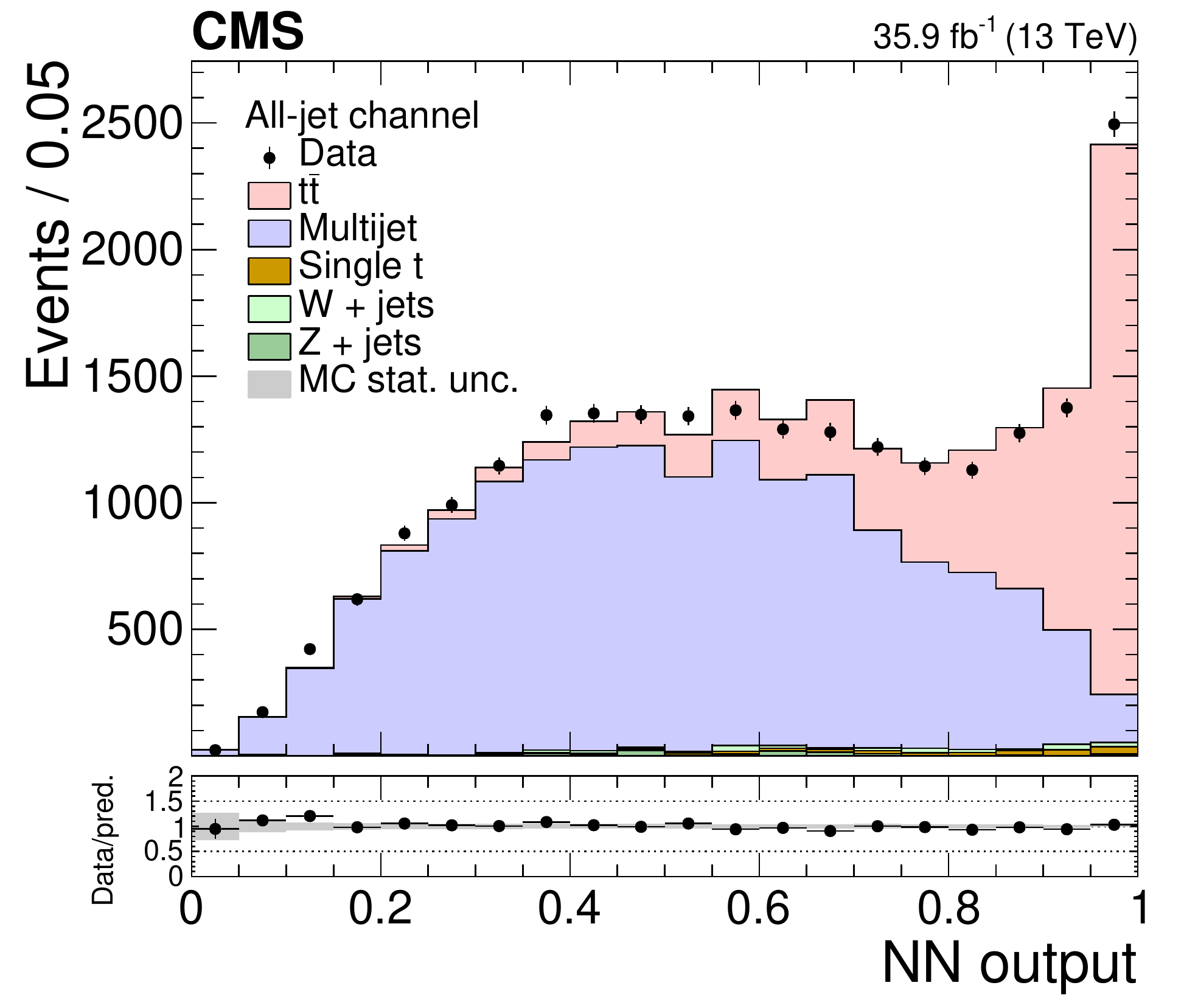}
    \caption{Comparison between data and prediction in the signal region SR$_{\mathrm{B}}$ (same as the SR, but without an NN requirement) of the NN output distribution for the all-jet channel. The contributions from \ttbar and multijet production are normalized according to the fitted values of their respective yields and shown as stacked histograms. The data points are represented by filled circles, while the shaded band represents the statistical uncertainty in simulation. The lower panel shows the data divided by the sum of the predictions.}
    \label{fig:NN}
\end{figure}

\begin{figure*}[hbtp]
\centering
\includegraphics[width=\cmsFigWidth]{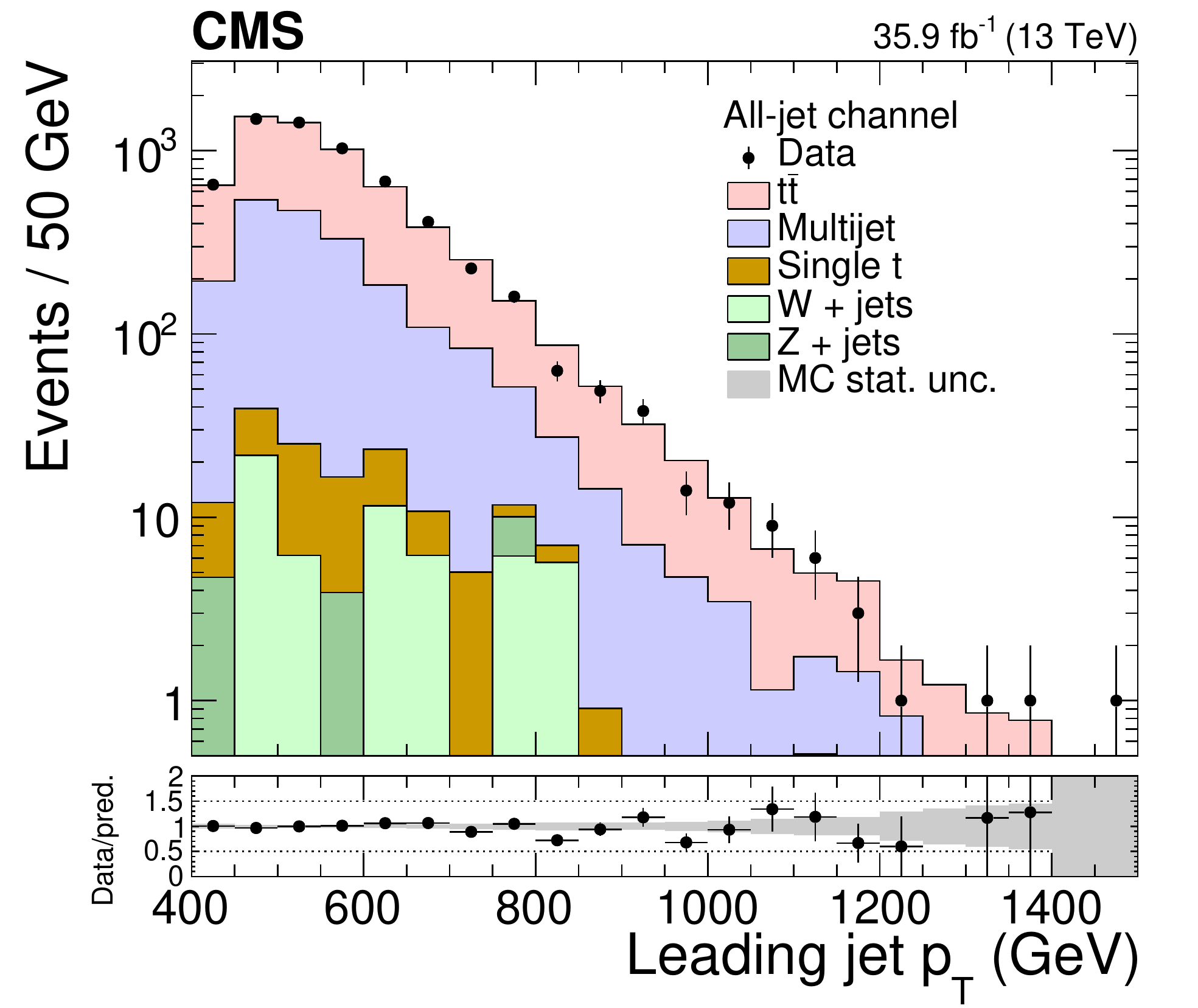}
\includegraphics[width=\cmsFigWidth]{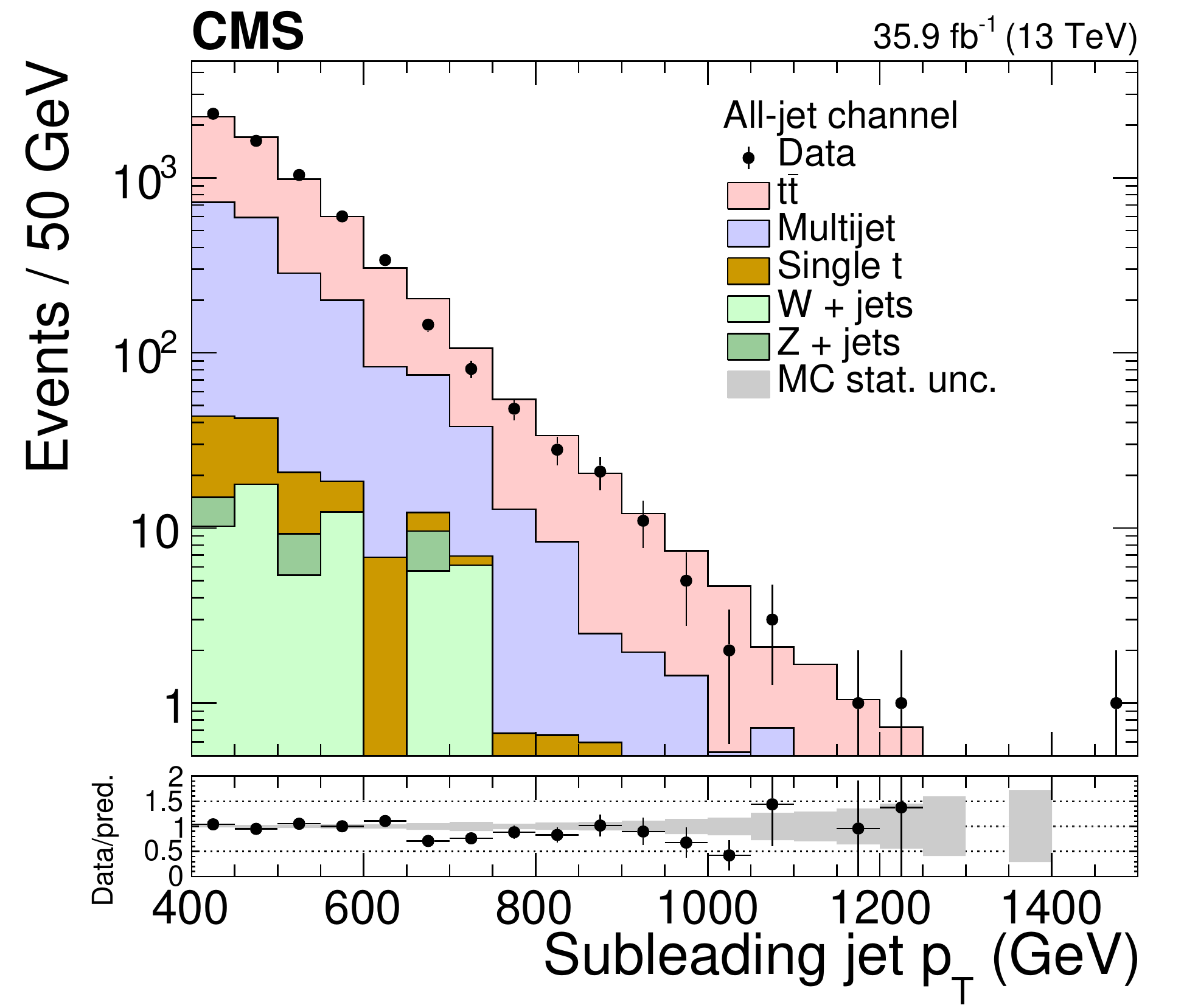}\\
\includegraphics[width=\cmsFigWidth]{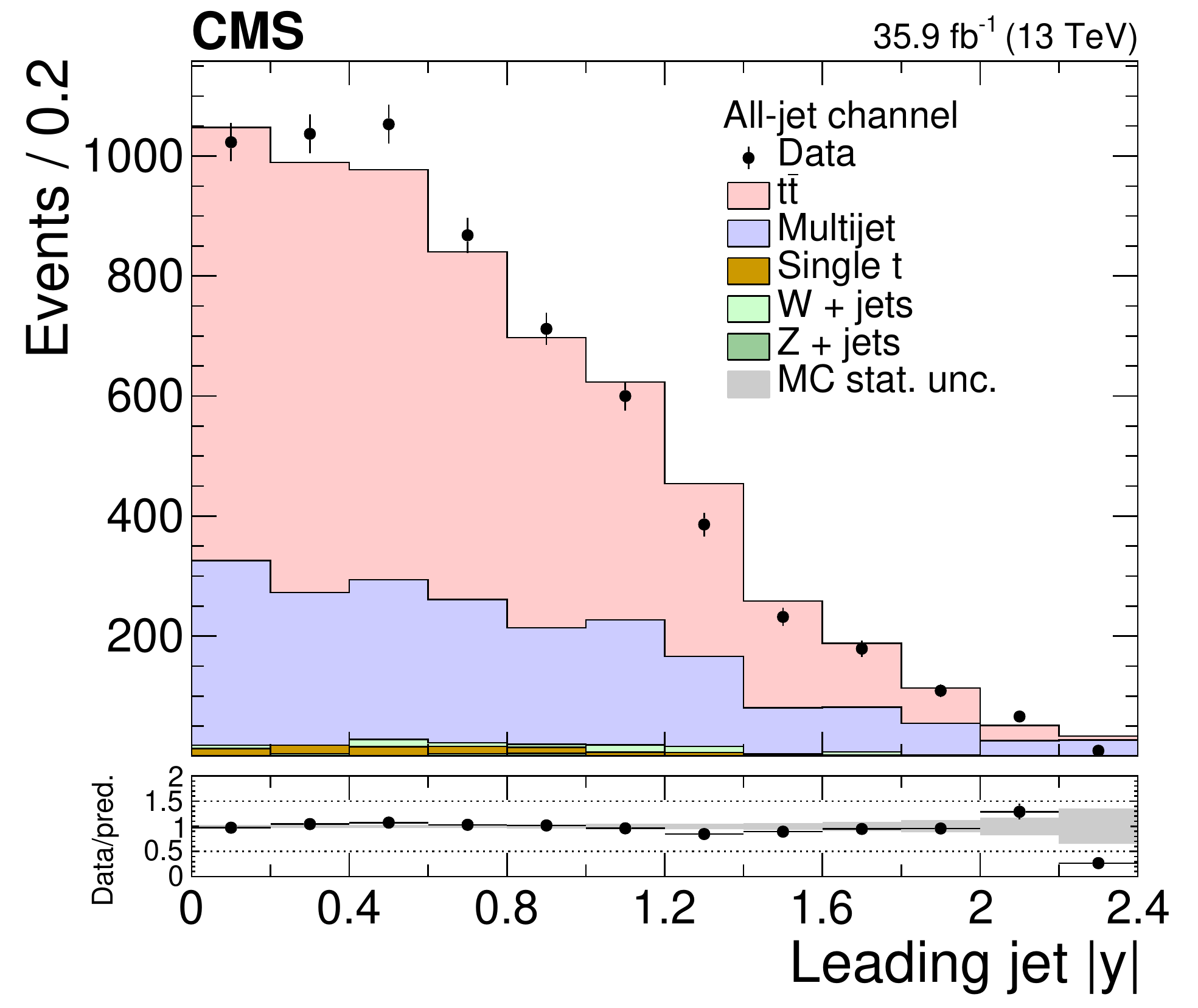}
\includegraphics[width=\cmsFigWidth]{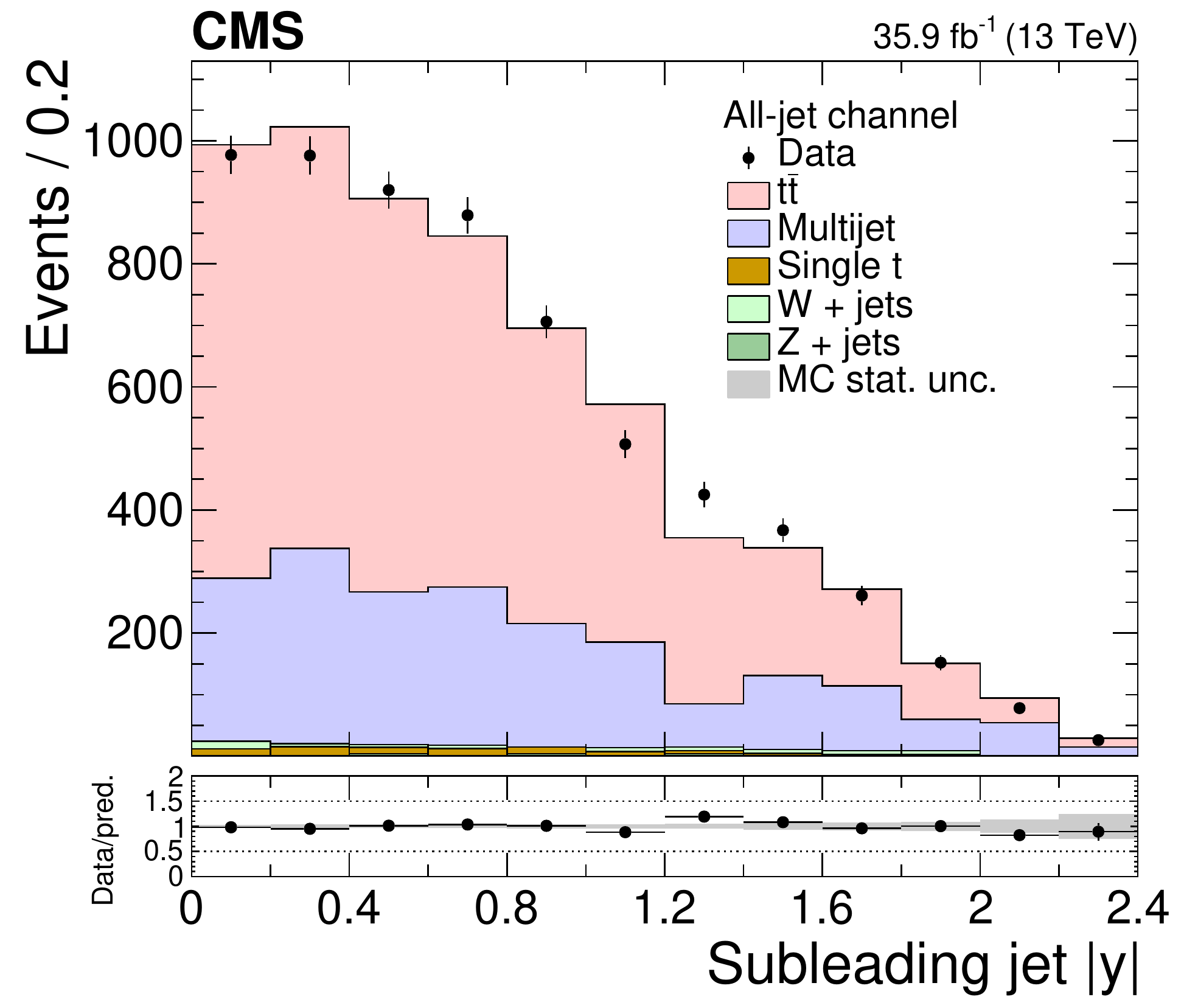}
\caption{Comparison between data and prediction in the signal region SR for the \pt (upper row) and absolute rapidity (lower row) of the leading (left column) and subleading (right column) large-$R$ jets in the all-jet channel. The contributions from \ttbar and multijet production are normalized according to the fitted values of the respective yields and are shown as stacked histograms. The data points are shown with filled circles, while the shaded band represents the statistical uncertainty in the simulation. The lower panel shows the data divided by the sum of the predictions.}
\label{fig:jet}
\end{figure*}

\begin{figure*}[hbtp]
\centering
\includegraphics[width=\cmsFigWidth]{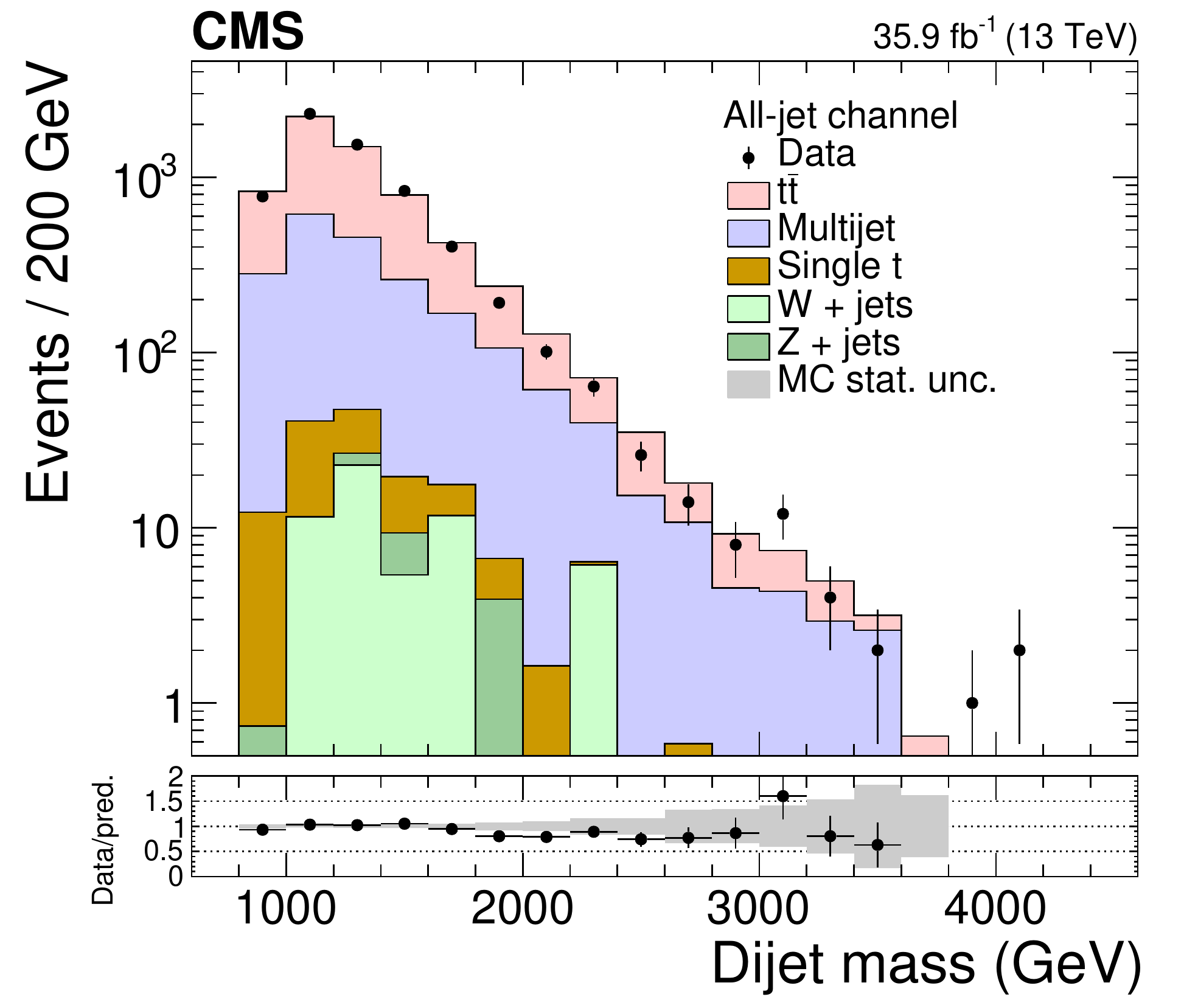}
\includegraphics[width=\cmsFigWidth]{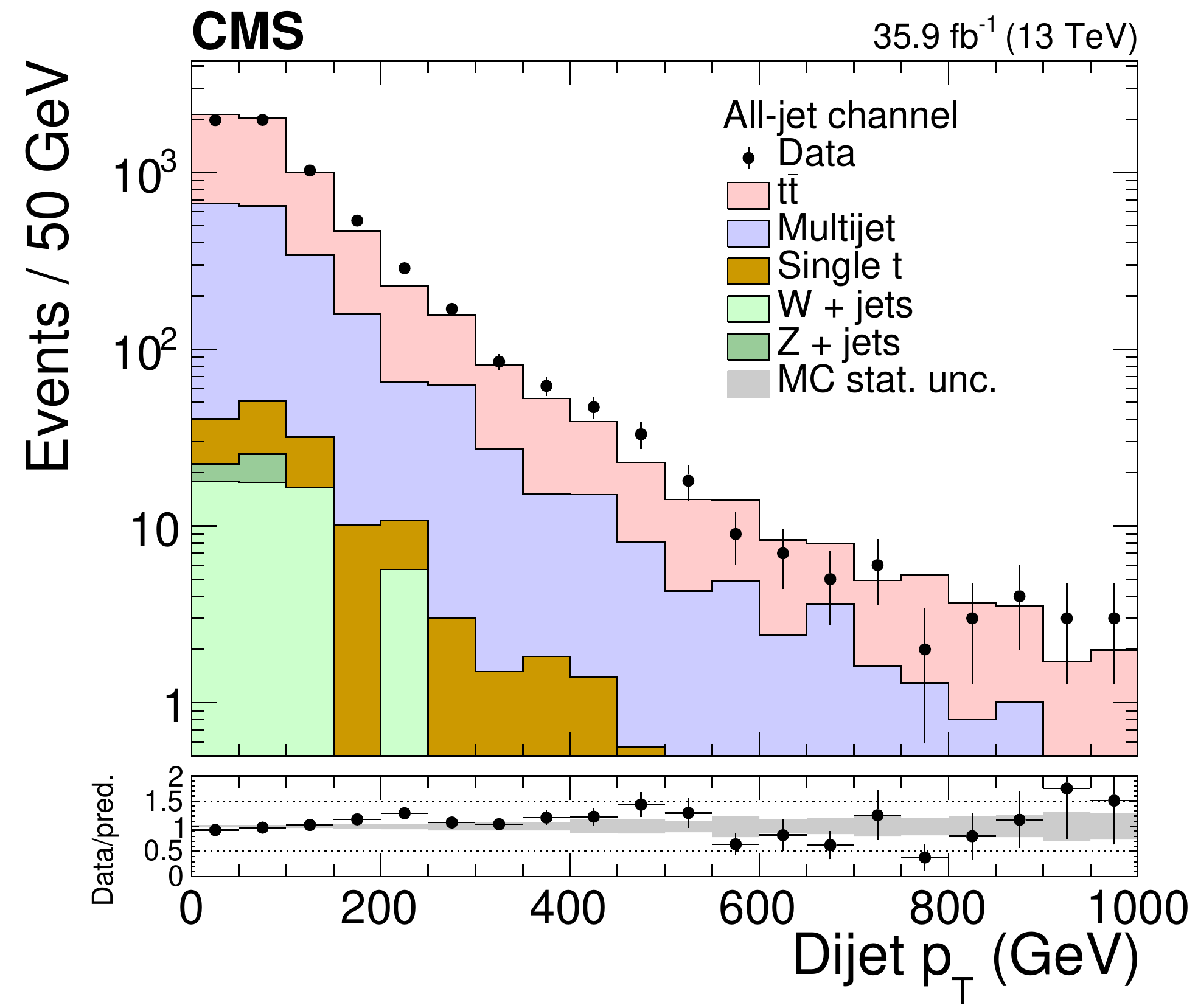}\\
\includegraphics[width=\cmsFigWidth]{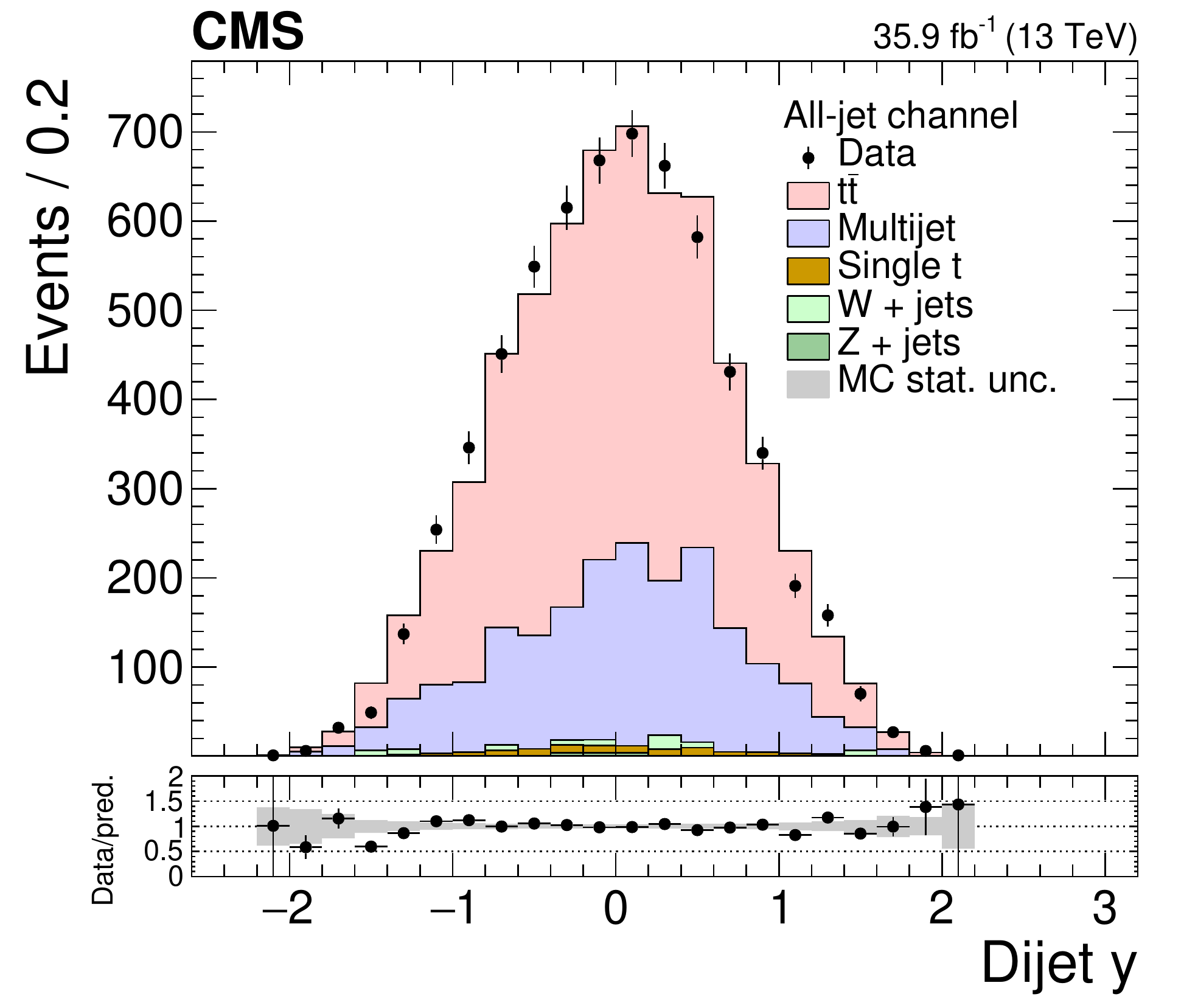}
\caption{Comparison between data and prediction in the signal region SR of the all-jet channel for the kinematic properties of the system of the two leading large-$R$ jets (\ttbar candidates). Specifically, the invariant mass (upper left), \pt (upper right), and rapidity (lower). The contributions from \ttbar and multijet production are normalized according to the fitted values of the respective yields and are shown as stacked histograms. The data points are shown with filled circles, while the shaded band represents the statistical uncertainty in the simulation. The lower panel shows the data divided by the sum of the predictions.}
\label{fig:JJ}
\end{figure*}

\begin{figure*}[hbtp]
\centering
    \includegraphics[width=\cmsFigWidth]{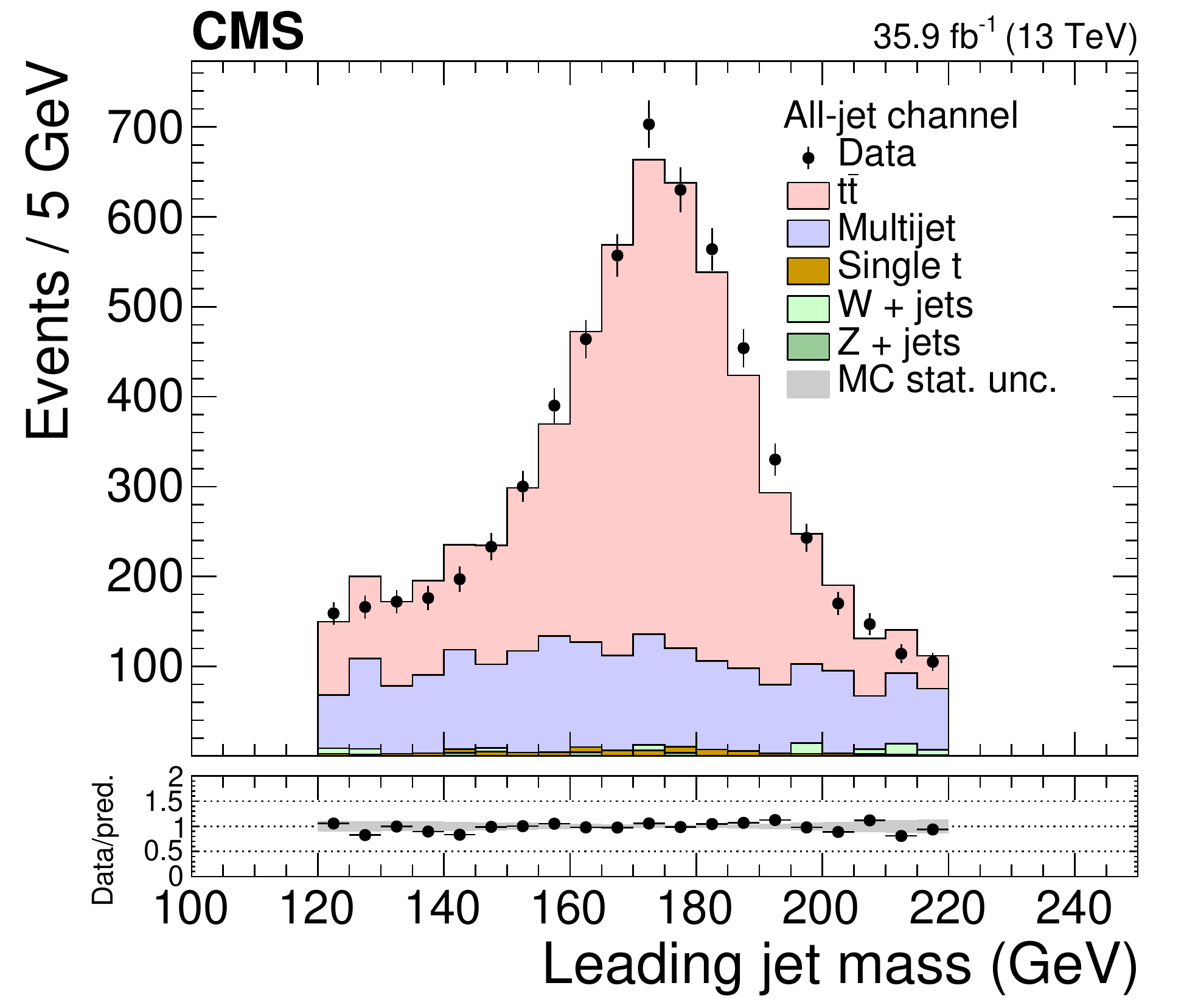}
    \includegraphics[width=\cmsFigWidth]{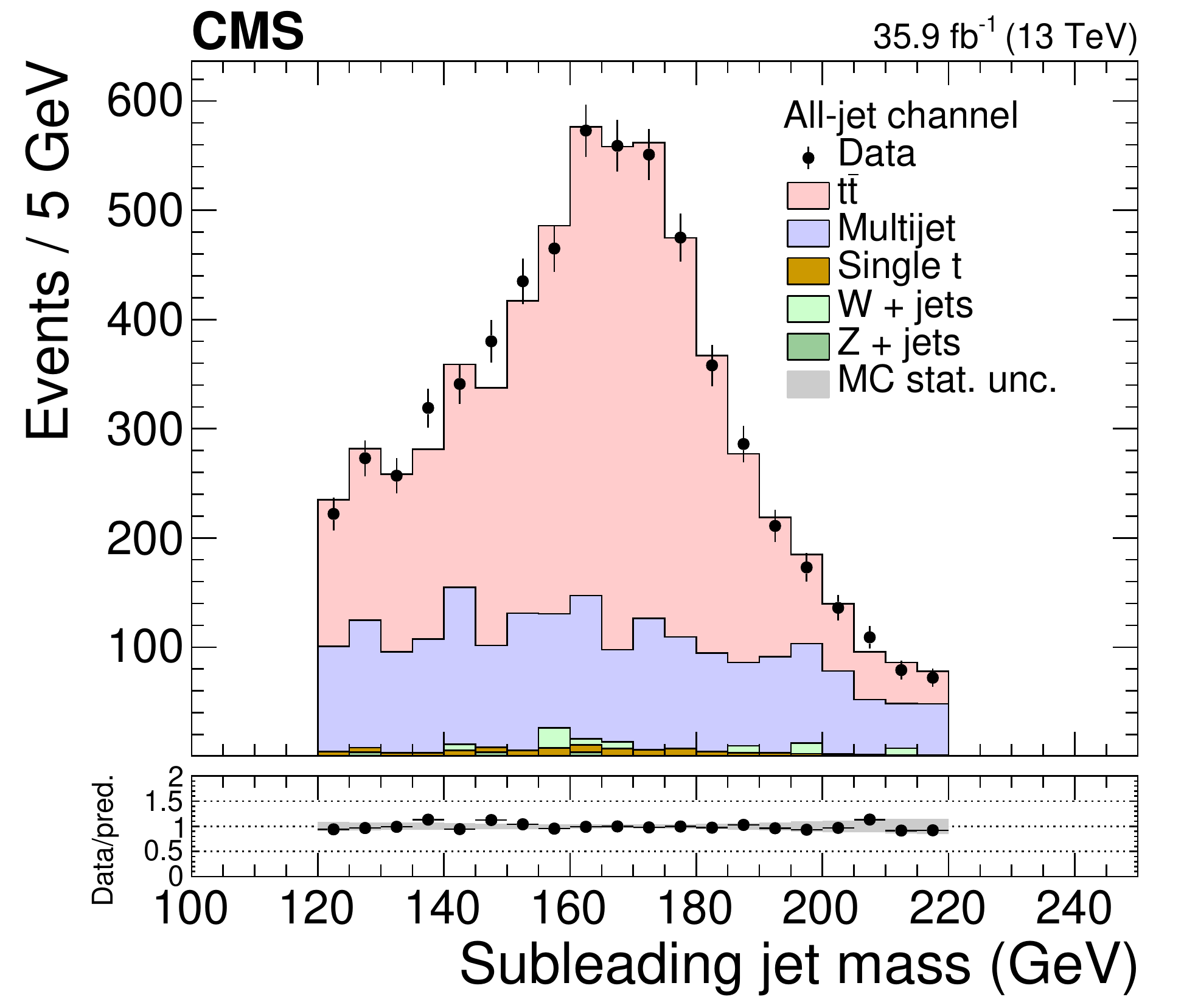}
    \caption{Comparison between data and prediction in the signal region SR for the mass of the leading (left) and subleading (right) large-$R$ jets in the all-jet channel. The \ttbar and multijet production are normalized according to the fitted values of the respective yields and are displayed as stacked histograms. The data points are shown with filled circles, while the shaded band represents the statistical uncertainty in the simulation. The lower panel shows the data divided by the sum of the predictions.}
    \label{fig:jetmass}
\end{figure*}

Figure~\ref{fig:NN} shows the distribution in the NN output in the SR$_{\mathrm{B}}$, and Figs.~\ref{fig:jet} and~\ref{fig:JJ} show the \pt and absolute rapidity $\abs{y}$ of the two top quark candidates and the mass, \pt, and rapidity $y$ of the \ttbar system, respectively. Also, the \mSD values of the two jets are shown in Fig.~\ref{fig:jetmass}. The \ttbar and multijet processes are normalized according to the results of the fit in SR$_{\mathrm{A}}$ described above, while the yields in subdominant backgrounds are taken from simulation. Table~\ref{tab:yields_SR} summarizes the event yields in the SR.

\begin{table}[h]
\setlength{\tabcolsep}{1pt}
\centering
\topcaption{Observed and predicted event yields with their respective statistical uncertainties in the signal region SR for the all-jet channel. The \ttbar and multijet yields are obtained from the fit in SR$_{\mathrm{A}}$.}
\begin{scotch}{lrl}
Process     & \multicolumn{2}{c}{Number of events} \\ [\cmsTabSkip] \hline \noalign{\vskip\cmsTabSkip}
\ttbar          &  ~~~~~~$4244$ & $\pm\,127$ \\
Multijet &  $1876$ & $\pm\,102$\\
Single \PQt      &    $83$   & $\pm\,41$ \\
$\PW$+jets       &    $58$   & $\pm\,29$ \\
$\PZ$+jets        &    $12$   & $\pm\,6$ \\ [\cmsTabSkip]
Total          & $6273$ & $\pm\,171$ \\ [\cmsTabSkip]
Data     & \multicolumn{2}{c}{6274} \\ 
\end{scotch}\label{tab:yields_SR}
\end{table}

In the $\ell$+jets channel, background events from \ttbar other, single top quark, $\PV$+jets, and diboson production are estimated from simulation. The multijet background is modeled using a data sideband region defined by inverting the isolation requirement on the lepton and relaxing the lepton identification criteria. The predicted contributions from signal and other background events are subtracted from the data distribution in the sideband region to obtain the kinematic distributions for multijet events. 
The normalization of the multijet background is extracted from a maximum likelihood fit, discussed in Section~\ref{sec:SignalExtractionLJets}; an initial estimate of its normalization is taken as the simulated prediction. 
The normalizations of the other background processes are also constrained via the fit.

\section{Signal extraction}\label{sec:SignalExtraction}

\subsection{All-jet channel}\label{sec:SignalExtractionAllHad}

In the all-jet channel, the \ttbar signal is extracted from data by subtracting the contribution from the background. The signal is extracted as a function of seven separate variables: \pt and $\abs{y}$ of the leading and subleading \PQt jet, as well as the mass, \pt, and $y$ of the \ttbar system, according to: 
\begin{equation}\label{eq:signal}
S(x)=D(x)-R_\text{yield}N_\text{multijet}Q(x)-B(x),
\end{equation} 
where $x$ corresponds to one of the variables {$\pt^{{\PQt}_i}$, $\abs{y^{{\PQt}_i}}$, $m^{\ttbar}$, $\pt^{\ttbar}$, or $y^{\ttbar}$}, $S(x)$ is the \ttbar signal distribution, $D(x)$ is the measured distribution in data, $Q(x)$ is the multijet distribution, and $B(x)$ is the contribution from the subdominant backgrounds (for which both the distribution and the normalization are taken from simulation). These distributions refer to the SR. The variable $N_\text{multijet}$ is the fitted number of multijet events in the SR$_{\mathrm{A}}$. The factor $R_\text{yield}$ is used to extract the number of multijet events in the SR from $N_\text{multijet}$ and it is found (in simulation) to be independent of the \PQb tagging requirement. This allows its estimate from the multijet control data as $R_\text{yield}\equiv N_\text{multijet}^{\mathrm{SR}}/N_\text{multijet}^{\mathrm{SR_A}}=N_\text{multijet}^{\mathrm{CR}}/N_\text{multijet}^{\mathrm{CR_A}} = 0.38\pm 0.02$. The uncertainty in $R_\text{yield}$ includes the statistical uncertainty of the data and the systematic uncertainty of the method as obtained with simulated events.

\subsection{\texorpdfstring{$\ell$+jets}{ell+jets} channel}\label{sec:SignalExtractionLJets}

In the $\ell$+jets channel, the \ttbar signal strength, the scale factor for the \PQt tagging efficiency, and the background normalizations are extracted through a simultaneous binned maximum-likelihood fit to the data across the different analysis categories. 
The 0t, 1t0b, and 1t1b categories are fitted simultaneously, normalizing each background component to the same cross section in all categories. The resulting fit is expressed in terms of a multiplicative factor, the signal strength~\textit{r}, applied to the input \ttbar cross section. Different variables are used to discriminate the \ttbar signal from the background processes. The small-$R$ jet $\eta$ distribution is used in the 0t and 1t0b categories, while the large-$R$ jet \mSD distribution is used in the 1t1b region.  These distributions were chosen as they provide good discrimination between \ttbar, $\PW$+jets, and multijet production, as \ttbar events tend to be produced more centrally than the background, and the \mSD distribution peaks near the top quark mass. 
The \ttbar signal and \ttbar background contributions merge into a single distribution in the fit, essentially constraining the leptonic branching fraction to equal that provided in the simulation.

Background normalizations and experimental sources of systematic uncertainty are treated as nuisance parameters in the fit. The uncertainties from the pileup reweighting, lepton scale factors, JES, JER, and \PQb and \PQt tagging efficiencies are treated as uncertainties in the input distributions. Two separate nuisance parameters are used to describe the \PQt tagging uncertainty: one for the \PQt tagging scale factor applied to the \ttbar and single top quark ($\PQt\PW$) events, where we expect the $\PQt$-tagged jet to correspond to a genuine top quark, while the \PQt misidentification scale factor is applied to the remaining background. The uncertainties in the integrated luminosity and background normalizations are treated as uncertainties in the production cross sections of the backgrounds. The event categories in the fit are designed such that the \PQt tagging efficiency is constrained by the relative population of events in the three categories. The different admixtures of the signal and background events between the categories provide constraints on the background normalizations. The measurement of the signal strength is correlated with various nuisance parameters, with the strongest correlation being with the \PQt tagging efficiency, as expected.
To determine the uncertainties in distributions, the nuisance parameter is used to interpolate between the nominal distribution and distributions corresponding to ${\pm}1$ standard deviation changes in the given uncertainty.
The uncertainties from theoretical modeling are evaluated independently from the fit.

\begin{figure*}[hbtp]
\centering 
\includegraphics[width=\cmsFigWidth]{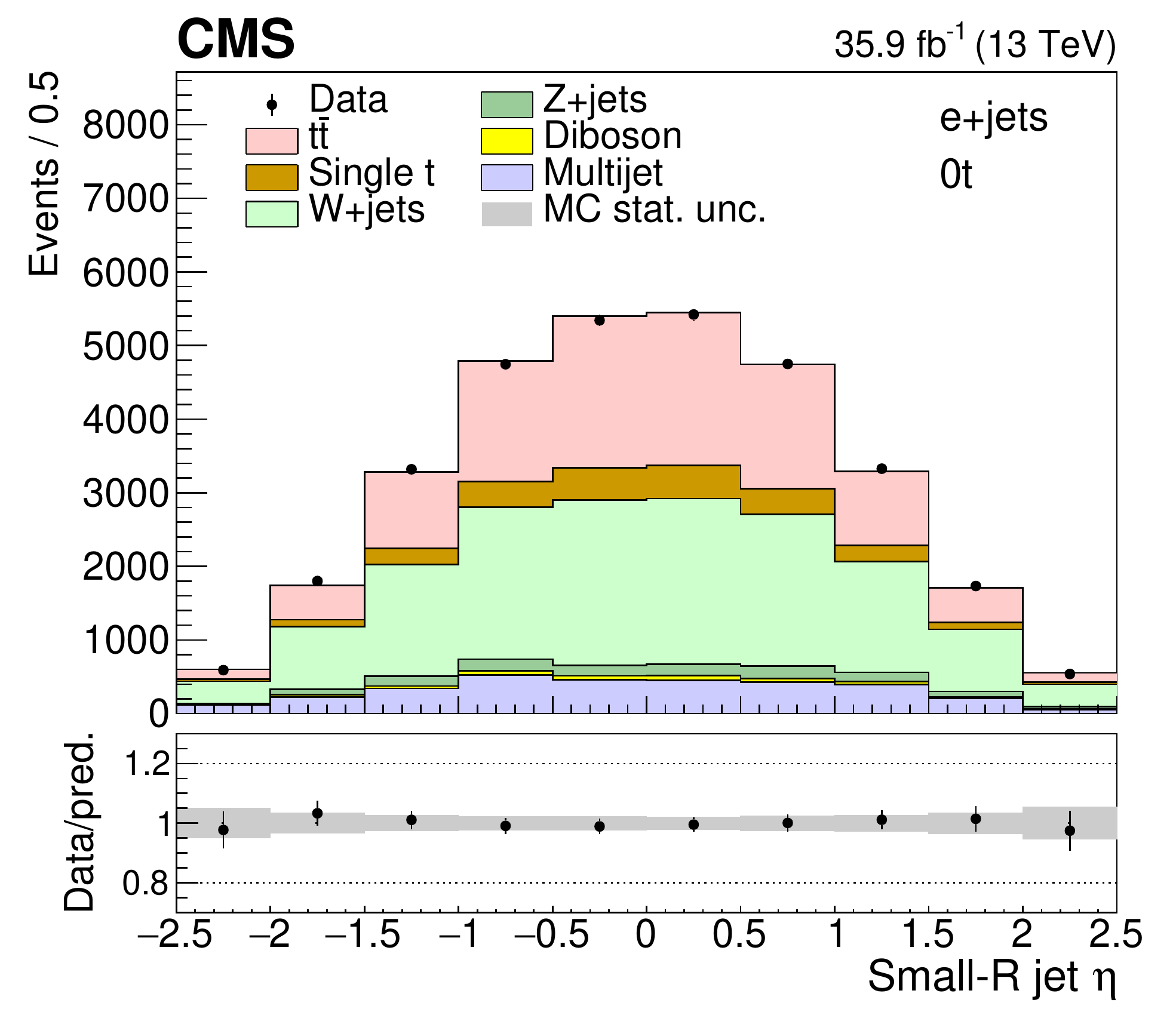}
\includegraphics[width=\cmsFigWidth]{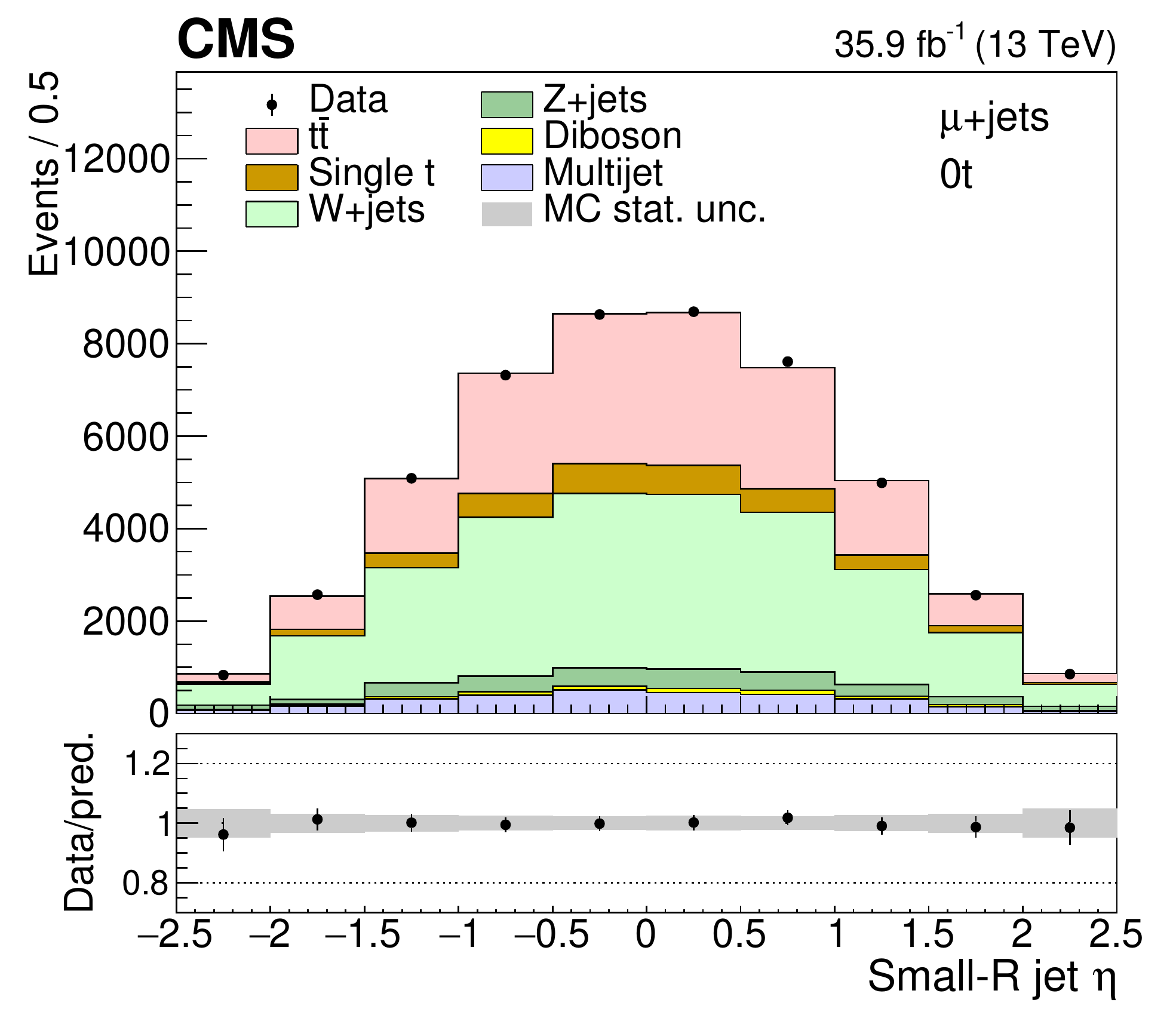} \\
\includegraphics[width=\cmsFigWidth]{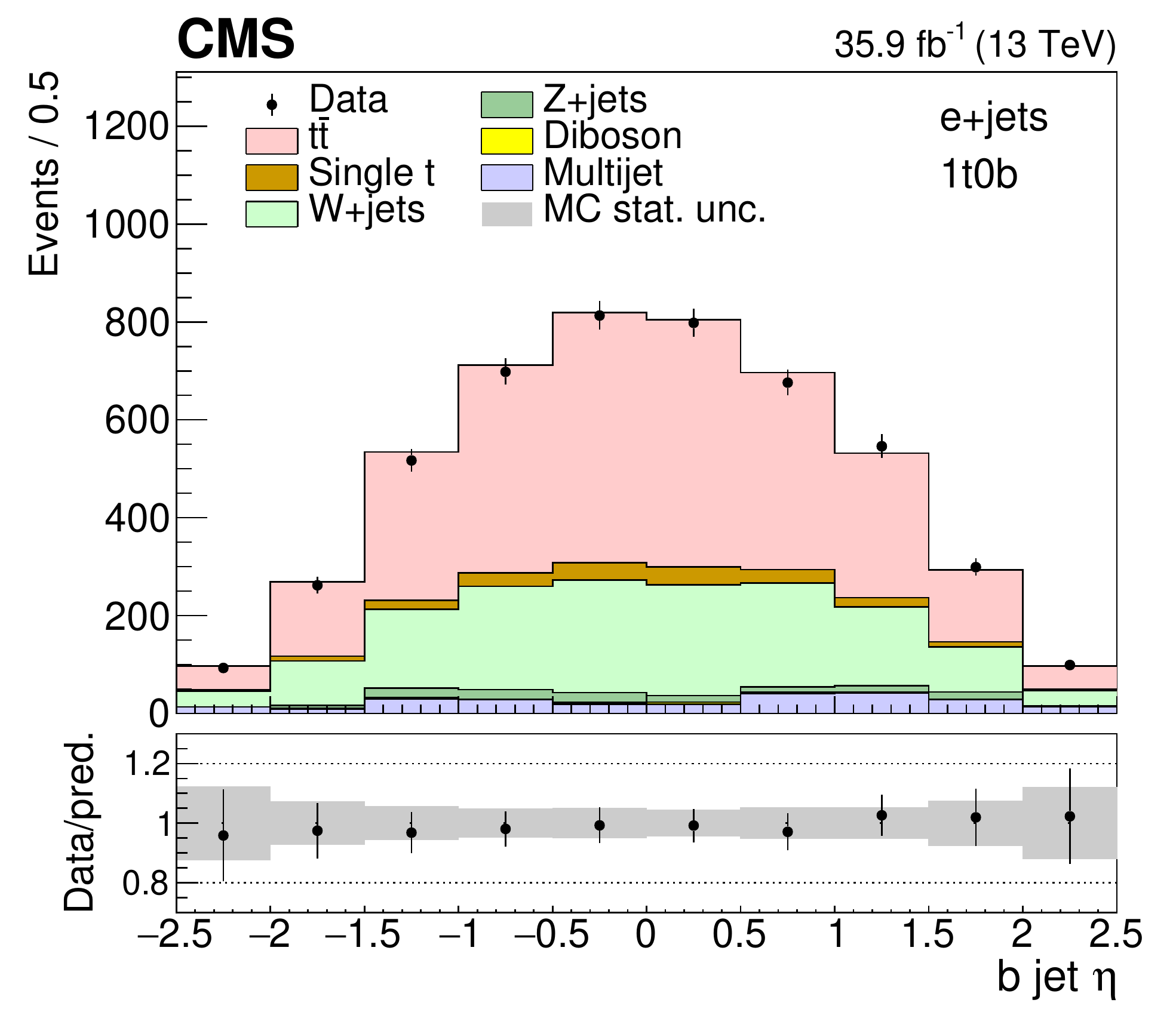}
\includegraphics[width=\cmsFigWidth]{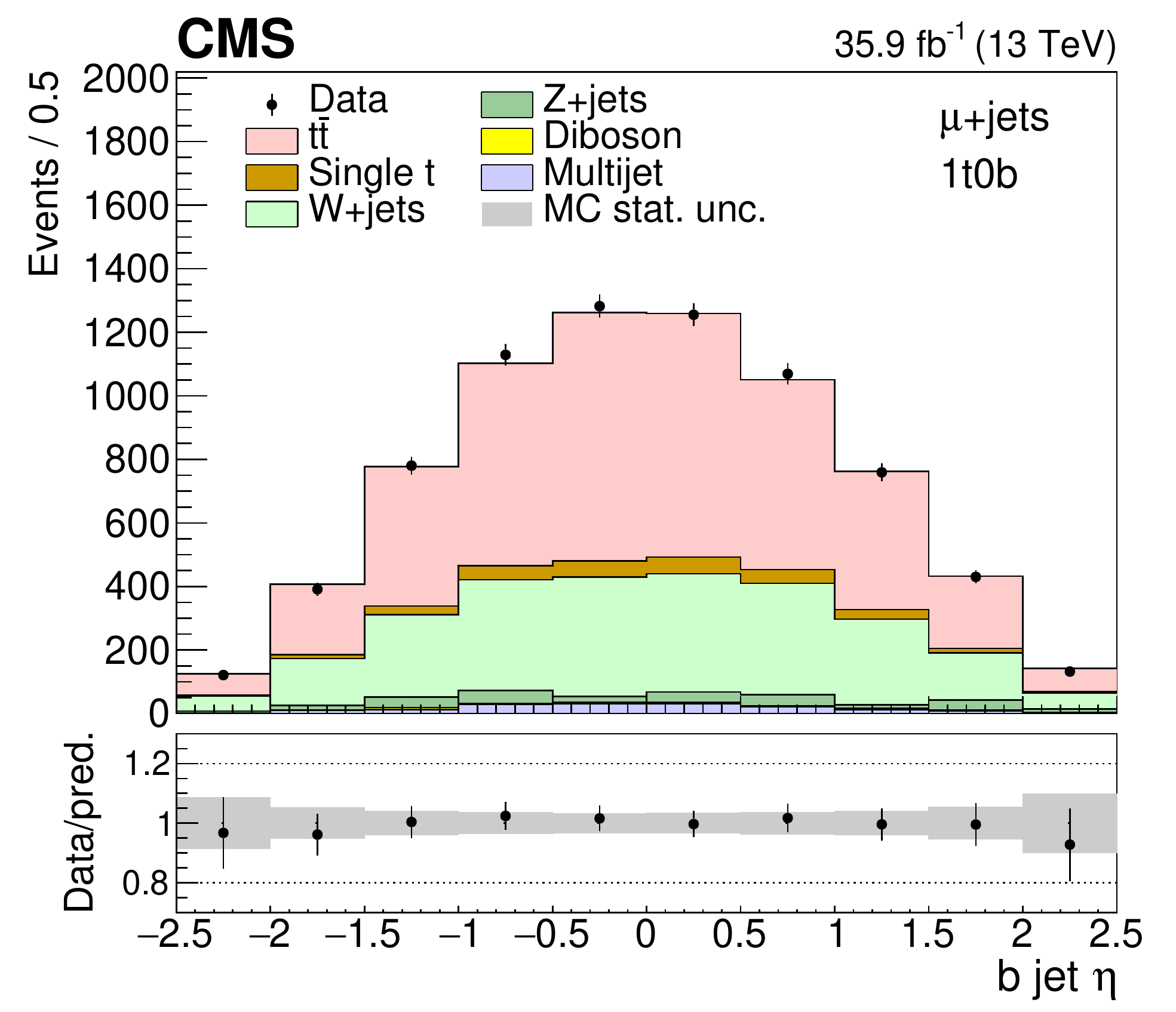} \\
\includegraphics[width=\cmsFigWidth]{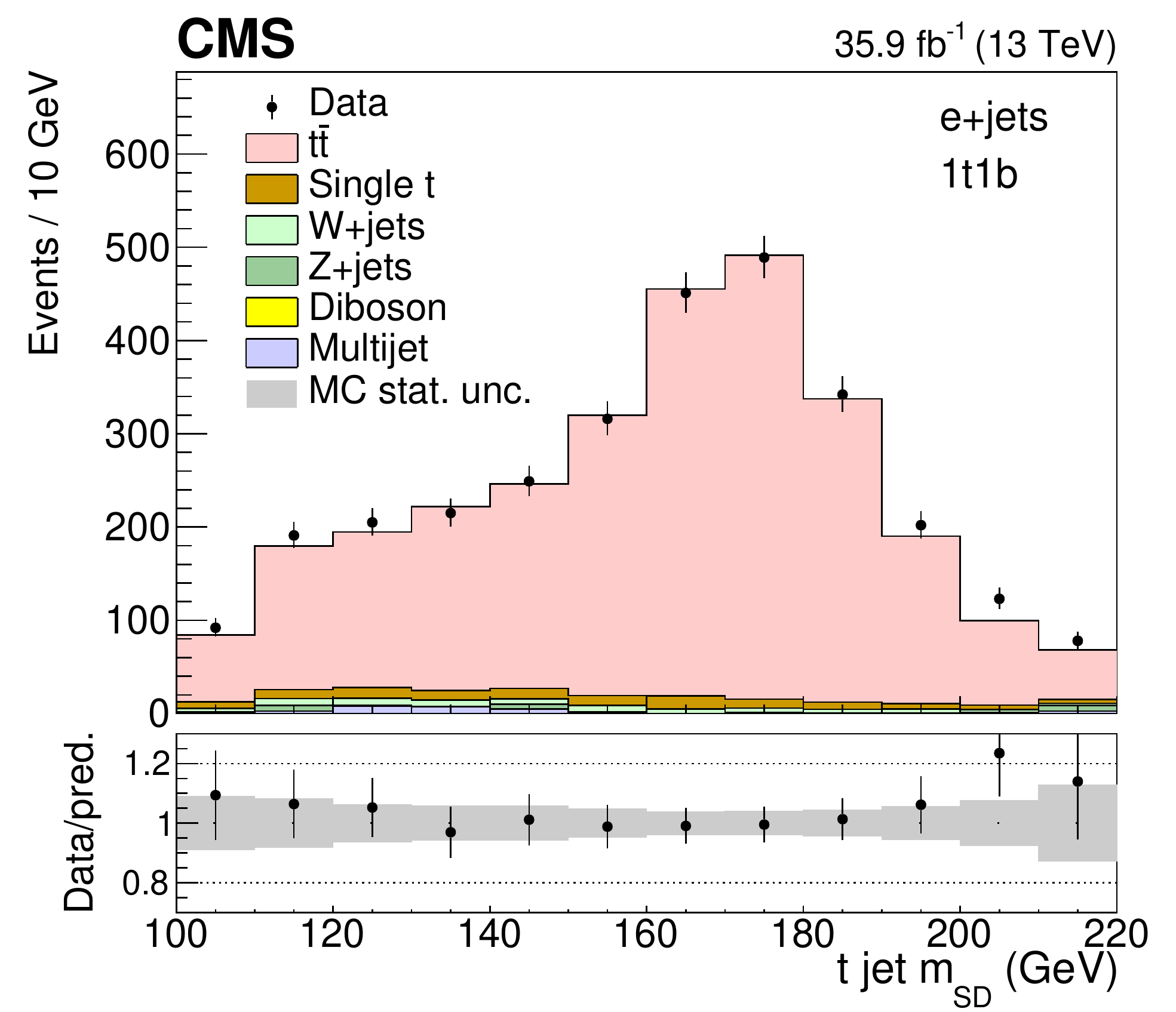}
\includegraphics[width=\cmsFigWidth]{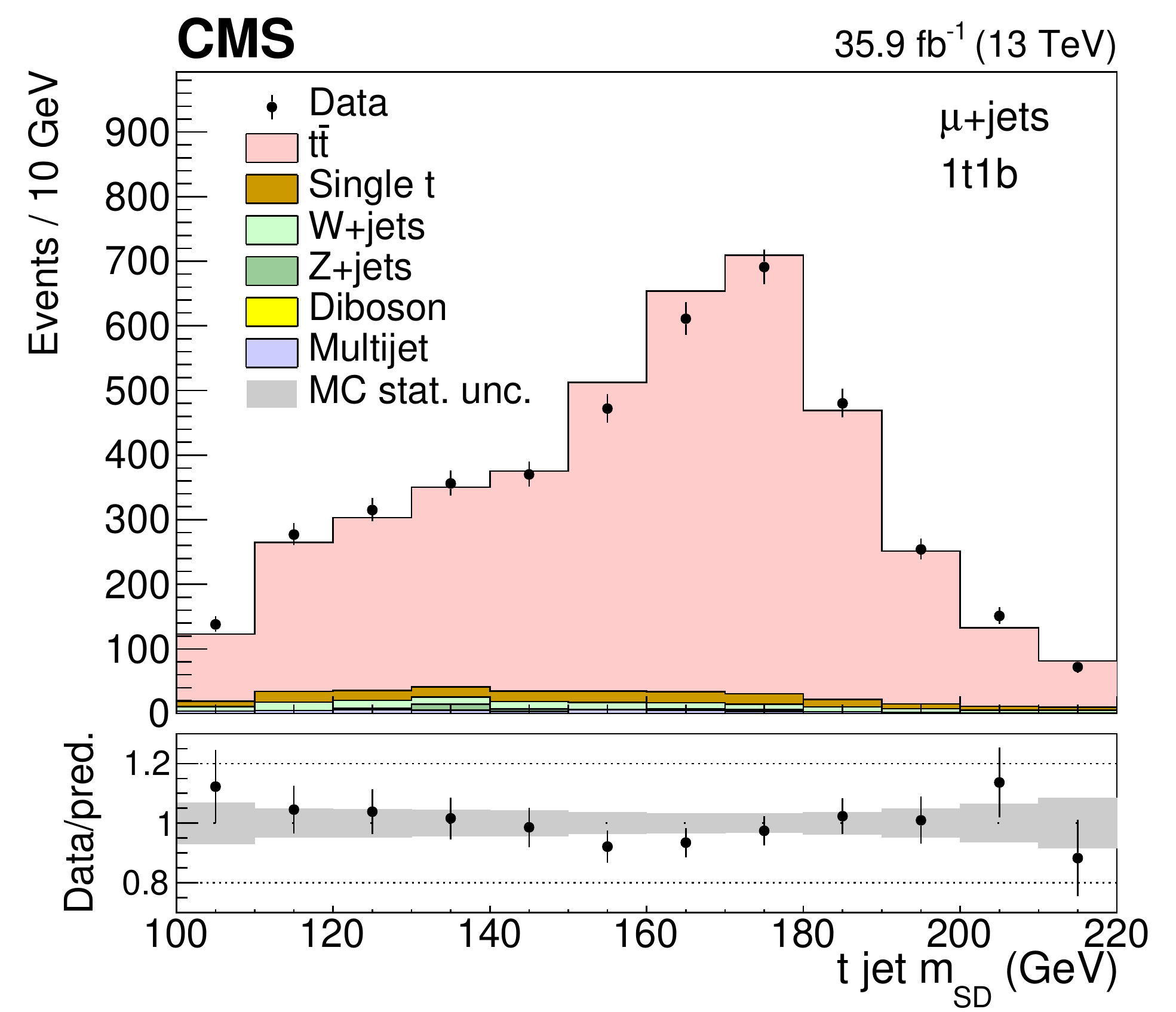}
\caption{\label{fig:combPostFit} Posterior kinematic distributions in the maximum-likelihood fit. Different event categories and variables are fitted: $\eta$ distribution for small-$R$ jets in 0t events (upper row), $\eta$ distribution of the \PQb jet candidate in 1t0b events (middle row), and \mSD of the \PQt jet candidate in 1t1b events (lower row), in the $\Pe$+jets (left column) and $\PGm$+jets (right column) channels. The data points are indicated by filled circles, while the signal and background predictions are shown as stacked histograms. The lower panels show data divided by the sum of the predictions and their systematic uncertainties as obtained from the fit (shaded band).}
\end{figure*}

The fit is performed by minimizing a joint binned likelihood constructed from the kinematic distributions in the $\Pe$+jets and $\PGm$+jets channels, with most nuisance parameters constrained to be identical in both channels. The nuisance parameters associated with the electron and muon scale factors are treated separately, as are the normalizations of the multijet background in the electron and muon channels.
The event yields that account for shifts in all nuisance parameters are given in Table~\ref{tab:combFit_counts}. 
The posterior kinematic distributions for the three event categories are shown in Fig.~\ref{fig:combPostFit}. 

\begin{table}[ht]
\topcaption{\label{tab:combFit_counts} Posterior signal and background event yields in the 0t, 1t0b, and 1t1b categories, together with the observed yields in data. The uncertainties include all posterior experimental contributions.}
\setlength{\tabcolsep}{1pt}
\centering
\begin{scotch}{lrlrlrl}
\multirow{2}{*}{Process} & \multicolumn{6}{c}{Number of events ($\Pe$+jets channel)} \\
            & \multicolumn{2}{c}{0t} & \multicolumn{2}{c}{1t0b} & \multicolumn{2}{c}{1t1b} \\ [\cmsTabSkip] \hline \noalign{\vskip\cmsTabSkip}
\ttbar &  10710 &$\pm$  940 &  2840 &$\pm$ 120 &  2670 &$\pm$  66 \\ 
Single $\PQt$ &  2270 &$\pm$  400 &   191 &$\pm$   47 &   107 &$\pm$   24 \\ 
$\PW$+jets & 13950 &$\pm$ 1740 &  1450 &$\pm$  190 &    62 &$\pm$   12 \\ 
$\PZ$+jets &  1070 &$\pm$  300 &   118 &$\pm$   37 &    17 &$\pm$   15 \\ 
Diboson &  370 &$\pm$  110 &    22 &$\pm$    7 &     2 &$\pm$    1 \\ 
Multijet &  3200 &$\pm$  740 &   242 &$\pm$   80 &    31 &$\pm$   30 \\ [\cmsTabSkip]
Total & ~~31600 &$\pm$ 2200~~ & ~~4850 &$\pm$ 250~~ & ~~2889 &$\pm$ 79~~ \\ [\cmsTabSkip]
Data        &       \multicolumn{2}{c}{31559}      &       \multicolumn{2}{c}{4801}     &       \multicolumn{2}{c}{2953}    \\ 
\end{scotch}
~\\[+5mm]
\begin{scotch}{lrlrlrl}
\multirow{2}{*}{Process} & \multicolumn{6}{c}{Number of events ($\PGm$+jets channel)} \\
            & \multicolumn{2}{c}{0t} & \multicolumn{2}{c}{1t0b} & \multicolumn{2}{c}{1t1b} \\ [\cmsTabSkip] \hline \noalign{\vskip\cmsTabSkip}
\ttbar & 16800 &$\pm$ 1400 &  4250 &$\pm$ 170 & 3905 & $\pm$ 80 \\ 
Single $\PQt$ &   3290 &$\pm$  590 &   282 &$\pm$   68 &   153 &$\pm$   34 \\ 
$\PW$+jets & 23100 &$\pm$ 2900 &  2370 &$\pm$  320 &   105 &$\pm$   20 \\ 
$\PZ$+jets &  2580 &$\pm$  680 &   234 &$\pm$   69 &    19 &$\pm$   10 \\
Diboson &  560 &$\pm$  160 &    31 &$\pm$   10 &     2 &$\pm$    1 \\ 
Multijet &  2800 &$\pm$ 1200 &   159 &$\pm$   76 &    43 &$\pm$   22  \\ [\cmsTabSkip]
Total & ~~49100 &$\pm$ 3500~~ & ~~7320 &$\pm$ 380~~ & ~~4228 &$\pm$ 93~~  \\ [\cmsTabSkip]
Data        &       \multicolumn{2}{c}{49137}      &      \multicolumn{2}{c}{7348}      &       \multicolumn{2}{c}{4187}     \\ 
\end{scotch}
\end{table}

\begin{figure*}[hbtp]
\centering
\includegraphics[width=\cmsFigWidth]{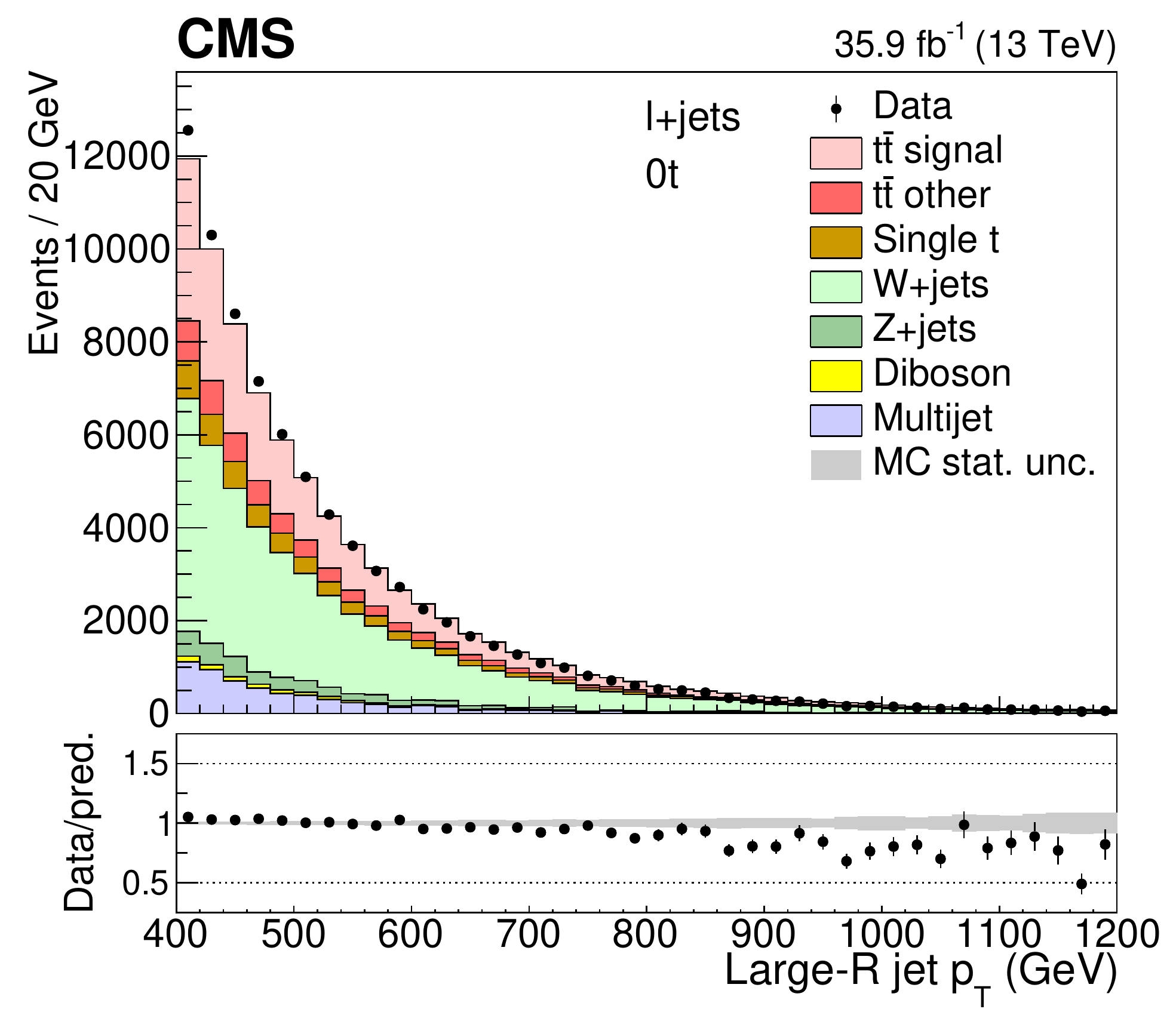}
\includegraphics[width=\cmsFigWidth]{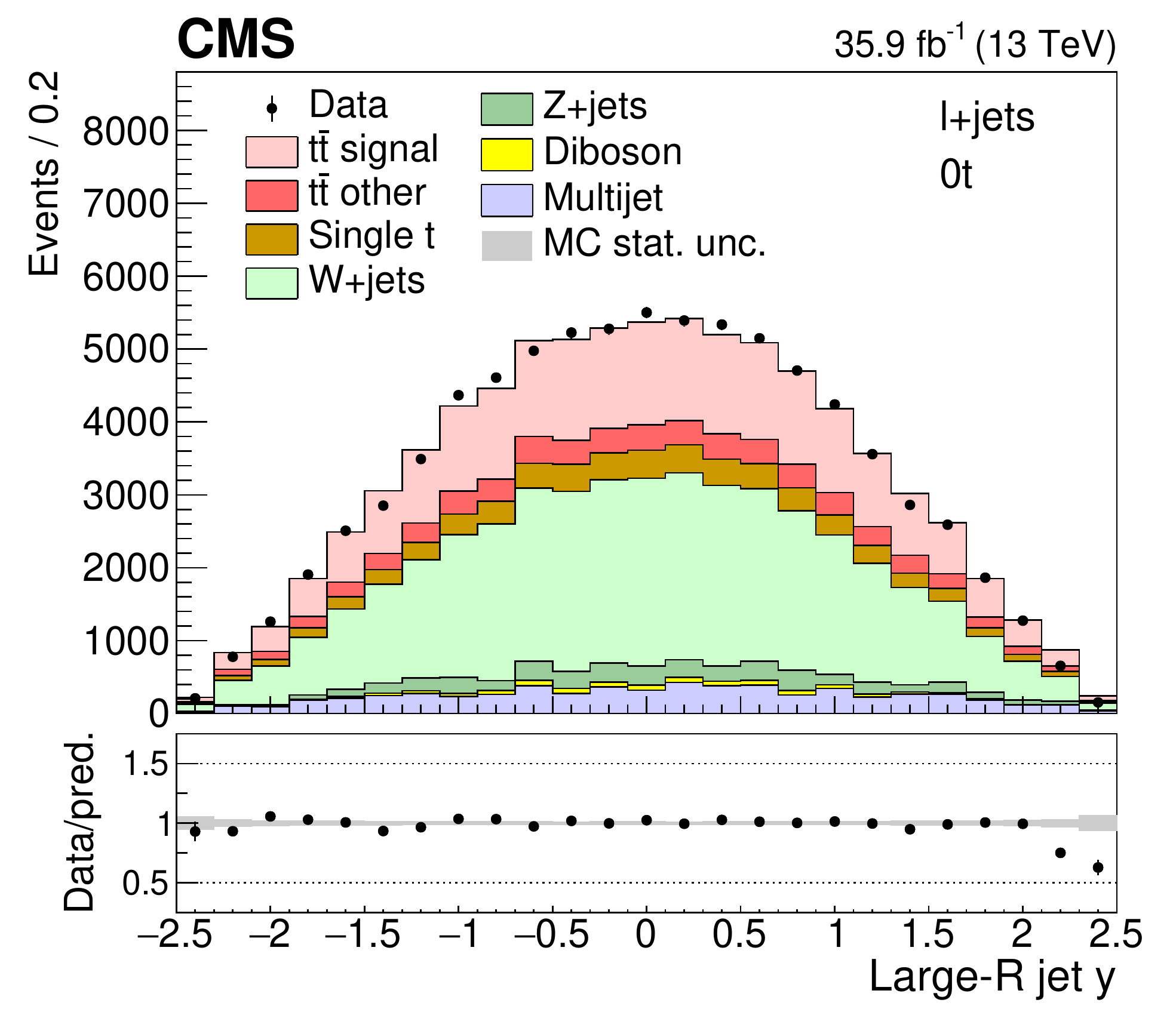} \\
\includegraphics[width=\cmsFigWidth]{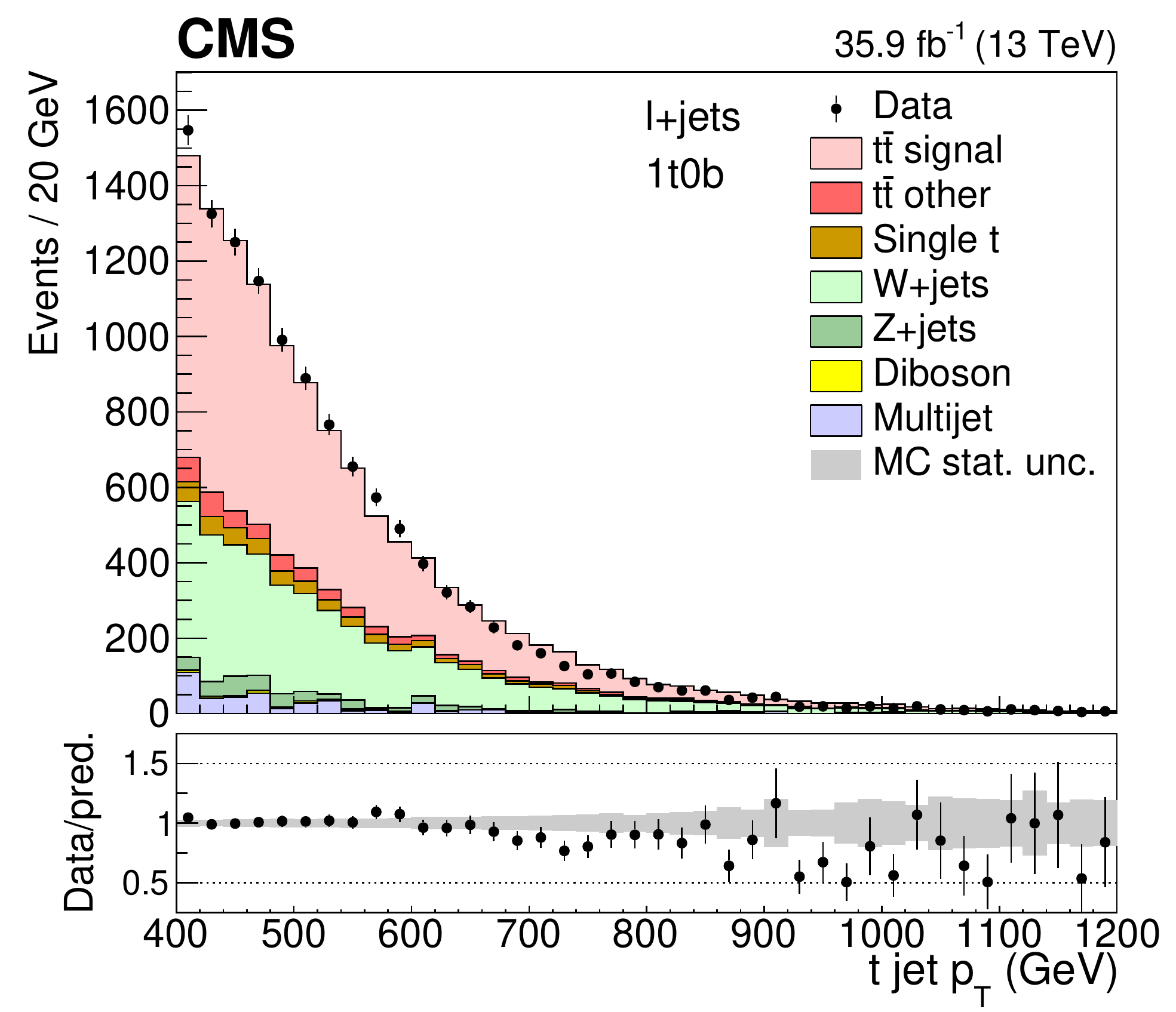}
\includegraphics[width=\cmsFigWidth]{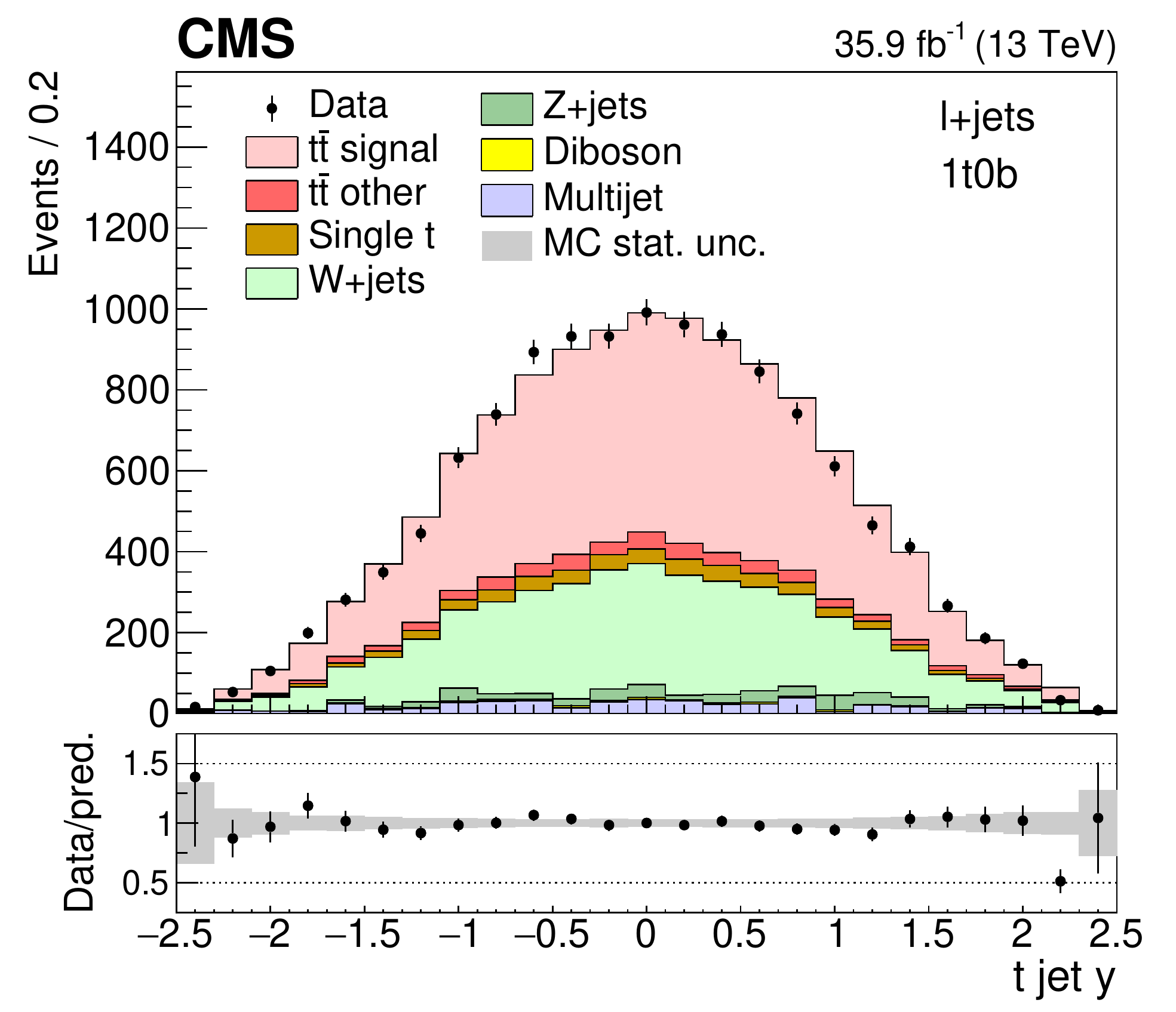}\\
\includegraphics[width=\cmsFigWidth]{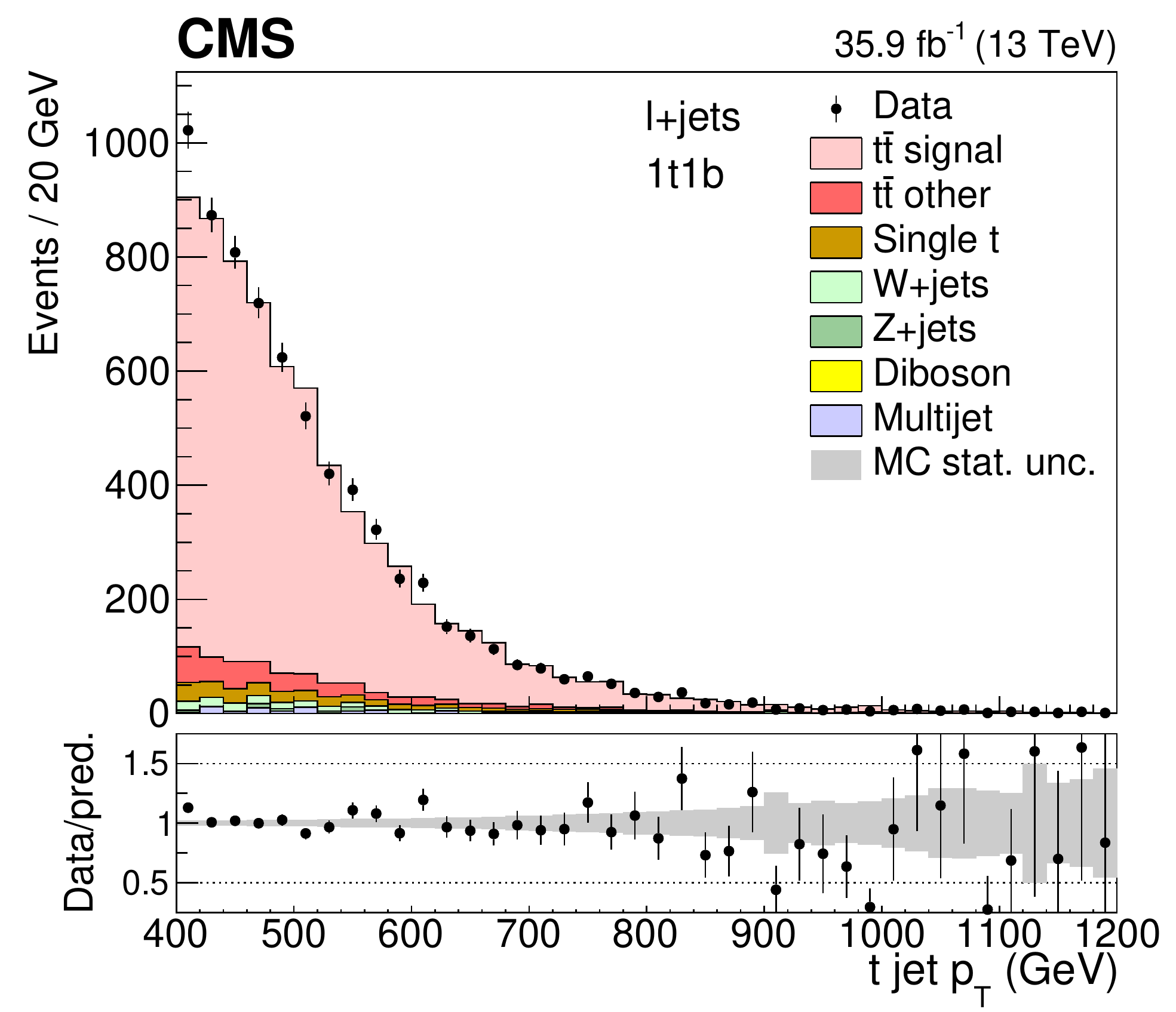}
\includegraphics[width=\cmsFigWidth]{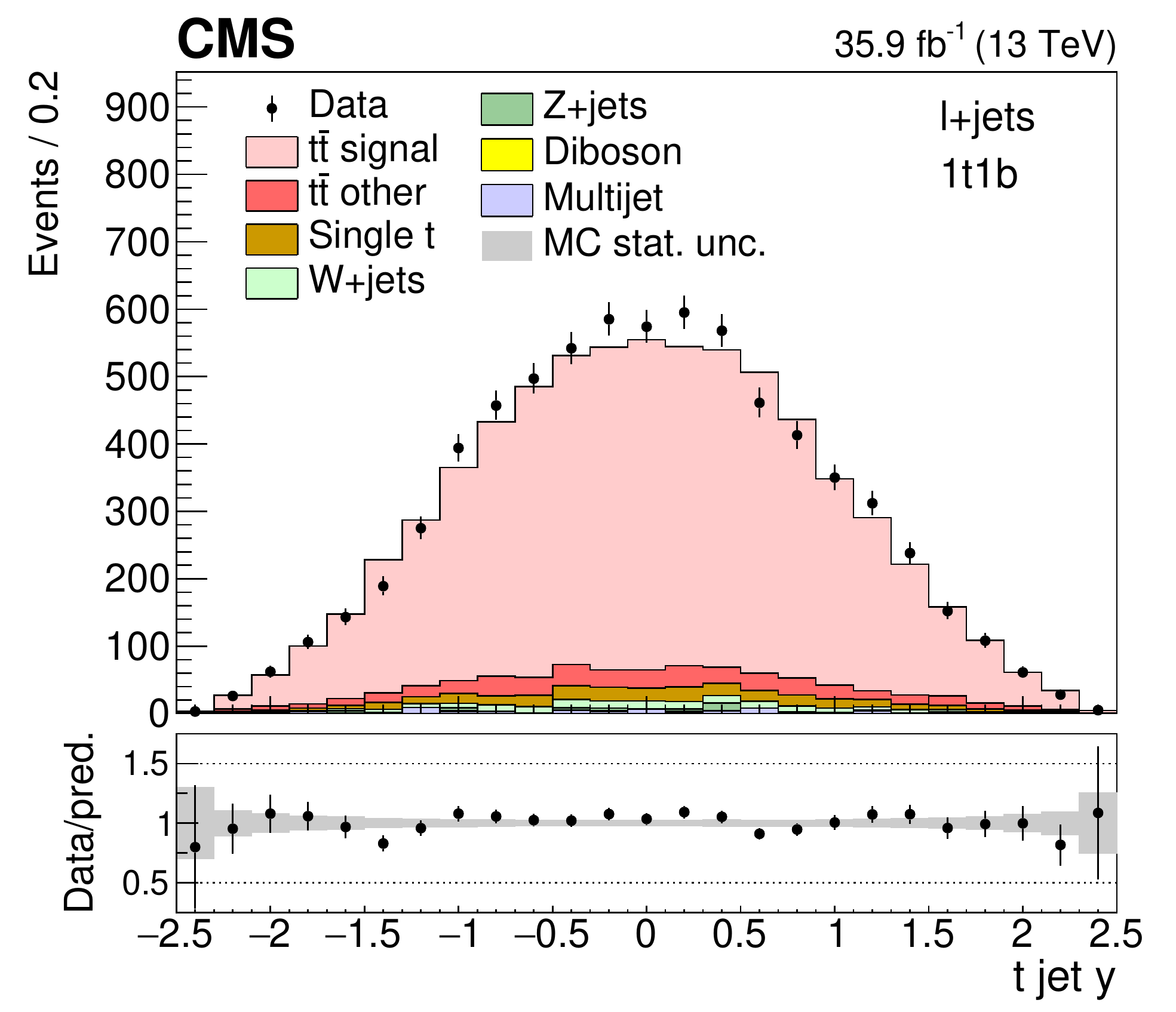}
\caption{\label{fig:topPostFit_lep} Distributions of the \pt (left column) and $y$ (right column) of the \PQt jet candidate for the 0t (upper row), 1t0b (middle row), and 1t1b (lower row) events in the combined $\ell$+jets channel that use the posterior \PQt tag scale factors and background normalizations. The data points are given by the filled circles, while the signal and background predictions are shown as stacked histograms. The lower panels show data divided by the sum of the predictions and their systematic uncertainties as obtained from the fit (shaded band).}
\end{figure*}

Figure~\ref{fig:topPostFit_lep} shows the \pt and $y$ distributions for the \PQt jet candidate in each of the three event categories for the combined $\ell$+jets channel. 
All distributions use the posterior \PQt tagging scale factors and background normalizations, but not the posterior values of other nuisance parameters. 
The posterior \PQt tagging efficiency and misidentification scale factors are $1.04\pm0.06$ and $0.79\pm 0.06$, with an additional \pt- and $\eta$-dependent uncertainty in the ranges of 1--8 and 1--13\%. 
The fitted background normalizations are generally in good agreement with their corresponding pre-fit values.

The posterior signal strength determined in the fit is $0.81 \pm 0.05$, \ie, the \ttbar simulation is observed to overestimate the data by roughly 25\% in the region of the fiducial phase space. The measured signal strength extrapolated from the fit serves as an indicator of the level of agreement between the measured integrated \ttbar cross section and the prediction from simulation.

\section{Systematic uncertainties}\label{sec:Sys} 

The systematic uncertainties originate from both experimental and theoretical sources. The former include all those related to differences in performance in particle reconstruction and identification between data and simulation, as well as in the modeling of background. The latter are related to the MC simulation of the \ttbar signal process and affect, primarily, the unfolded results through the acceptance, efficiency, and migration matrices. Each systematic variation produces a change in the measured differential cross section and that difference, relative to the nominal result, defines the effect of this variation on the measurement. 

The dominant experimental sources of the systematic uncertainty in the all-jet channel are the JES and the subjet \PQb tagging efficiency. In the $\ell$+jets channel, the efficiencies in \PQt and \PQb tagging provide the largest contributions to the uncertainties. The different sources are described below: 
\begin{enumerate}
\item \textit{Multijet background (all-jet):} The fitted multijet yield as well as the uncertainty in $R_\text{yield}$ in Eq.~(\ref{eq:signal}) impact the distribution of the signal events as a function of each variable of interest. These are estimated to be about $1\%$ from a comparison of the distribution in each variable of the SR with its CR (as described in Section~\ref{sec:Reco}) in simulated events, as well as for different pileup profiles in data collected with the control trigger relative to the signal trigger. The uncertainty in $R_\text{yield}$ is dominated by the assumption of the extraction method (estimated through simulated events), while the statistical contribution is smaller.
\item \textit{Subdominant backgrounds (all-jet):} The expected yield from the subdominant backgrounds estimated from simulation (single top quark production and vector bosons produced in association with jets) is changed by ${\pm} 50\%$, leading to a negligible uncertainty (${<} 1\%$).
\item \textit{Background estimate ($\ell$+jets):} An a priori uncertainty of 30\% is applied to the single top quark and $\PW$+jets background normalizations, to cover a possible mismodeling of these background sources in the region of phase space probed in the analysis. An additional uncertainty in flavor composition of the $\PW$+jets process is estimated by changing the light- and heavy-flavor components independently by their 30\% normalization uncertainties. For the multijet normalization, an a priori uncertainty of 50\% is used to reflect the combined uncertainty in the normalization and the extraction of the kinematic contributions from the sideband region in data. These background sources and the corresponding systematic uncertainties are all constrained in the maximum likelihood fit. 
\item \textit{JES:} The uncertainty in the energy scale of each reconstructed large-$R$ jet is a leading experimental contribution in the all-jet channel. It is divided into 24 independent sources~\cite{Khachatryan:2016kdb} and each change is used to provide a new jet collection that affects the repeated event interpretation. This results not only in changes in the \pt scale, but can also lead to different \PQt jet candidates. The \pt- and $\eta$-dependent JES uncertainty is about $1$--$2\%$ per jet. The resulting uncertainty in the measured cross section is typically about $10\%$ but can be much larger at high top quark \pt. 
For the $\ell$+jets channel, the uncertainty in JES is estimated for both small-$R$ and large-$R$ jets by shifting the jet energy in simulation up or down by their \pt- and $\eta$-dependent uncertainties, with a resulting impact on the differential cross section of 1--10\%. 
\item \textit{JER:} The impact on the JER is determined by smearing the jets according to the JER uncertainty~\cite{Khachatryan:2016kdb}. The effect on the cross section is relatively small, at the level of 2\%.
\item \textit{\PQt tagging efficiency ($\ell$+jets):} The \PQt tagging efficiency and its associated uncertainty are extracted simultaneously with the signal strength and background normalizations in the likelihood fit of the $\ell$+jets analysis, discussed in Section~\ref{sec:SignalExtraction}. The uncertainty in the \PQt tagging efficiency is in the range 6--10\%, while for the misidentification rate it is 8--15\%, depending on the \pt and $\eta$ of the \PQt jet. 
\item \textit{Subjet \PQb tagging efficiency (all-jet):} The uncertainty in the identification of \PQb subjets within the large-$R$ jets (estimated in Ref.~\cite{BTV-16-002}) is the leading experimental uncertainty in the all-jet channel. The effect on the cross sections is about 10\%, relatively independent of the observables. Unlike the uncertainty associated with JES, the $\PQb$-subjet tagging uncertainty largely cancels in the normalized cross sections.
\item \textit{\PQb tagging efficiency ($\ell$+jets):} For the $\ell$+jets channel, the small-$R$ jet \PQb tagging efficiency in the simulation is corrected to match that measured in data using \pt- and $\eta$-dependent scale factors~\cite{BTV-16-002}. The resulting uncertainty in the differential cross sections is about 1--2\%. The \PQb tagging efficiency and non-$\PQb$ jet misidentification uncertainties are treated as fully correlated.
\item \textit{Pileup:} The uncertainty related to the pileup modeling is subdominant. The impact on the measurement is estimated by changing the total inelastic cross section used to reweight the simulated events by ${\pm} 4.6\%$~\cite{Aaboud:2016mmw}. The effect on the cross sections is negligible (${<} 1\%$).
\item \textit{Trigger (all-jet):} The uncertainty associated with the trigger, accounting for the difference between the simulated and observed trigger efficiency, is well below 1\% in the phase space of the all-jet channel. The measurement of the trigger efficiency is performed in events collected with an orthogonal trigger that requires the presence of an isolated muon with \pt greater than 27\GeV.
\item \textit{Lepton identification and trigger ($\ell$+jets):} The performance of the lepton identification, reconstruction, trigger, and isolation constitutes a small source of systematic uncertainty. Correction factors used to modify the simulation to match the efficiencies observed in data are estimated through a tag-and-probe method using $\PZ \to \ell\ell$ decays. The corresponding uncertainty is determined by changing the correction factors up or down by their uncertainties. The resulting systematic uncertainties depend on lepton \pt and $\eta$, and are in the range 1--7 (1--5)\% for electrons (muons).
\item \textit{Integrated luminosity:} The uncertainty in the measurement of the integrated luminosity is 2.5\%~\cite{CMS-PAS-LUM-17-001}. 
\end{enumerate}

The theoretical uncertainties are divided into two sub-categories: sources of systematic uncertainty related to the matrix element calculations of the hard scattering process and sources related to the modeling of the parton shower and the underlying event. The first category (consisting of the first three sources below) is evaluated using variations of the simulated event weights, while the second category is evaluated with dedicated, alternative MC samples with modified parameters. These sources are:
\begin{enumerate}
\item \textit{Parton distribution functions:} The uncertainty from PDFs is estimated by applying event weights corresponding to the 100 replicas of the NNPDF PDFs~\cite{Ball:2014uwa}. For each observable we compute its standard deviation from the 100 variants.
\item \textit{QCD renormalization and factorization scales:} This source of systematic uncertainty is estimated by applying event weights corresponding to different renormalization and factorization scale options. Both scales are changed independently by a factor of two up or down in the event generation, omitting the two cases where the scales are changed in opposite directions, and taking the envelope of the six results.
\item \textit{Strong coupling (\alpS):} The uncertainty associated with \alpS is estimated by applying event weights corresponding to higher or lower values of \alpS for the matrix element using the changed NNPDF PDFs~\cite{Ball:2014uwa} values of $\alpS=0.117$ or 0.119, compared to the nominal value 0.118.
\item \textit{ISR and FSR:} The uncertainty in the ISR and FSR is estimated from alternative MC samples with reduced or increased values of \alpS used in \PYTHIA to generate that radiation. The scale in the ISR is changed by factors of 2 and 0.5, and the scale in the FSR by factors of $\sqrt{2}$ and $1/\sqrt{2}$~\cite{Skands_2014}. In the all-jet channel, the FSR uncertainty is constrained by a fit to the data in SR$_{\mathrm{B}}$, using the NN output that is sensitive to the modeling of FSR. This leads to a reduced uncertainty that is 0.3 times the variations from the alternative MC samples.  
\item \textit{Matching of the matrix element to the parton shower}: In the \POWHEG matching of the matrix element to the parton shower (ME-PS), the resummed gluon damping factor $h_{\text{damp}}$ is used to regulate high-\pt radiation. The nominal value is $h_{\text{damp}}=1.58m_{\PQt}$. Uncertainties in $h_{\text{damp}}$ are parameterized by considering alternative simulated samples with $h_{\text{damp}}=m_{\PQt}$ and $h_{\text{damp}}=2.24m_{\PQt}$~\cite{CMS-PAS-TOP-16-021}.
\item \textit{Underlying event tune:} This uncertainty is estimated from alternative MC samples using the CUETP8M2T4 parameters varied by ${\pm}1$ standard deviation~\cite{CMS-PAS-TOP-16-021}.
\end{enumerate}

\section{Cross section measurements}\label{sec:Meas}

Here, we discuss the differential \ttbar production cross sections measured in the all-jet and $\ell$+jets channels as a function of different kinematic variables of the top quark or \ttbar system, corrected to the particle and parton levels using an unfolding procedure. The measurements are compared to predictions from different MC event generators. 

\subsection{Definition of particle and parton levels}
\label{sec:definitionLevels}

The parton-level phase space to which the measurement is unfolded is constrained by the kinematic requirements of the detector-level fiducial region. Namely, in the all-jet decay channel, the \PQt and \PAQt must have $\pt>400\GeV$ and $\abs{\eta}<2.4$. In addition, $m^{\ttbar}>800\GeV$ is required to avoid extreme events with large top quark \pt and small $m^{\ttbar}$. 

The parton-level definition for the $\ell$+jets channel differs in that it is defined for $\ell$+jets events, where one top quark decays according to $\PQt \to \PW \PQb \to \PQq \PAQq' \PQb$ and has $\pt > 400\GeV$ to match the fiducial requirement at the detector level, and the other top quark decays as $\PQt \to \PW \PQb \to \ell \PGn \PQb$ without any \pt requirement. 

The so-called particle level represents the state of quasi-stable particles with a mean lifetime greater than 30\unit{ps} originating from the $\Pp\Pp$ collision after hadronization but before the interaction of these particles in the detector. The observables computed from the momenta of particles are typically better defined than those computed from parton-level information. Also, the associated phase space is closer to the fiducial phase space of the measurement at the detector level, which provides smaller theoretical uncertainties. In the context of this analysis, particle jets are reconstructed from quasi-stable particles, excluding neutrinos, using the anti-\kt algorithm with a distance parameter of 0.8---identical to reconstruction at detector level---and just the particles originating from the primary interaction. Subsequently, jets that are geometrically matched to generated leptons within $\Delta R<0.4$ in $\eta$-$\phi$ (\ie, from the leptonic decays of \PW bosons) are removed from the particle jet collection. 

For the all-jet channel, the two particle jets with highest \pt are considered the particle-level \PQt jet candidates. To match the fiducial phase space as closely as possible, the same kinematic selection criteria are applied as for the detector-level events. In particular, the particle-level jets must have $\pt>400\GeV$ and $\abs{\eta}<2.4$, while the mass of each jet must be in the 120--220\GeV range and the invariant mass of the two jets be greater than 800\GeV. The matching efficiency between the particle-level \PQt jet candidates and the original top quarks at the parton level lies between 96 and 98\%.

The particle-level phase space for the $\ell$+jets channel is set up to mimic the kinematic selections at the detector level. Particle-level large-$R$ jets are selected if they fulfill $\pt > 400\GeV$, $\abs{\eta}<2.4$, and the jet mass is in the range 105--220\GeV, and are then referred to as particle-level \PQt jets. Particle-level small-$R$ jets are selected if they have $\pt > 50\GeV$, $\abs{\eta}<2.4$, and are flagged as \PQb jets (contain a \PQb hadron); these are referred to as particle-level \PQb jets. Particle-level electrons and muons are selected if they have $\pt > 50\GeV$ and $\abs{\eta}<2.1$. To fulfill the particle-level selection criteria, an event must contain at least one \PQt jet, at least one \PQb jet, and at least one electron or muon, all at the particle level. 

\begin{figure*}[hbtp]
\centering
    \includegraphics[width=\cmsFigWidth]{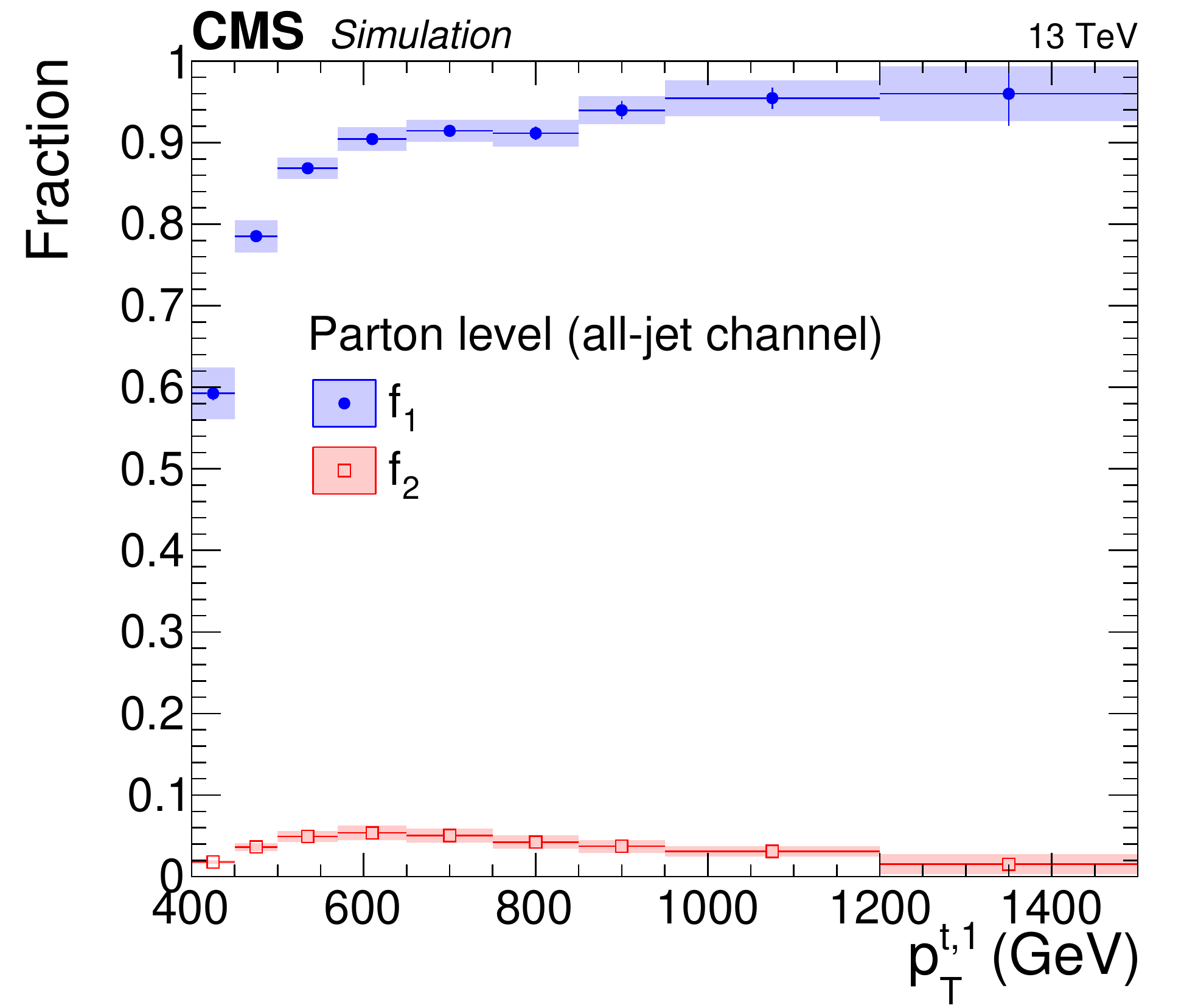}
    \includegraphics[width=\cmsFigWidth]{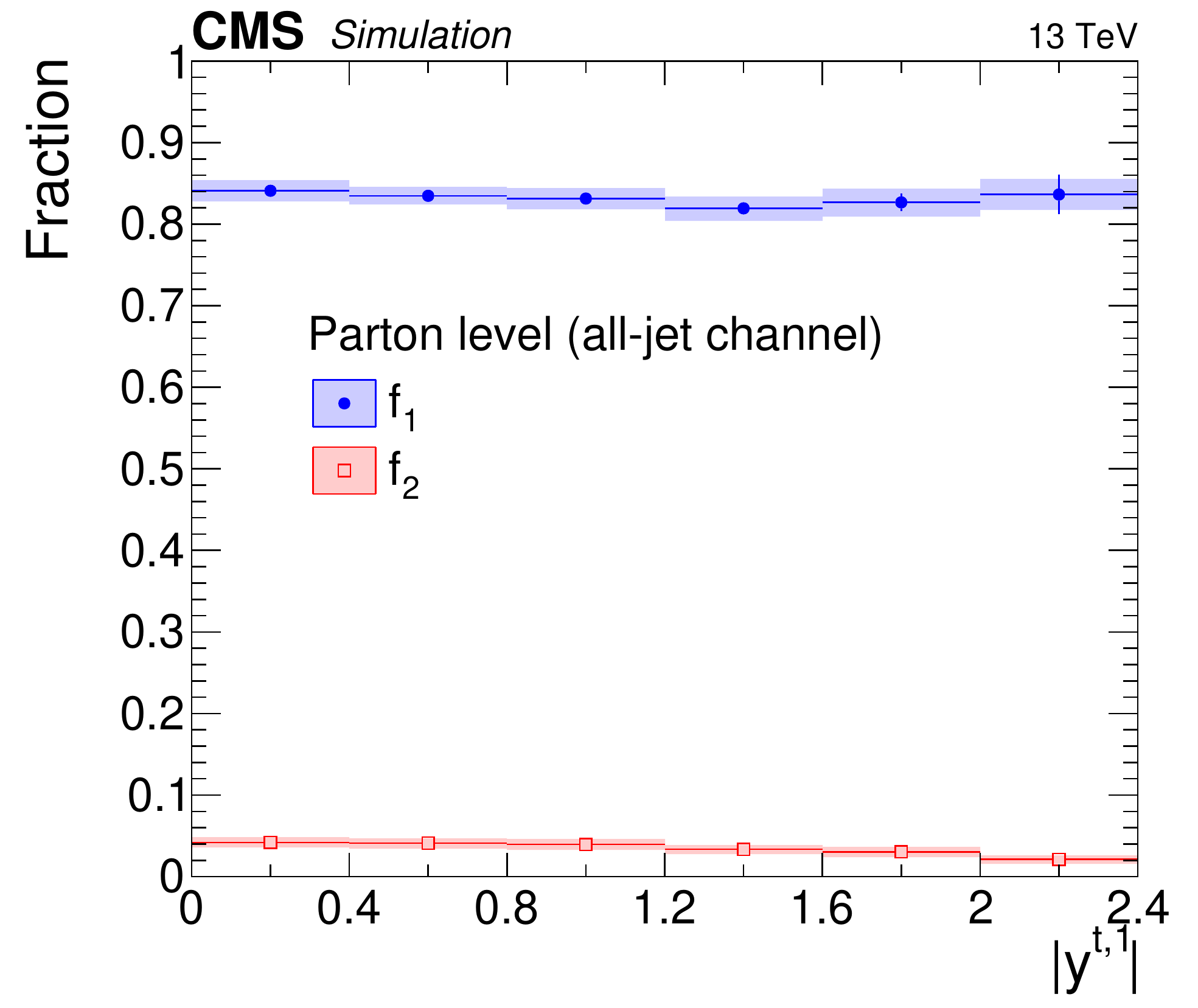} \\
    \includegraphics[width=\cmsFigWidth]{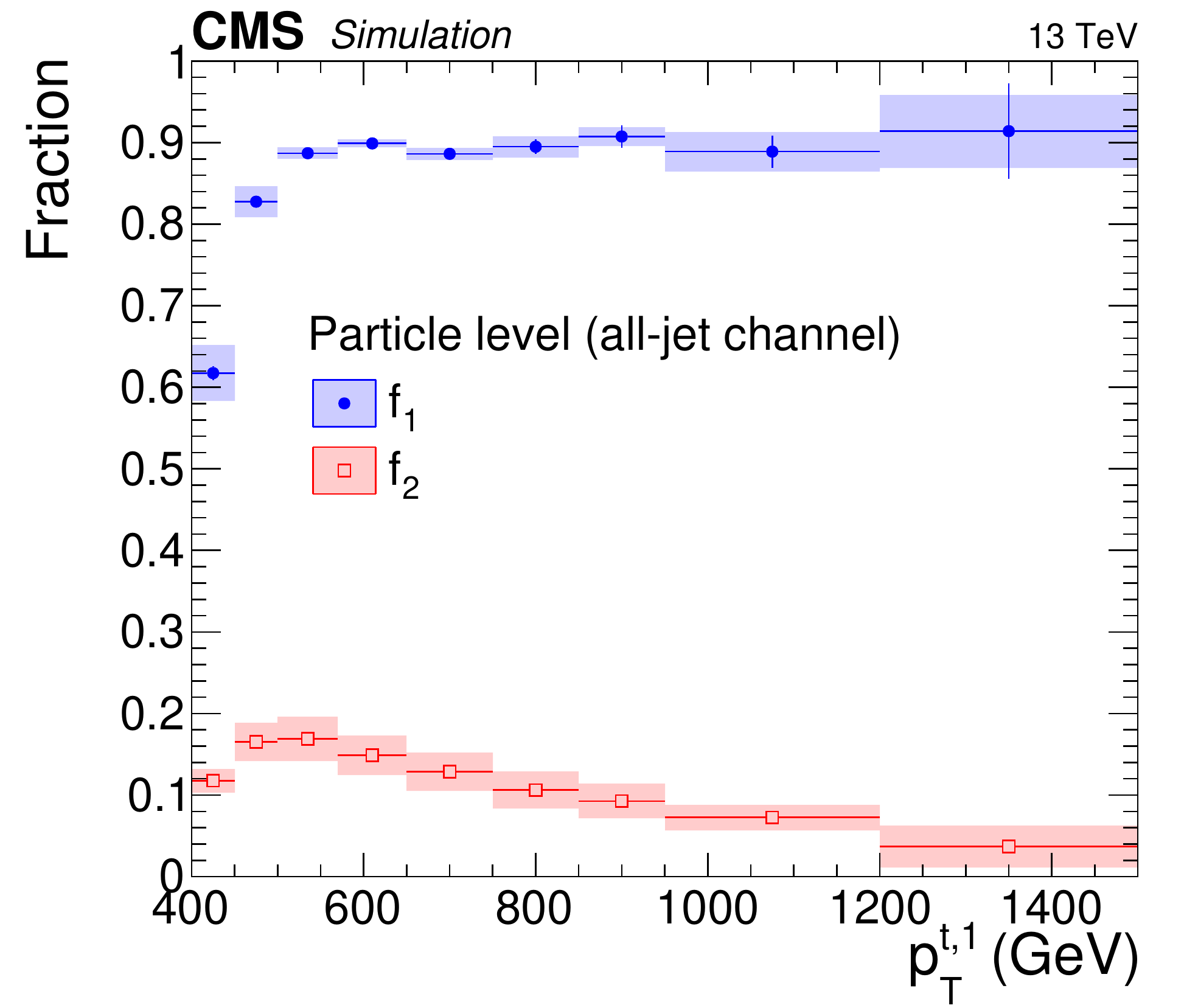}
    \includegraphics[width=\cmsFigWidth]{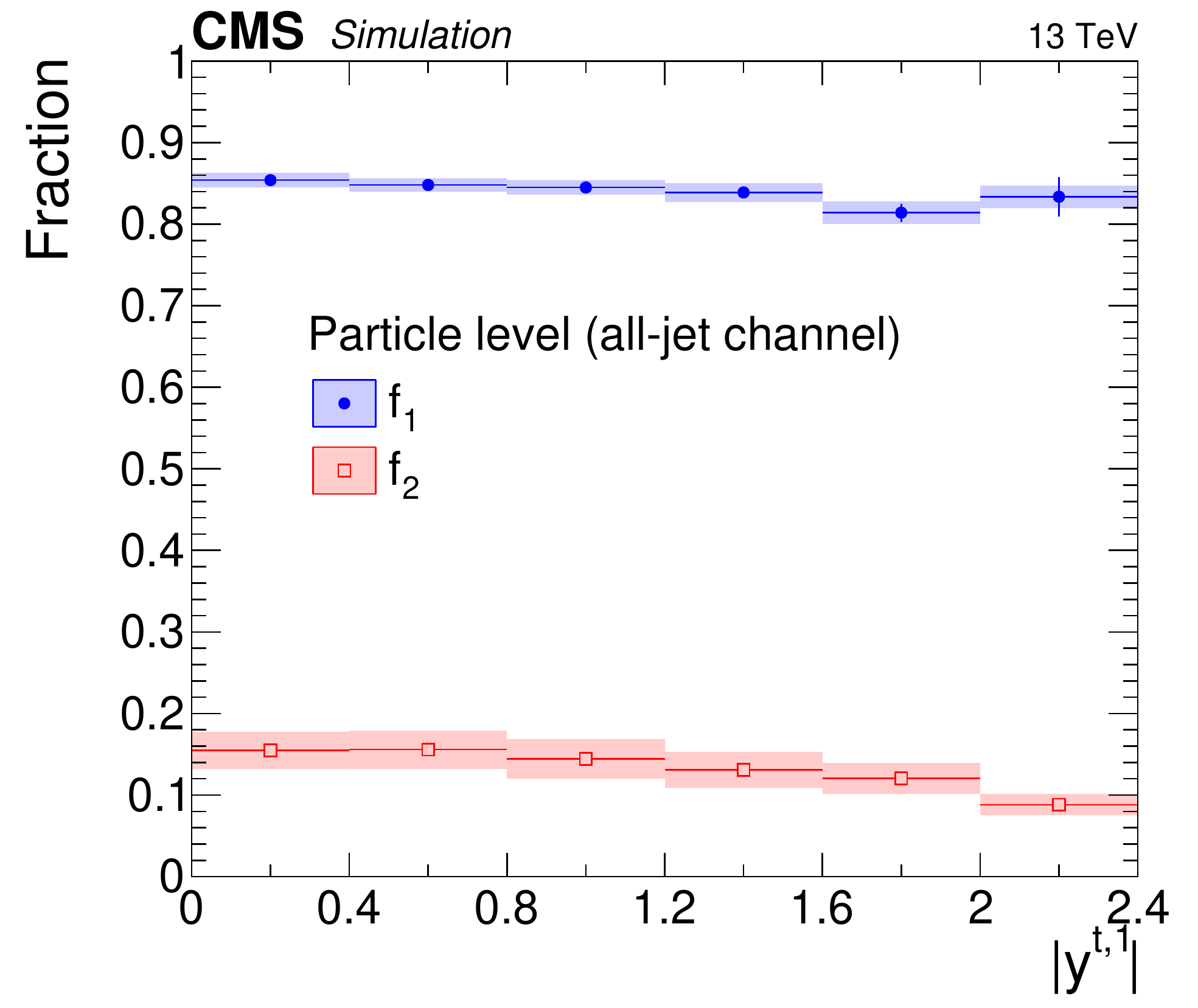}
    \caption{Simulated fractions $f_1$ and $f_2$ for the parton-level (upper row) and particle-level (lower row) selection in the all-jet channel as a function of the leading top quark \pt (left column) and $\abs{y}$ (right column). The fraction $f_1$ is a function of the leading reconstructed top quark and the $f_2$ is a function of the leading top quark at parton or particle level. The error band contains all uncertainty sources listed in Section 8.}
    \label{fig:efficiency_parton-particle}
\end{figure*}

To quantify the overlap in the definitions of \mbox{detector-,} particle-, and parton-level phase space, we define two fractions $f_{1,2}$, where $f_1$ is the fraction of reconstructed events that pass the selection at the unfolded level (parton or particle) in the same observable range, and $f_2$ is the fraction of generated events at the unfolded level that are selected at the reconstruction level. Figure~\ref{fig:efficiency_parton-particle} presents these fractions at the parton and particle levels for the all-jet channel, as a function of the leading top quark \pt and $\abs{y}$. The fraction $f_1$ is a function of the leading reconstructed top quark and the $f_2$ is a function of the leading top quark at parton or particle level. The distribution of $f_1$ vs. \pt shows a characteristic threshold behavior due to the resolution in \pt, while $f_1$ is independent of $\abs{y}$. The $f_2$ value decreases with \pt, primarily due to the inefficiency of subjet \PQb tagging and the NN output dependence on the \pt (at high jet \pt it is more difficult to differentiate between ordinary jets and highly boosted top quarks). Also, $f_2$ decreases at high $\abs{y}$ values due to the increased inefficiency in \PQb tagging at the edges of the CMS tracker.

\subsection{Unfolding}\label{sec:Unf}

\begin{figure*}[hbtp]
\centering
    \includegraphics[width=\cmsFigWidth]{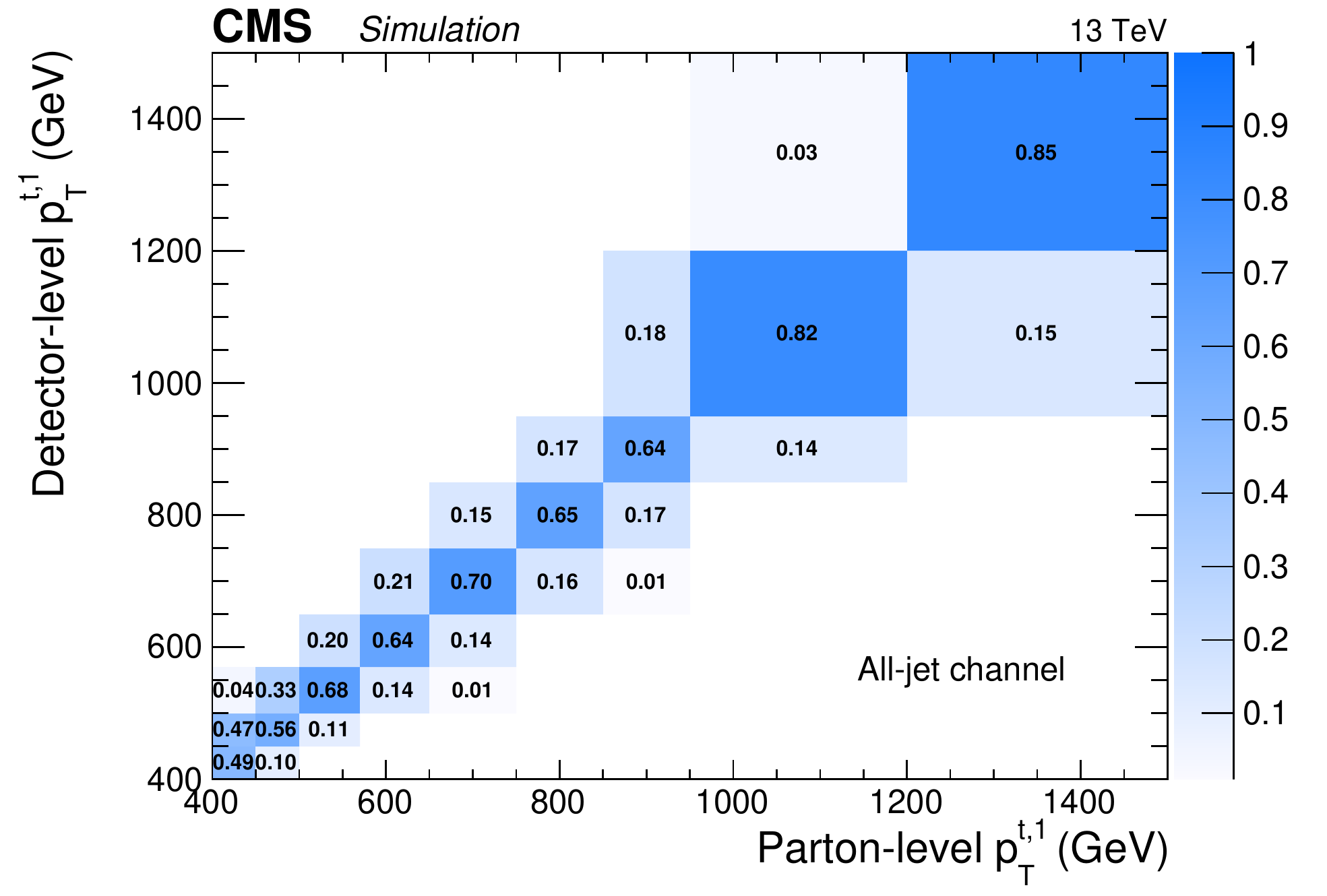}
    \includegraphics[width=\cmsFigWidth]{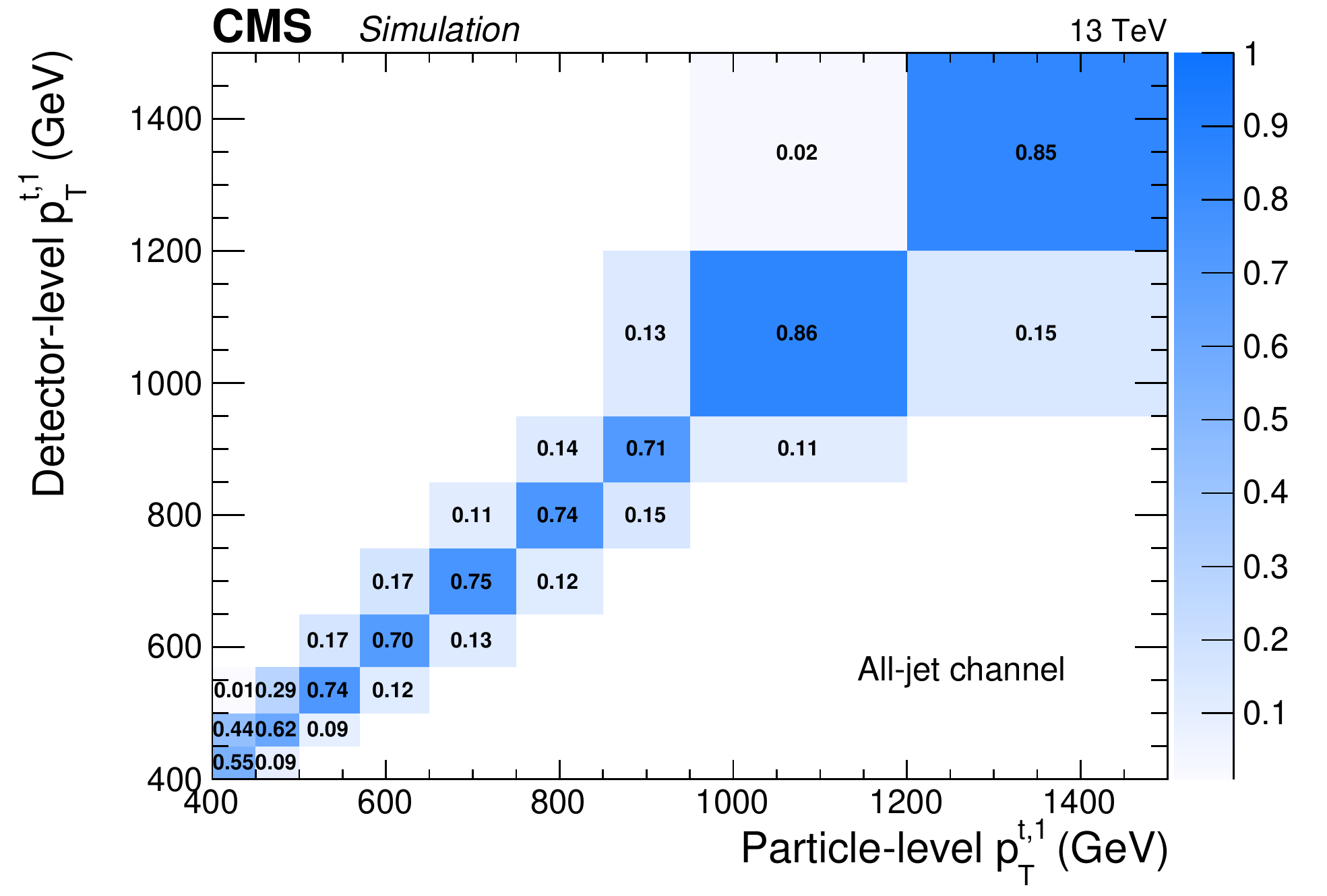} \\
    \includegraphics[width=\cmsFigWidth]{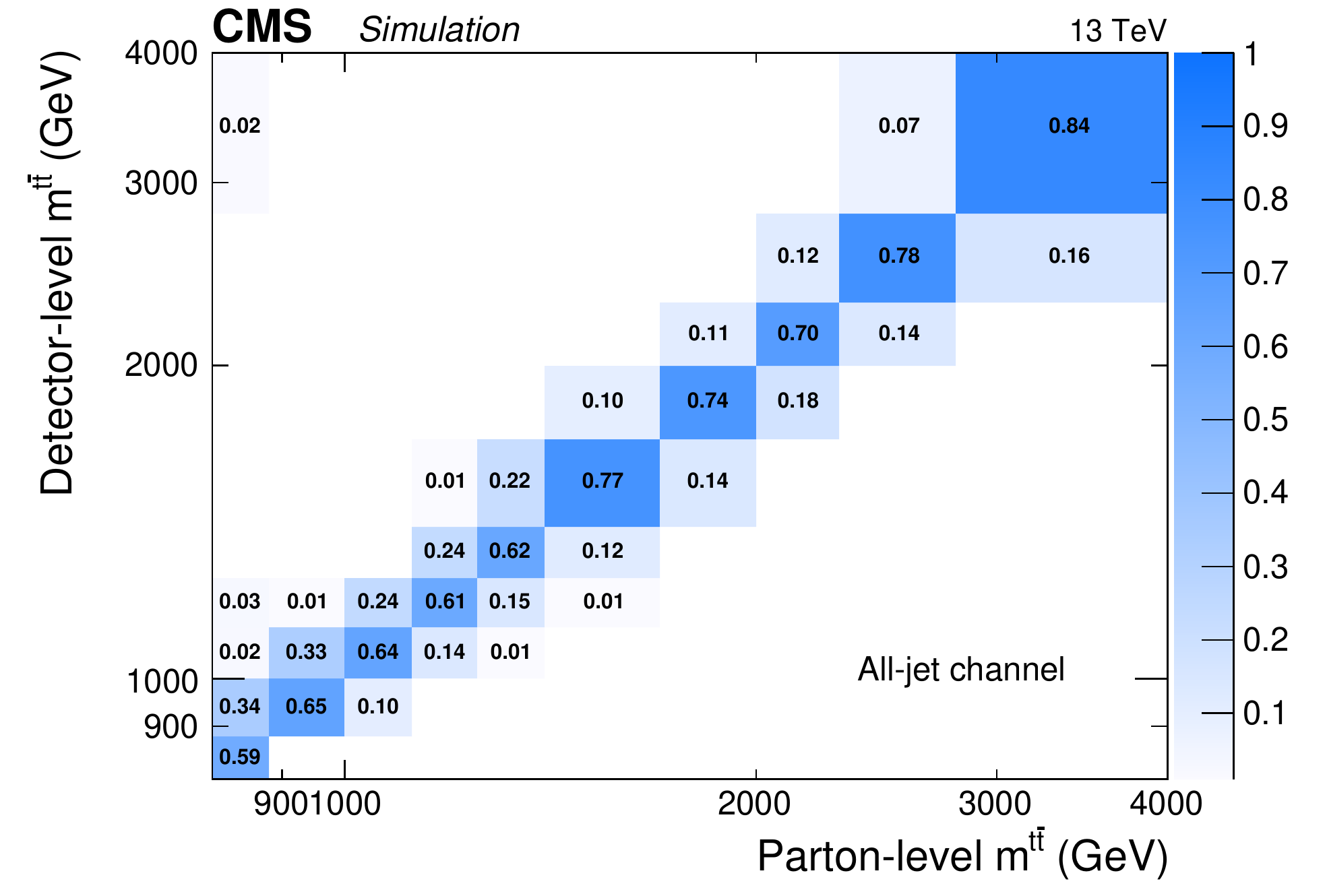}
    \includegraphics[width=\cmsFigWidth]{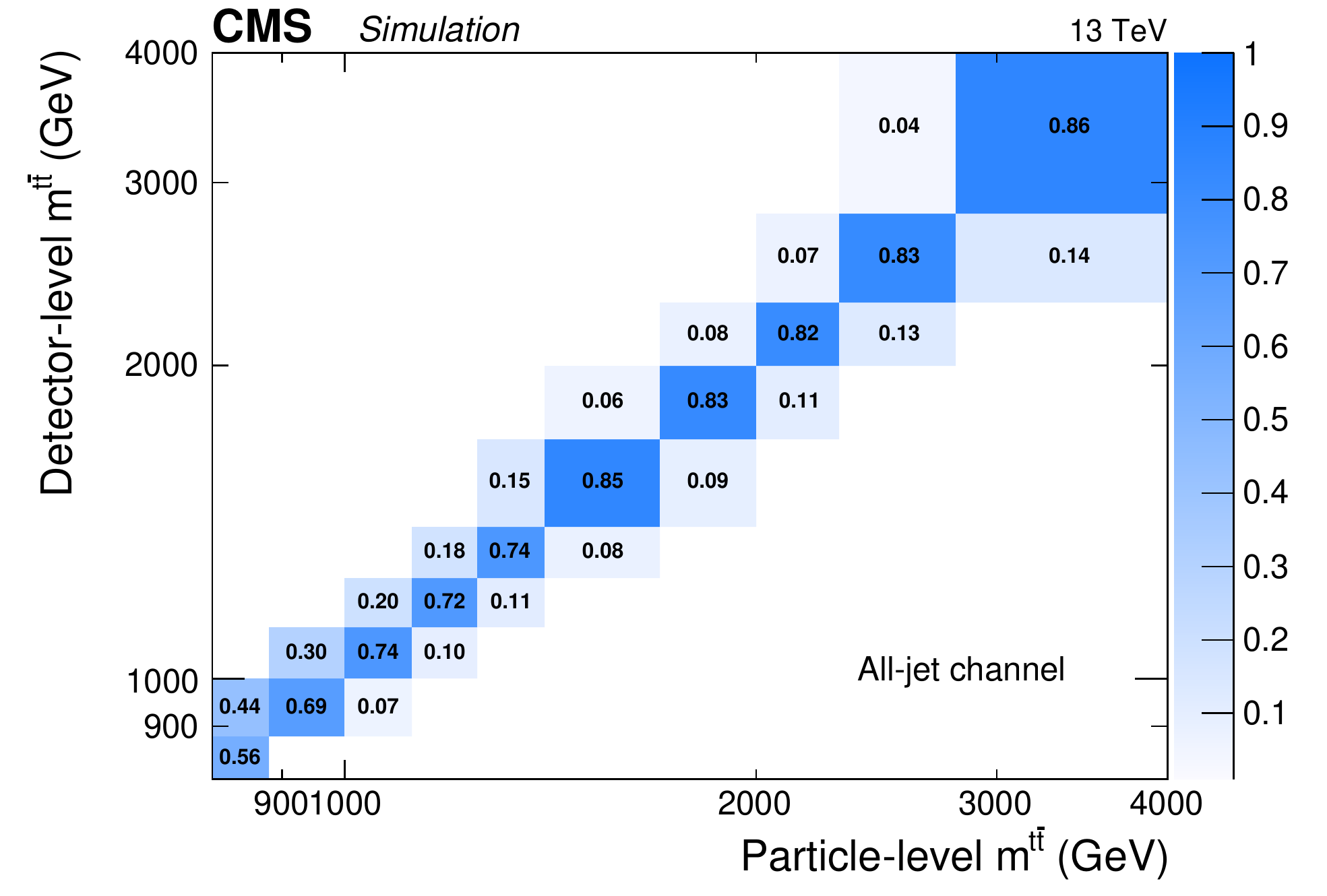}
    \caption{Migration matrices determined from simulation for the leading top quark \pt (upper row) and $m^{\ttbar}$ (lower row) at the parton level (left) and particle level (right) in the all-jet channel. The sum of the elements in each column is normalized to unity.}
    \label{fig:migration_parton-particle}
\end{figure*}

\begin{figure*}[hbtp]
\centering
    \includegraphics[width=\cmsFigWidth]{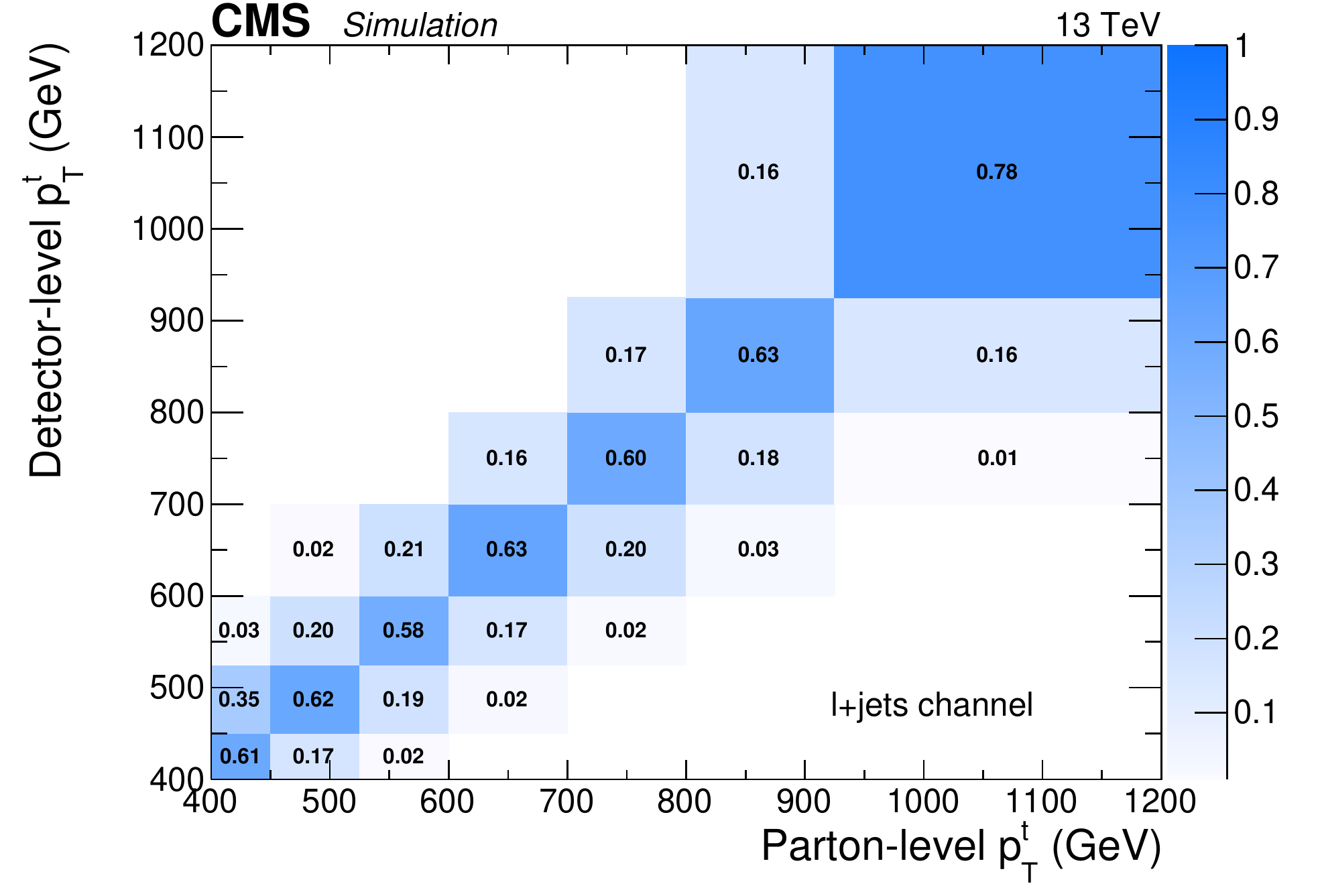}
    \includegraphics[width=\cmsFigWidth]{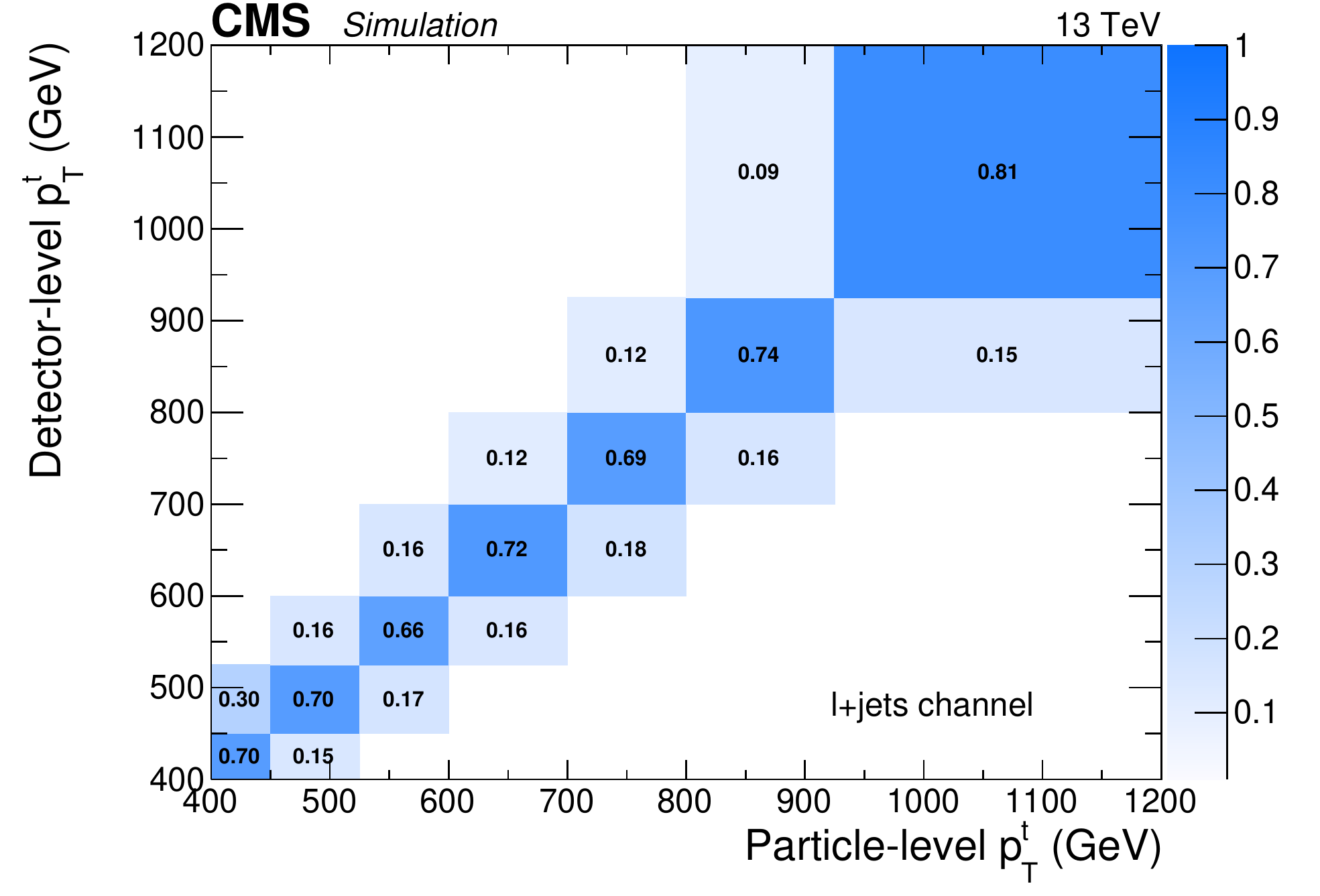} \\
    \includegraphics[width=\cmsFigWidth]{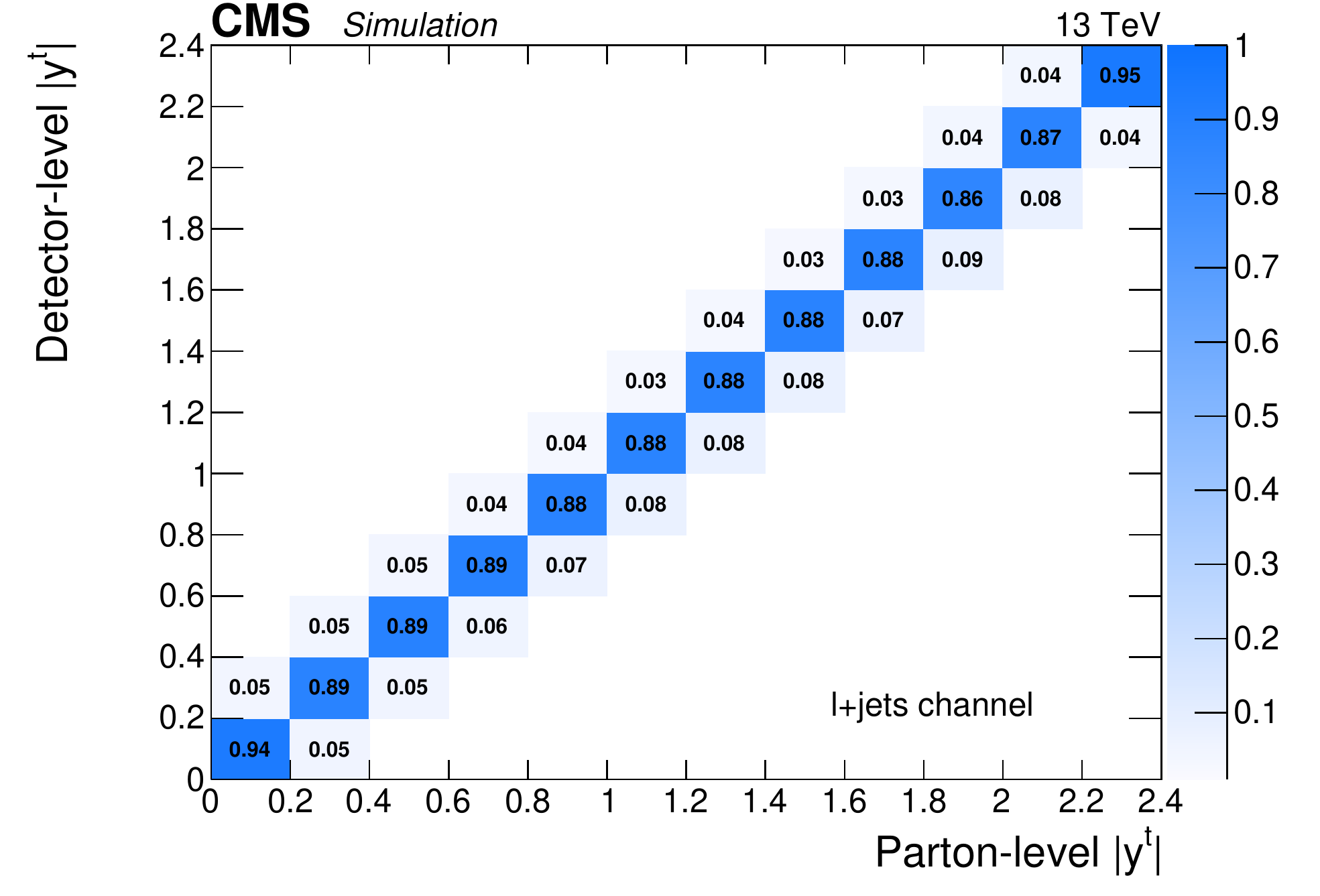}
    \includegraphics[width=\cmsFigWidth]{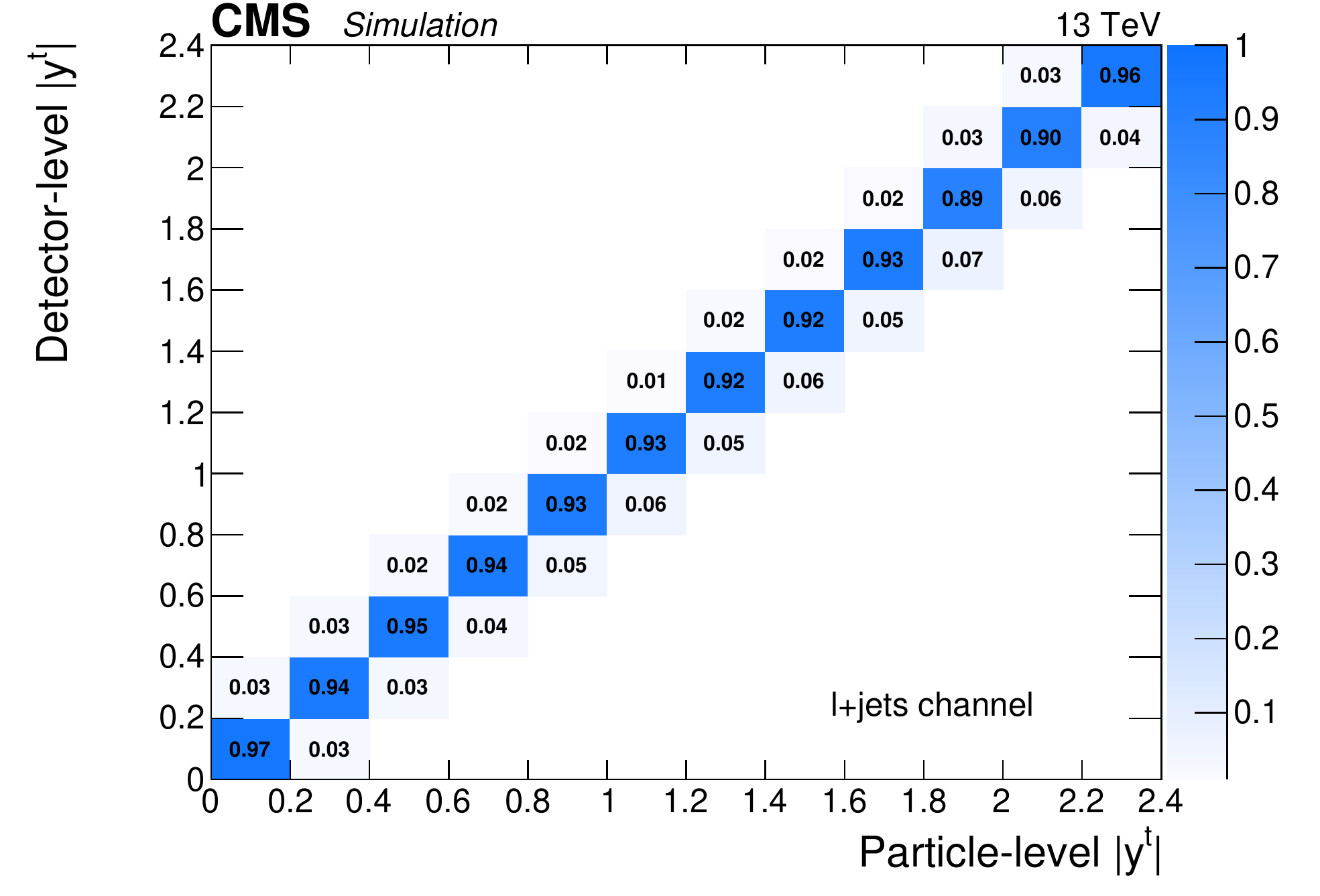}
    \caption{Migration matrices determined from simulation for top quark \pt (upper row) and rapidity (lower row) at the parton level (left) and particle level (right) in the $\ell$+jets channel. The sum of the elements in each column is normalized to unity.}
    \label{fig:migration_parton-particle_ljets}
\end{figure*}

We extract the differential cross sections by applying an unfolding procedure, which is necessary due to the finite resolution of the detector. The unfolded cross sections are evaluated as follows
\begin{equation}\label{eq:xs_unfolded}
\frac{\rd \sigma^\text{unf}_i}{\rd x}=\frac{1}{\mathcal{L} \Delta x_i} \frac{1}{f_{2,i}} \sum_{j}{\left(R^{-1}_{ij} f_{1,j} S_j\right)},
\end{equation}
where $\mathcal{L}$ is the total integrated luminosity and $\Delta x_i$ is the width of the $i$-th bin of the observable $x$. The quantity $R^{-1}_{ij}$ is the inverse of the migration matrix between the $i$- and $j$-th bins, and $S_j$ is the signal yield in the $j$-th bin computed from Eq.~(\ref{eq:signal}). The binning of the various observables is chosen such that the purity (fraction of reconstructed events for which the true value of the observable lies in the same bin) and the stability (fraction of true events where the reconstructed observable lies in the same bin) are well above 50\% for most of the bins. This choice results in migration matrices with suppressed nondiagonal elements, shown for the all-jet channel in Fig.~\ref{fig:migration_parton-particle} and for the $\ell$+jets channel in Fig.~\ref{fig:migration_parton-particle_ljets}. To minimize biases introduced by the various unfolding methods utilizing regularization, we use migration-matrix inversion, as written in Eq.~(\ref{eq:xs_unfolded}) and implemented in the \textsc{TUnfold} framework~\cite{Schmitt:2012kp}, for the price of a moderate increase in statistical uncertainty compared to unfolding methods utilizing regularization. For the all-jet channel, the fractions $f_1$ and $f_2$ in Eq.~(\ref{eq:xs_unfolded}) are determined independently from the unfolding, as described in Section~\ref{sec:definitionLevels} and shown in Fig.~\ref{fig:efficiency_parton-particle}. For the $\ell$+jets channel, both the reconstruction efficiencies and bin migrations are accounted for directly via \textsc{TUnfold}.

\subsection{All-jet channel}

\begin{figure*}[hbtp]
\centering
    \includegraphics[width=\cmsFigWidth]{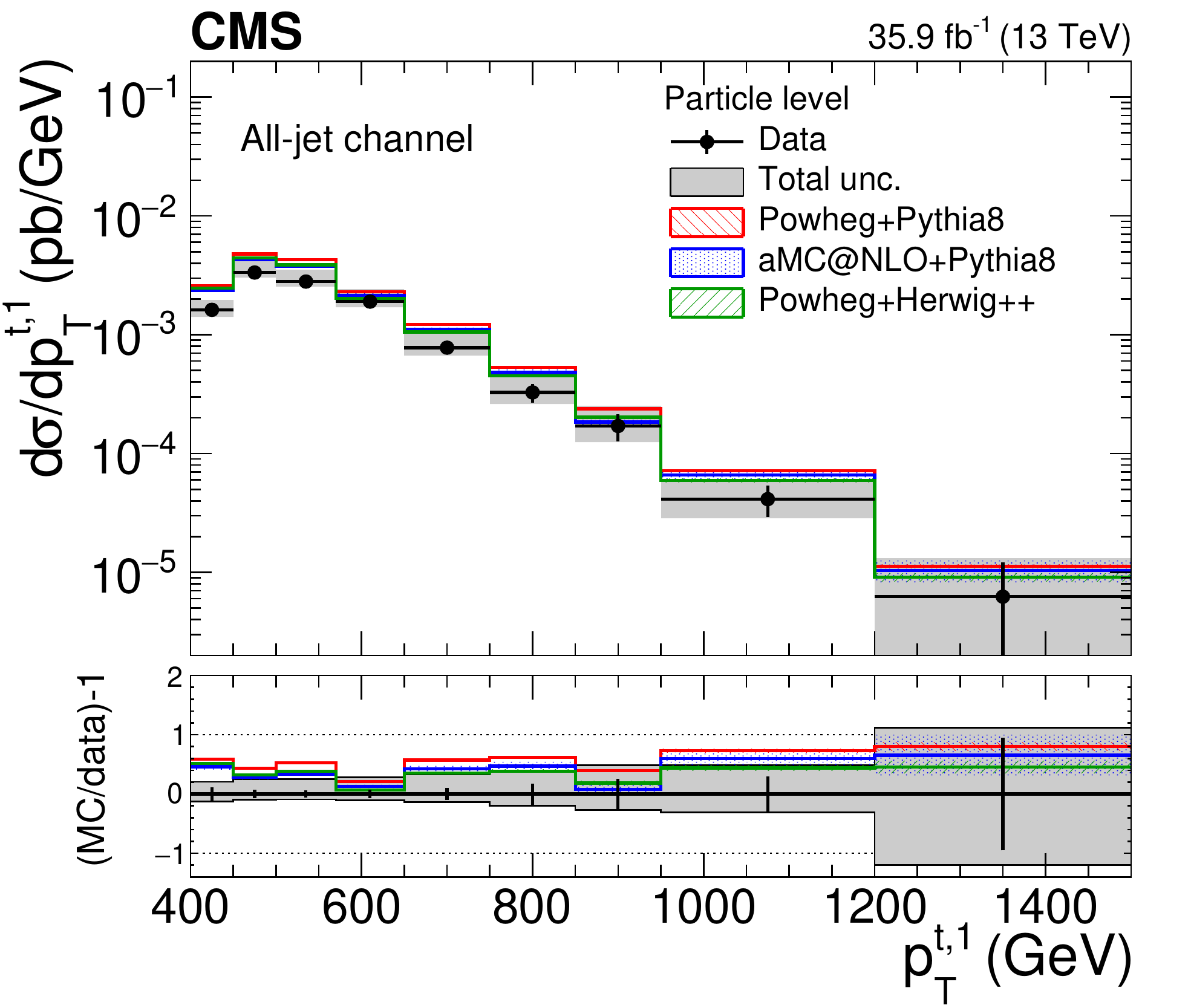}
    \includegraphics[width=\cmsFigWidth]{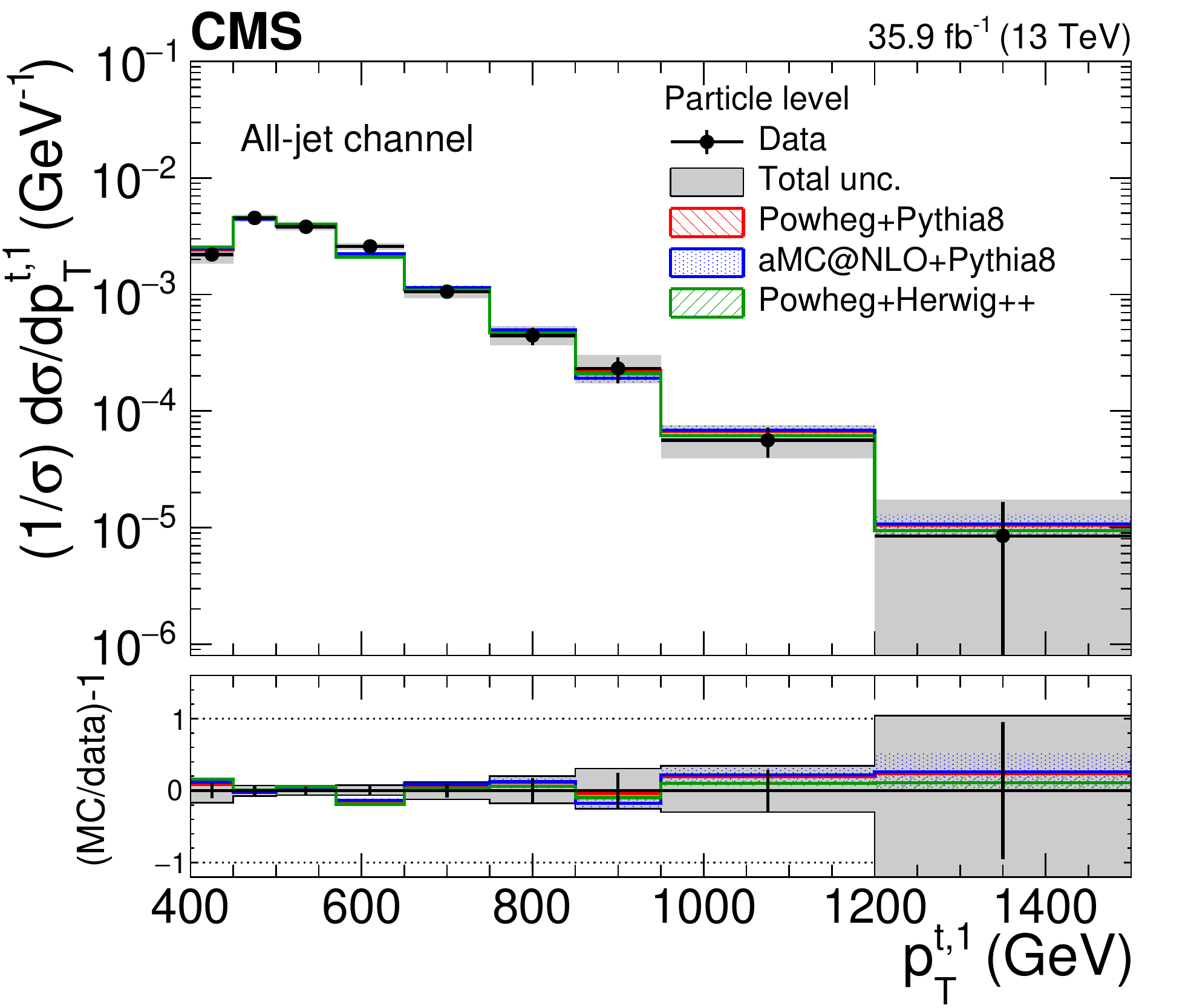} \\
    \includegraphics[width=\cmsFigWidth]{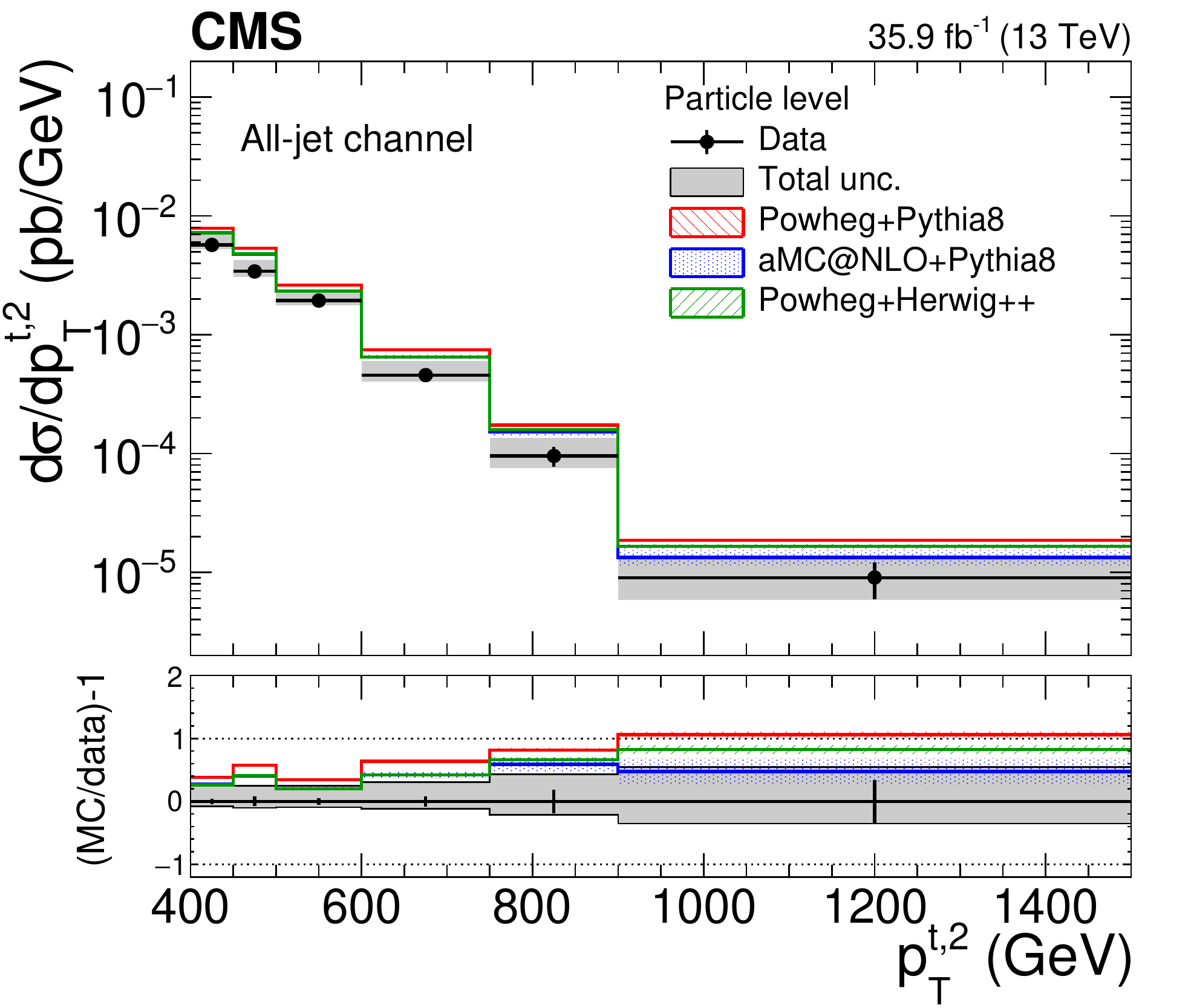}
    \includegraphics[width=\cmsFigWidth]{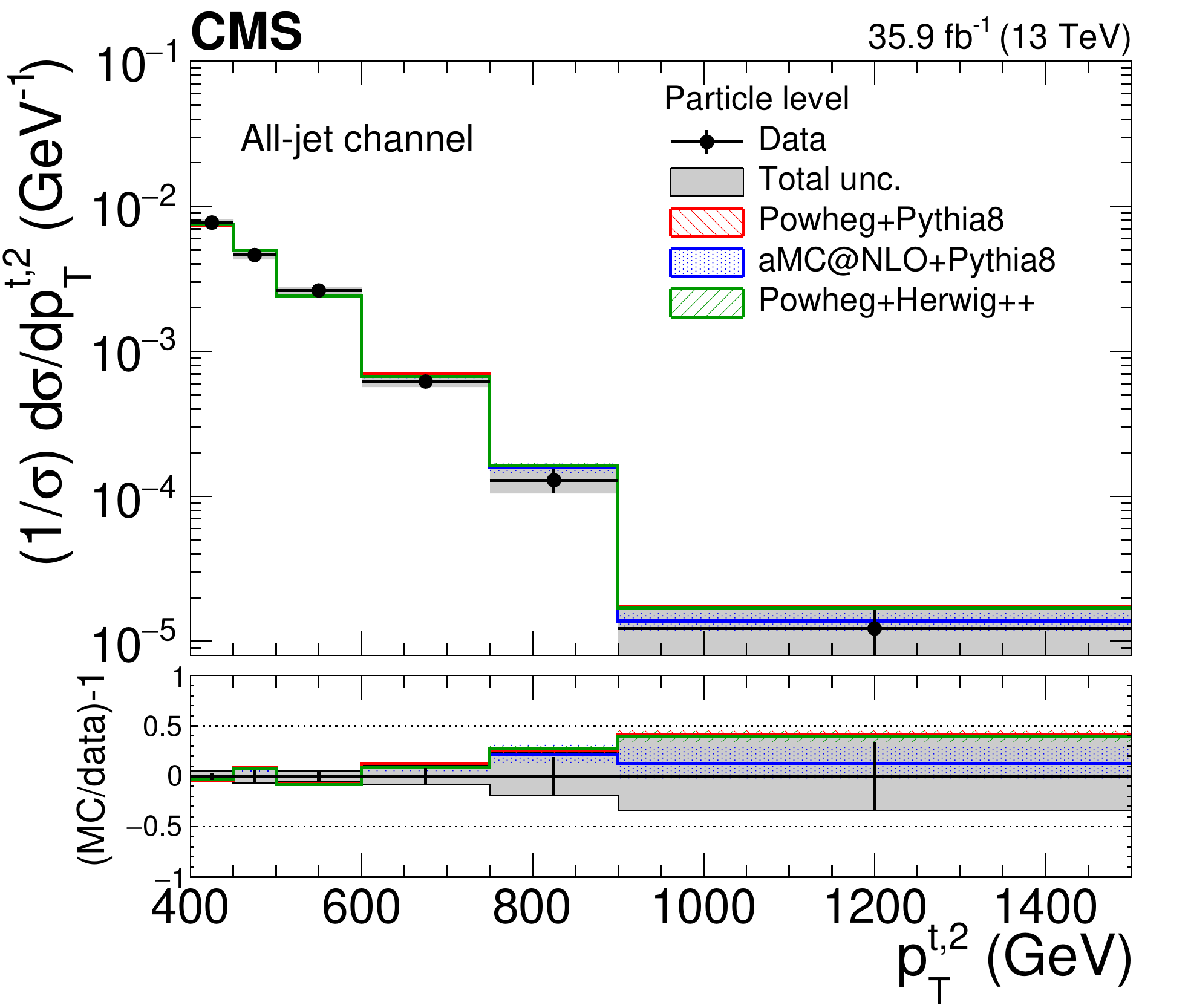}
    \caption{Differential cross section unfolded to the particle level, absolute (left) and normalized (right), as a function of the leading (upper row) and subleading (lower row) top quark \pt in the all-jet channel. The lower panel shows the ratio (MC/data)$-$1. The vertical bars on the data and in the ratio represent the statistical uncertainty in data, while the shaded band shows the total statistical and systematic uncertainty added in quadrature. The hatched bands show the statistical uncertainty of the MC samples.}
    \label{fig:Particle_jetPt01}
\end{figure*}

\begin{figure*}[hbtp]
\centering
    \includegraphics[width=\cmsFigWidth]{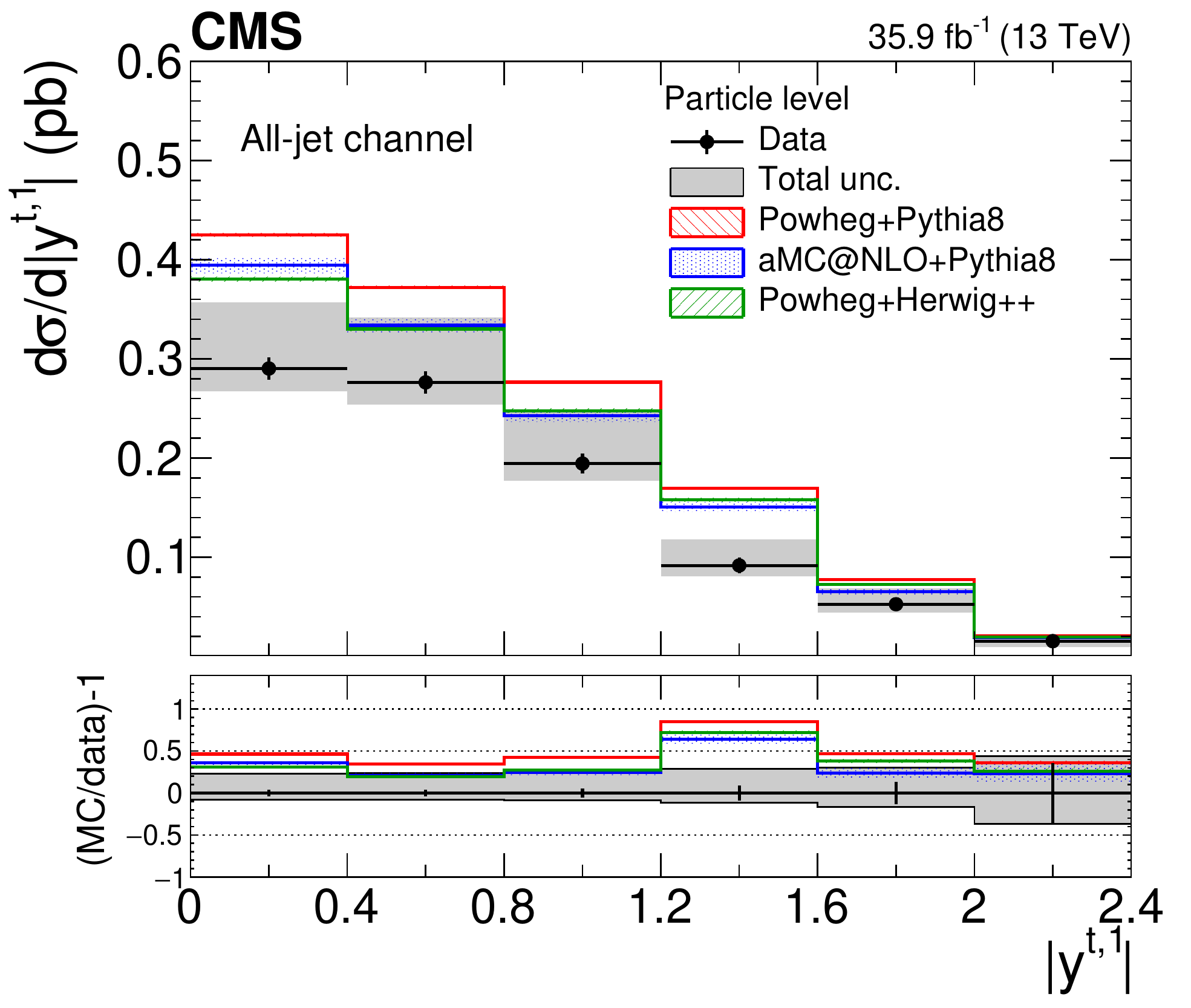}
    \includegraphics[width=\cmsFigWidth]{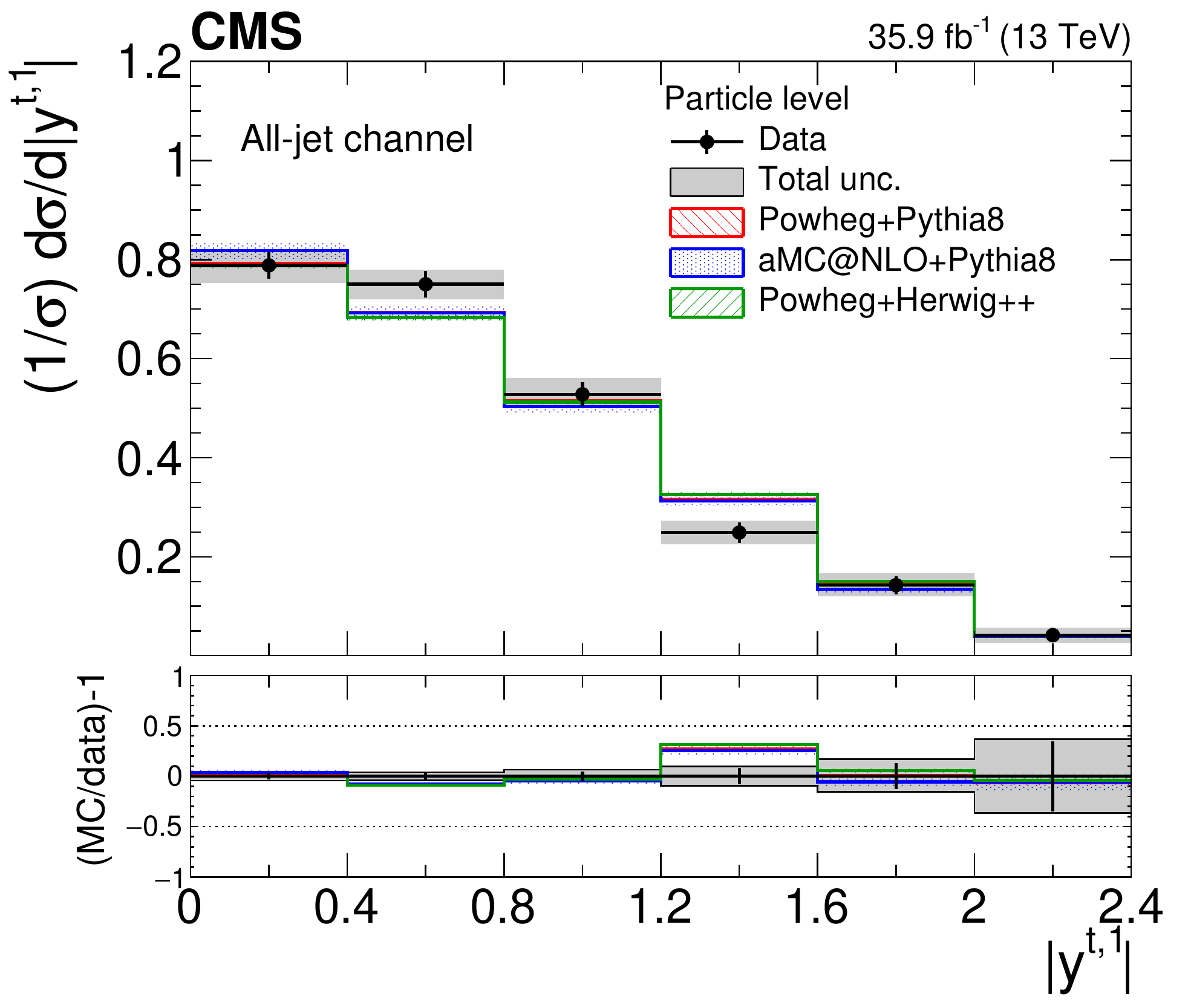}\\
    \includegraphics[width=\cmsFigWidth]{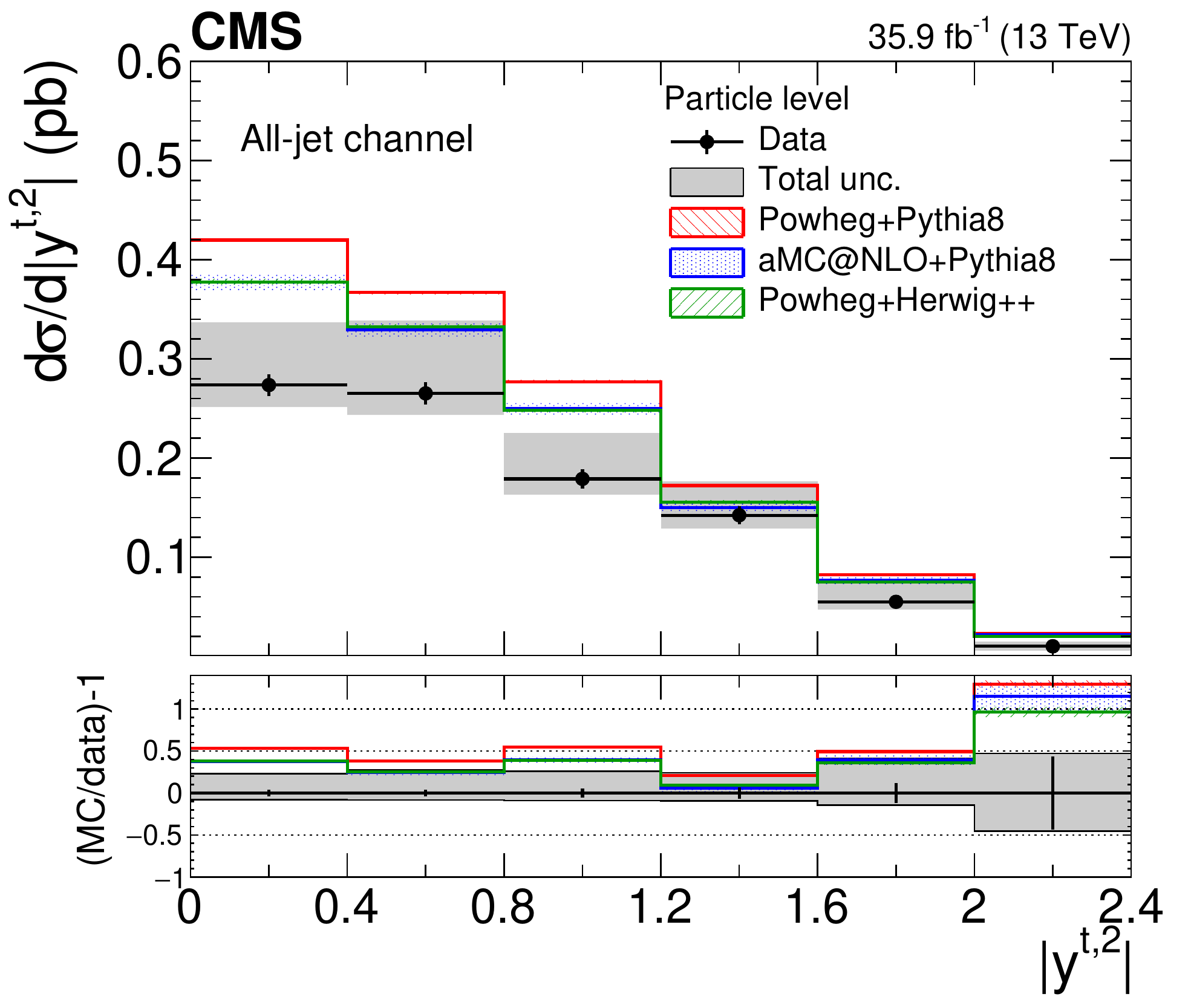}
    \includegraphics[width=\cmsFigWidth]{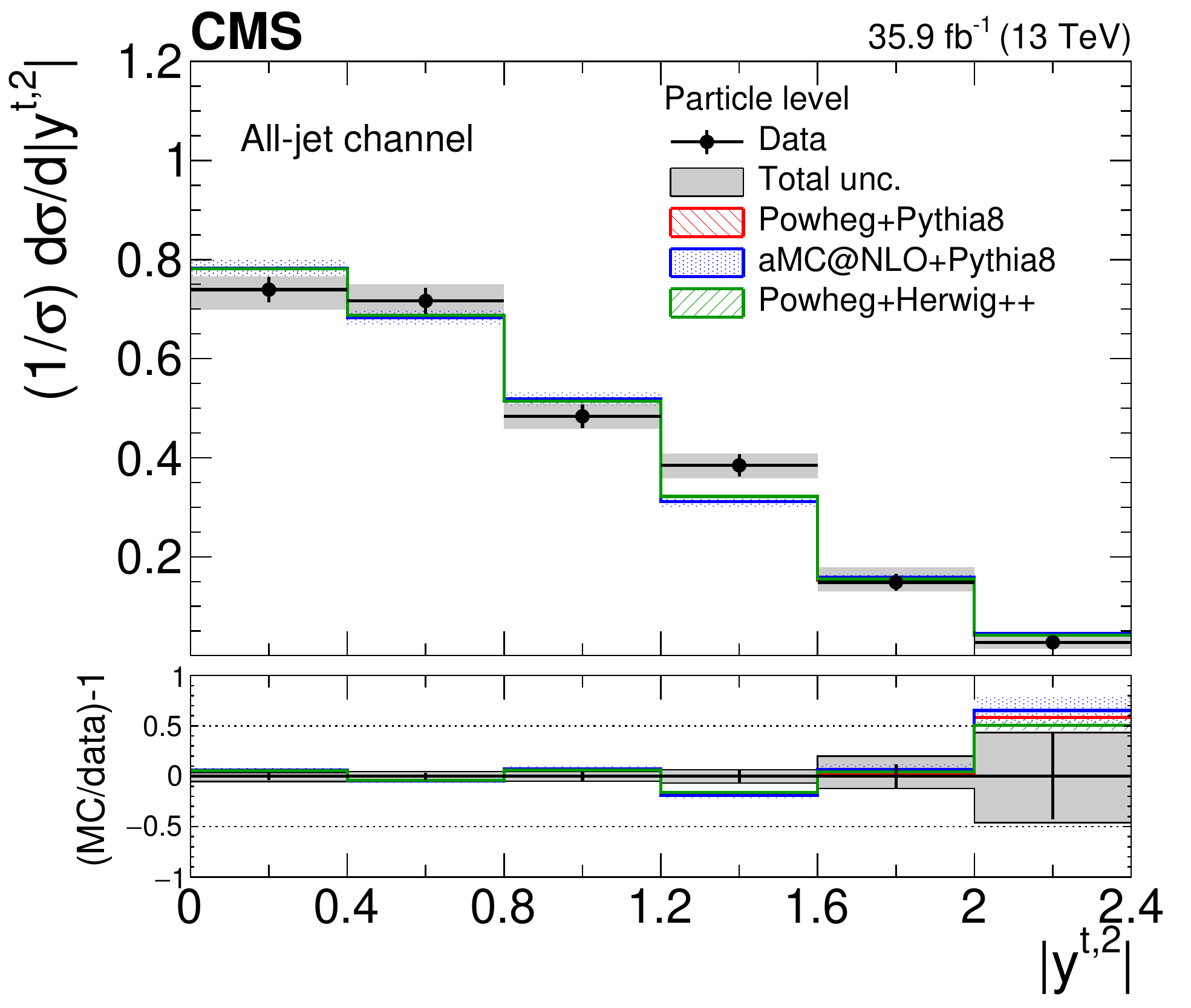}
    \caption{Differential cross section unfolded to the particle level, absolute (left) and normalized (right), as a function of the leading (upper row) and subleading (lower row) top quark $\abs{y}$ in the all-jet channel. The lower panel shows the ratio (MC/data)$-$1. The vertical bars on the data and in the ratio represent the statistical uncertainty in data, while the shaded band shows the total statistical and systematic uncertainty added in quadrature. The hatched bands show the statistical uncertainty of the MC samples.}
    \label{fig:Particle_jetY01}
\end{figure*}

\begin{figure*}[hbtp]
\centering
    \includegraphics[width=\cmsFigWidth]{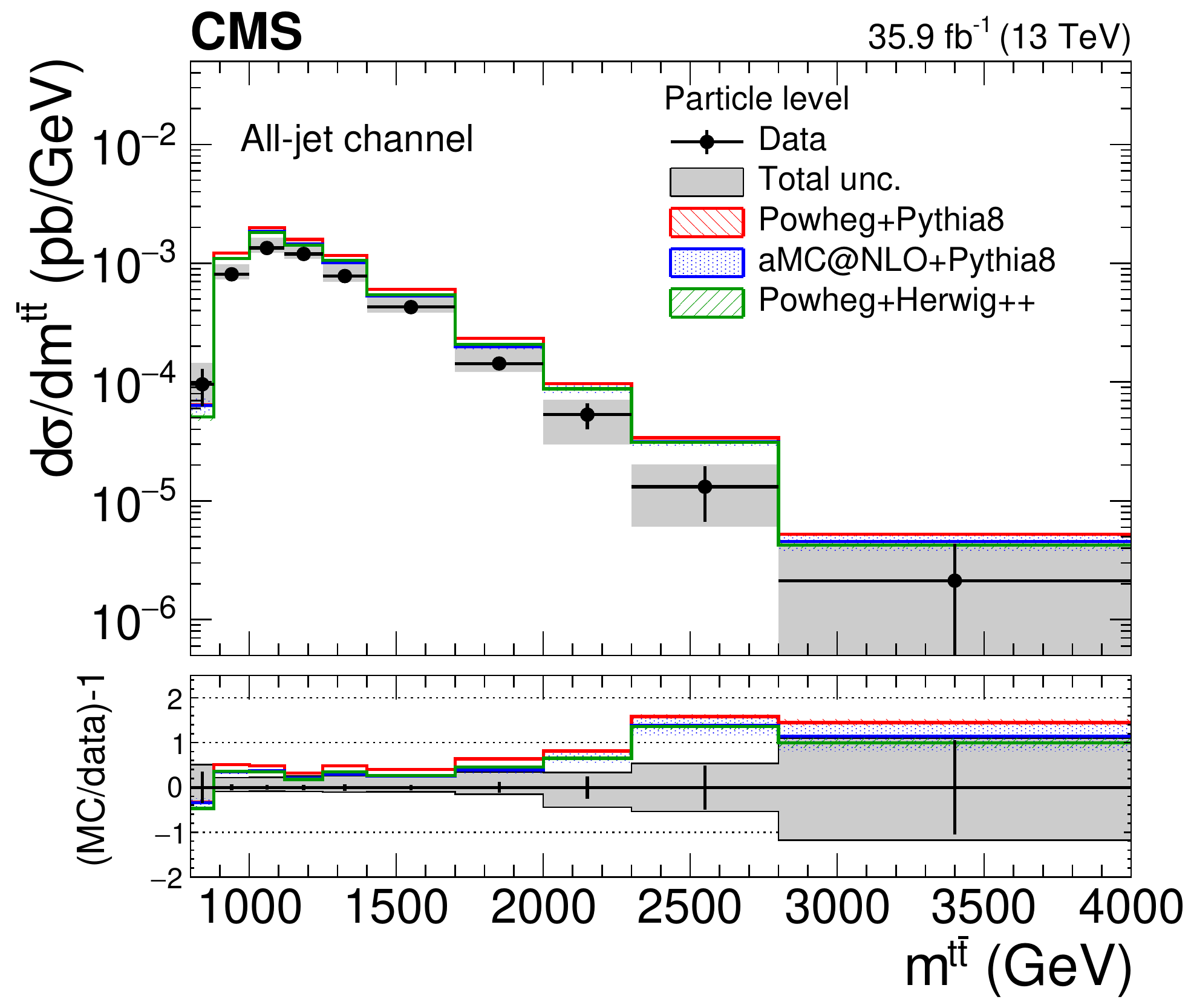}
    \includegraphics[width=\cmsFigWidth]{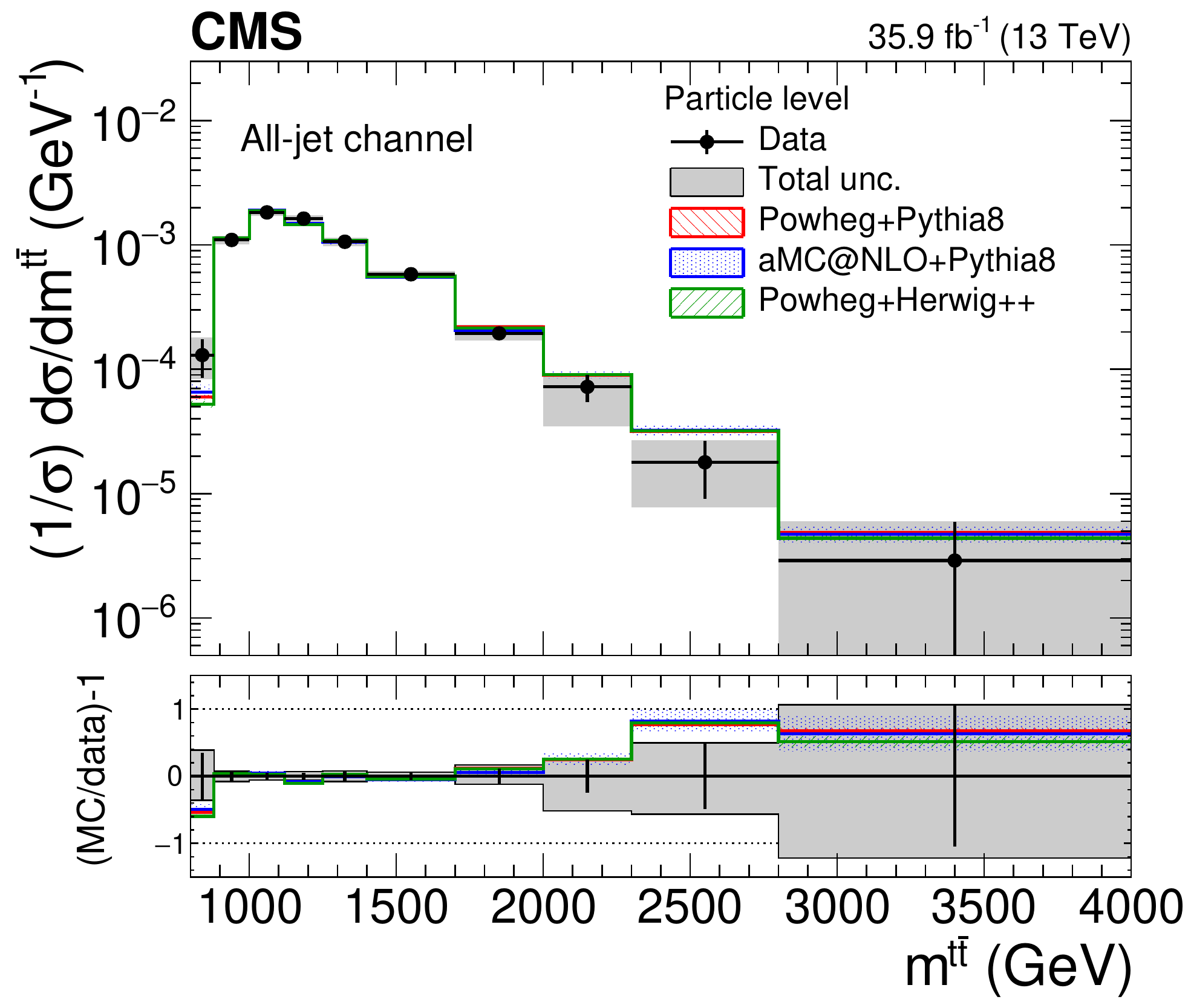} \\
    \includegraphics[width=\cmsFigWidth]{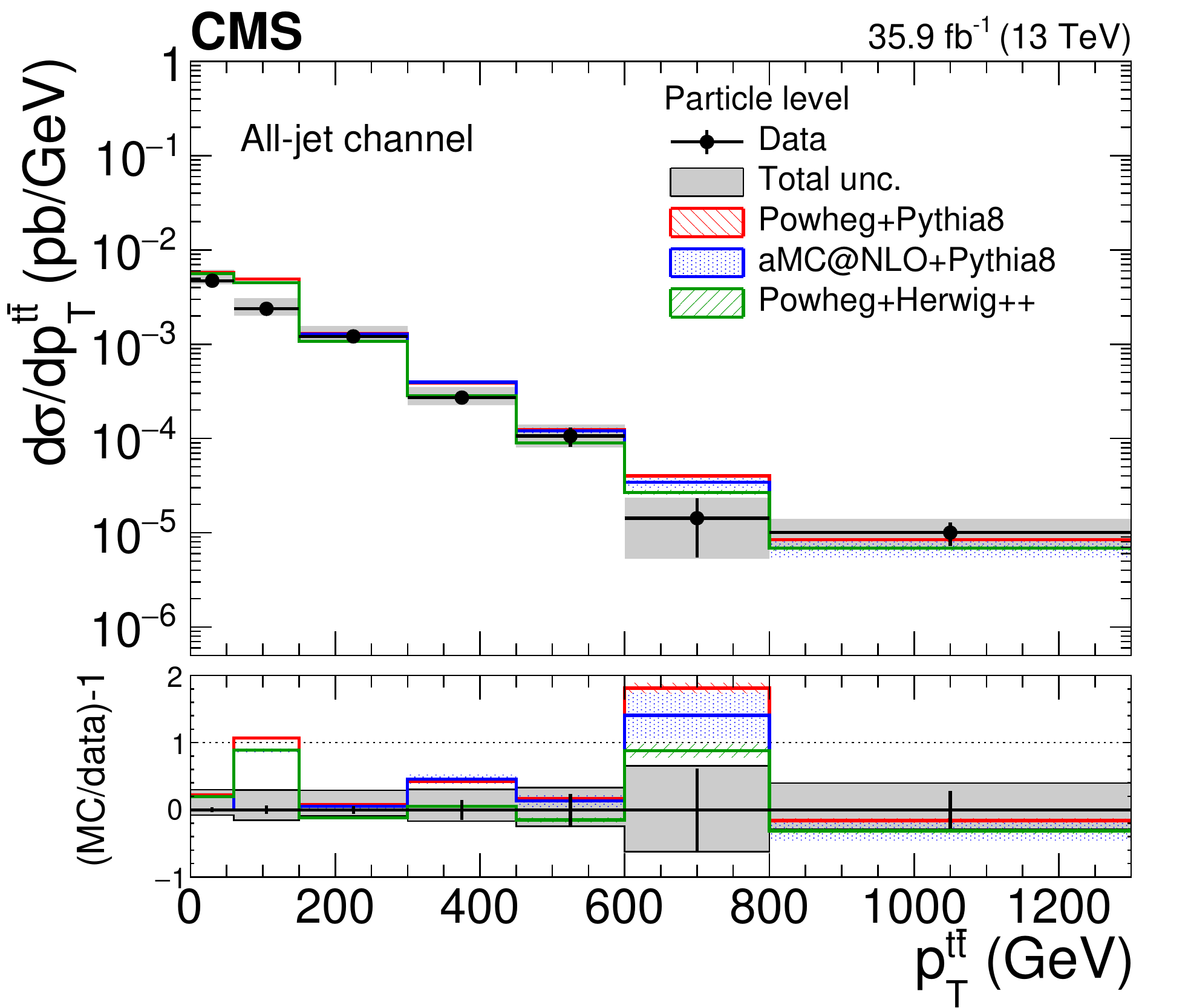}
    \includegraphics[width=\cmsFigWidth]{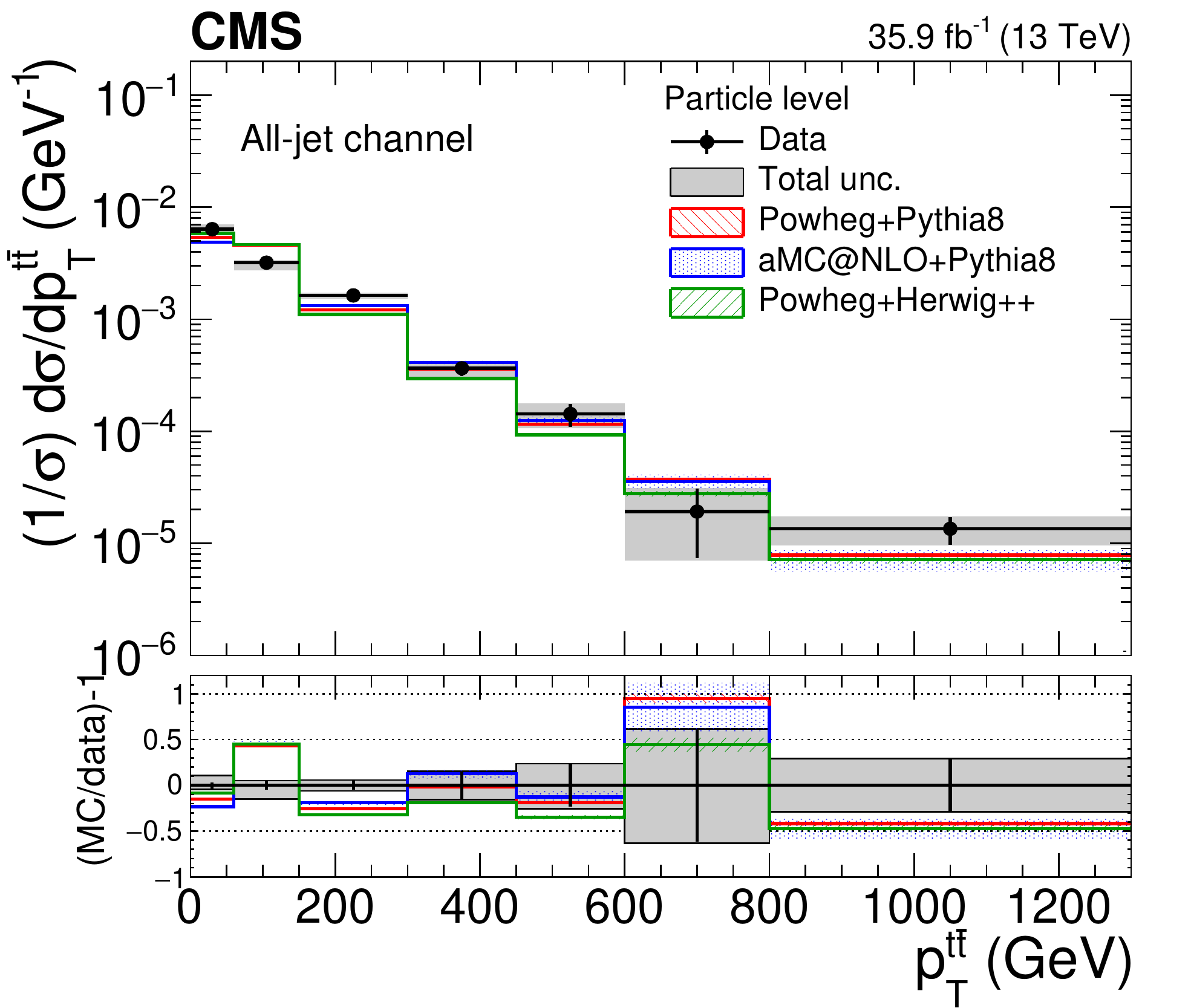} \\
    \includegraphics[width=\cmsFigWidth]{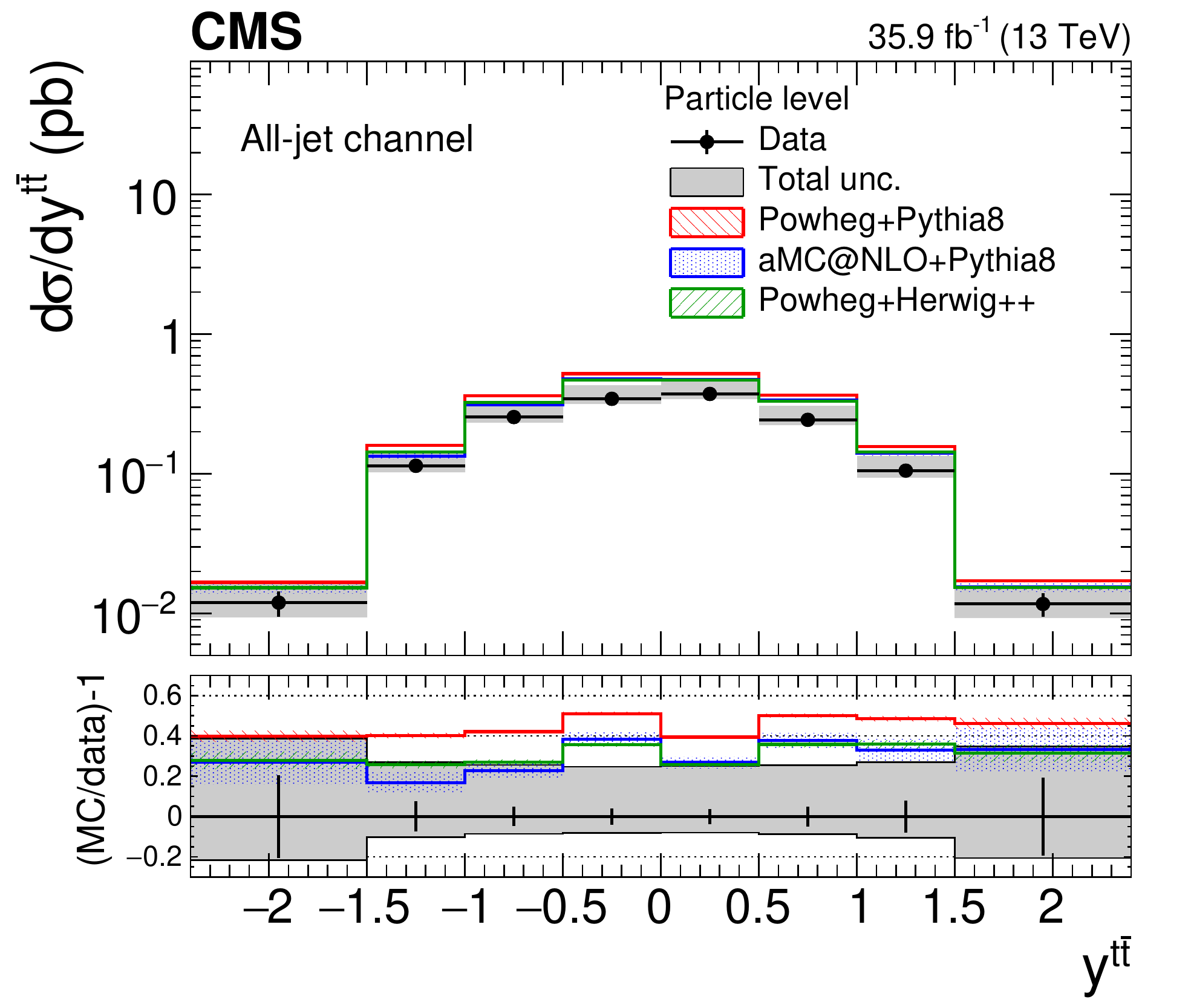}
    \includegraphics[width=\cmsFigWidth]{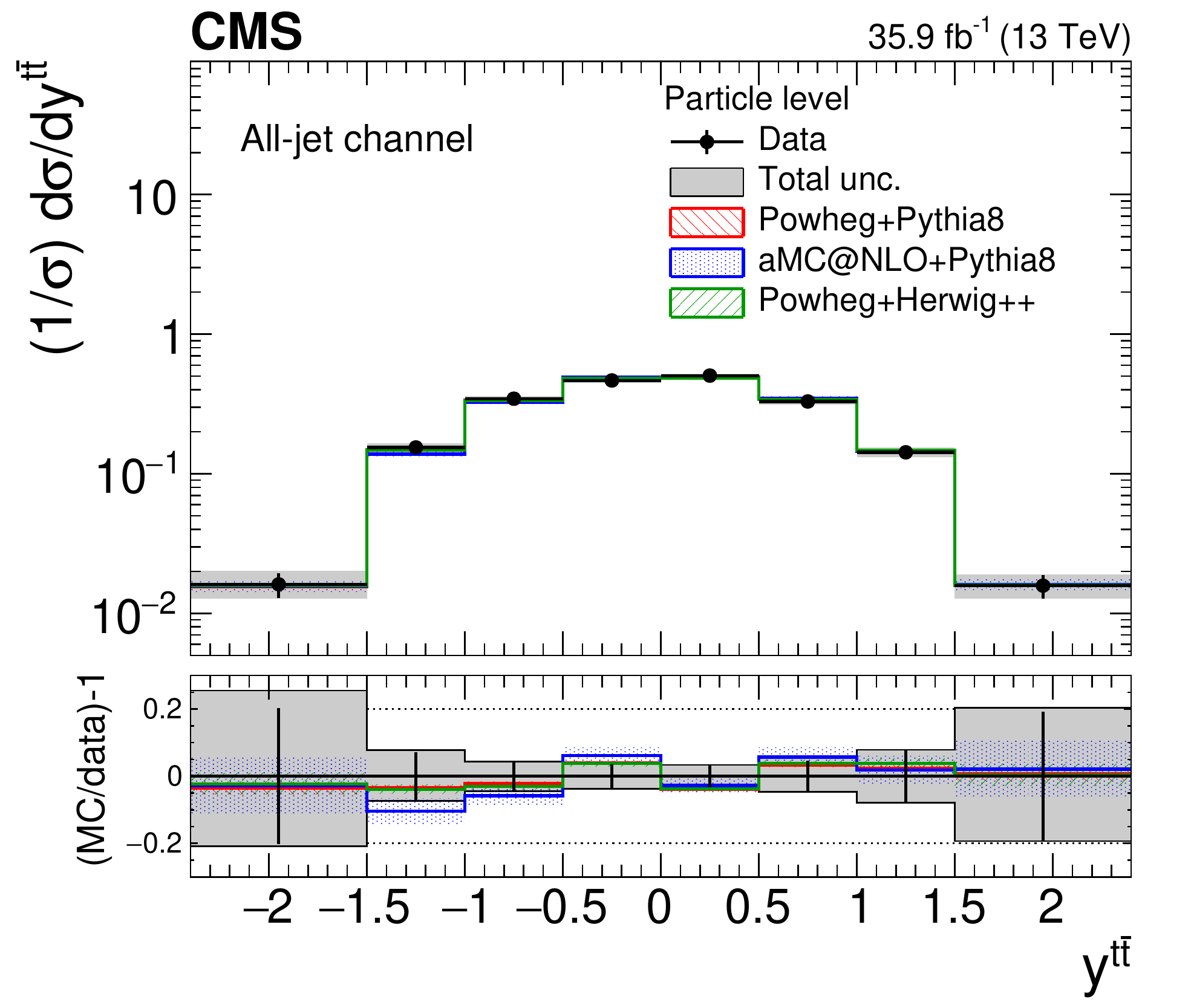}
    \caption{Differential cross section unfolded to the particle level, absolute (left) and normalized (right), as a function of $m^{\ttbar}$ (upper row), $\pt^{\ttbar}$ (middle row), and $y^{\ttbar}$ (lower row) in the all-jet channel. The lower panel shows the ratio (MC/data)$-$1. The vertical bars on the data and in the ratio represent the statistical uncertainty in data, while the shaded band shows the total statistical and systematic uncertainty added in quadrature. The hatched bands show the statistical uncertainty of the MC samples.}
    \label{fig:Particle_mJJ_ptJJ_yJJ}
\end{figure*}

\begin{figure*}[hbtp]
\centering
    \includegraphics[width=\cmsFigWidth]{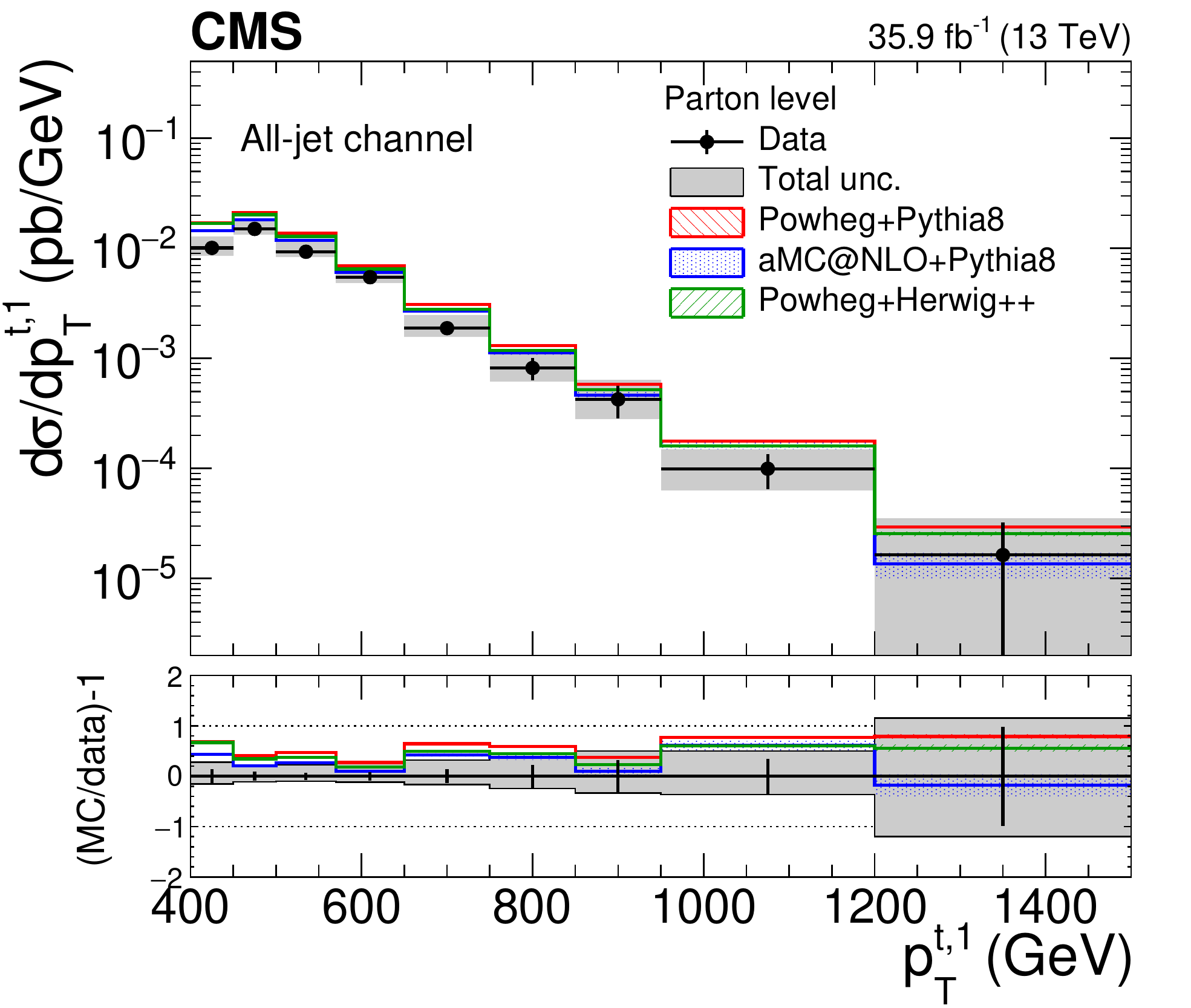}
    \includegraphics[width=\cmsFigWidth]{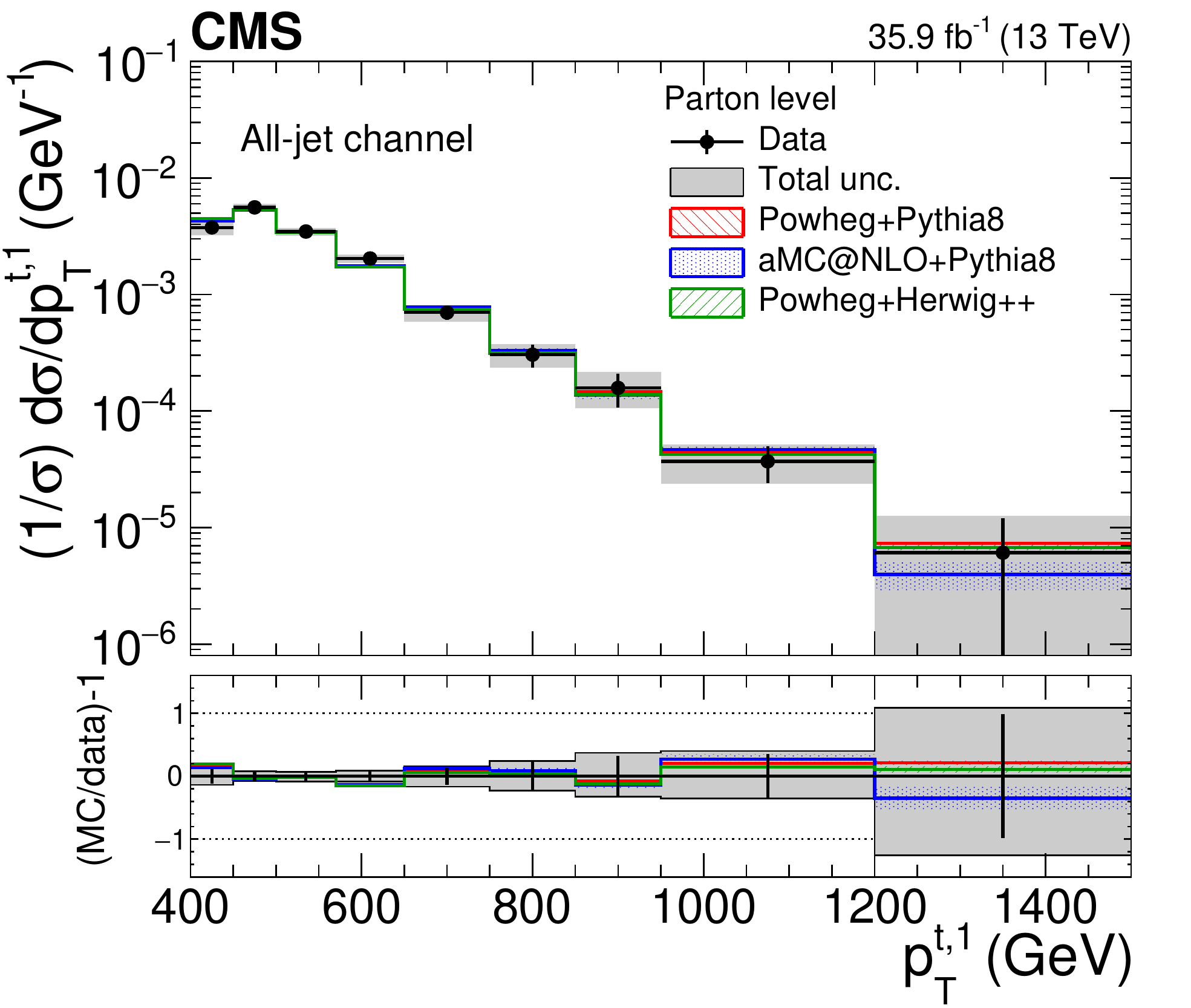}\\
    \includegraphics[width=\cmsFigWidth]{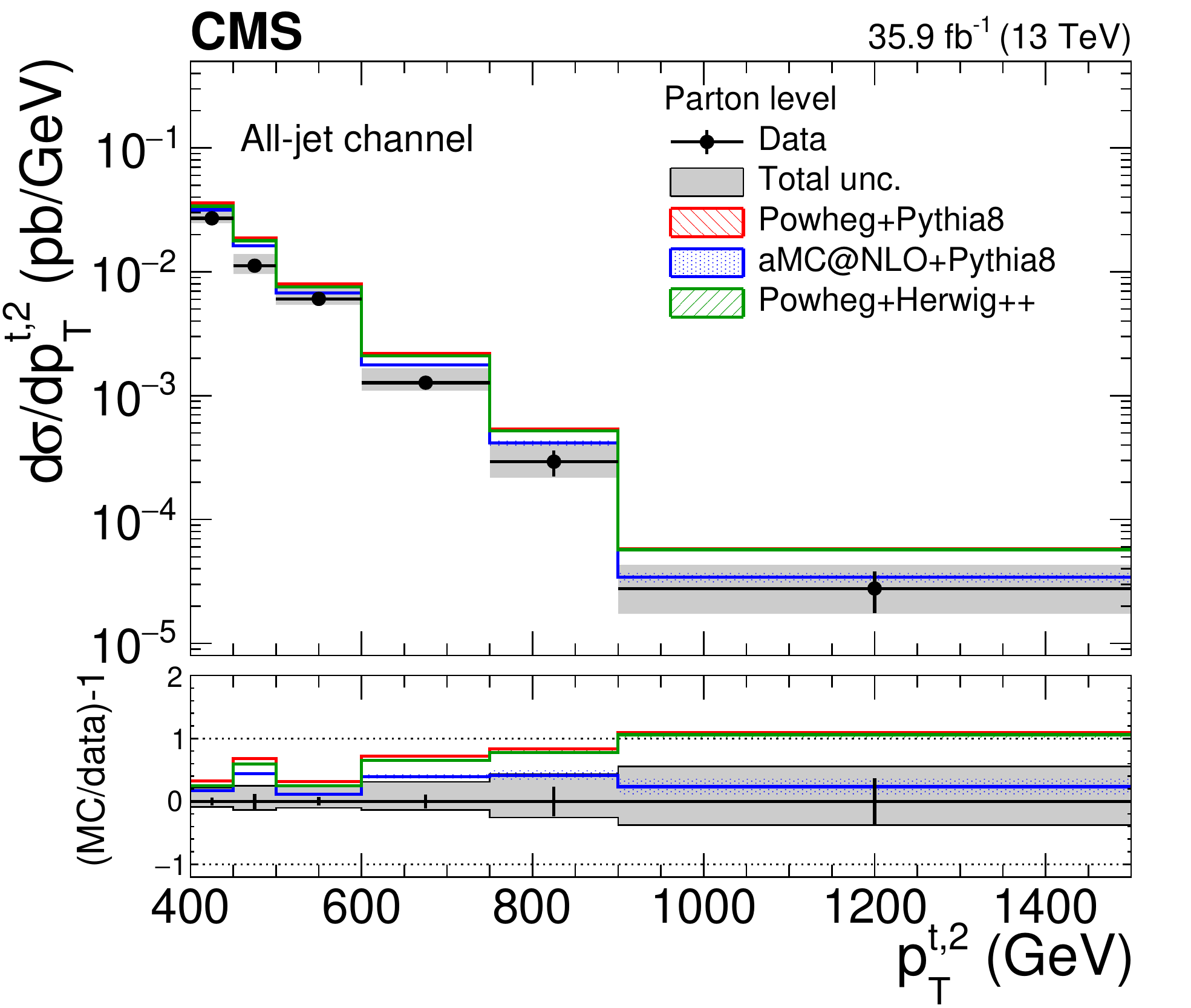}
    \includegraphics[width=\cmsFigWidth]{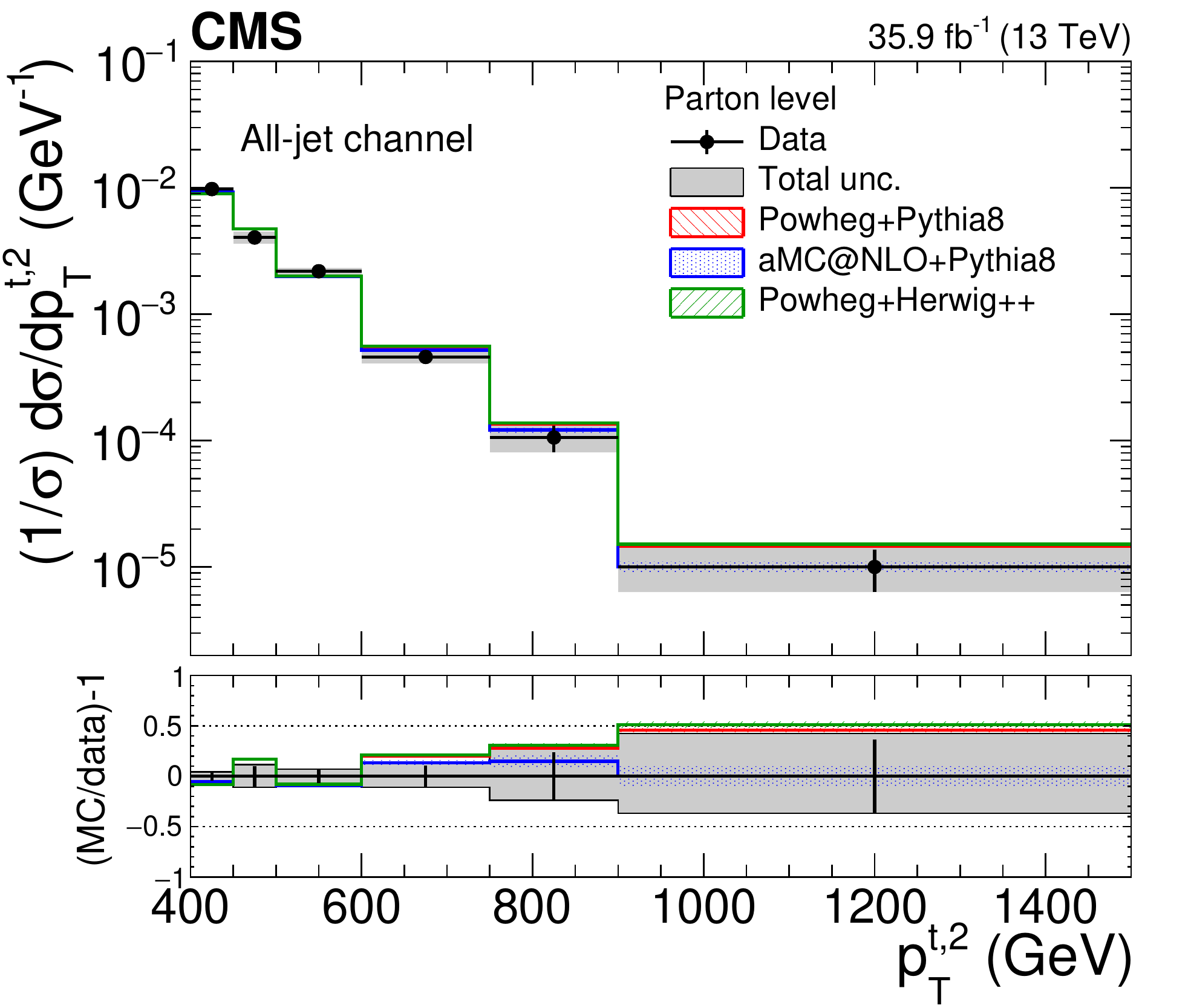}
    \caption{Differential cross section unfolded to the parton level, absolute (left) and normalized (right), as a function of the leading (upper row) and subleading (lower row) top quark \pt in the all-jet channel. The lower panel shows the ratio (MC/data)$-$1. The vertical bars on the data and in the ratio represent the statistical uncertainty in data, while the shaded band shows the total statistical and systematic uncertainty added in quadrature. The hatched bands show the statistical uncertainty of the MC samples.}
    \label{fig:Parton_jetPt01}
\end{figure*}

\begin{figure*}[hbtp]
\centering
    \includegraphics[width=\cmsFigWidth]{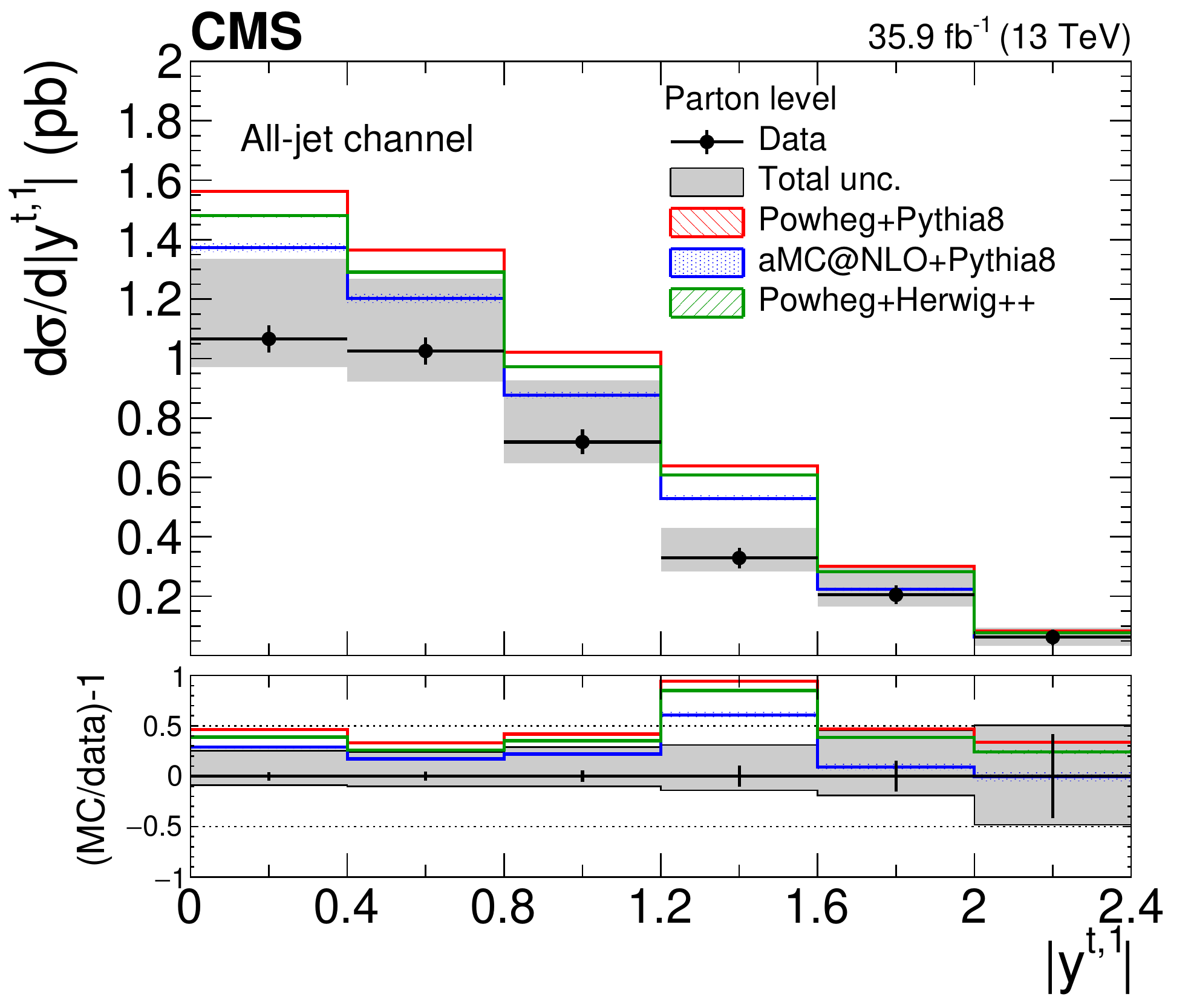}
    \includegraphics[width=\cmsFigWidth]{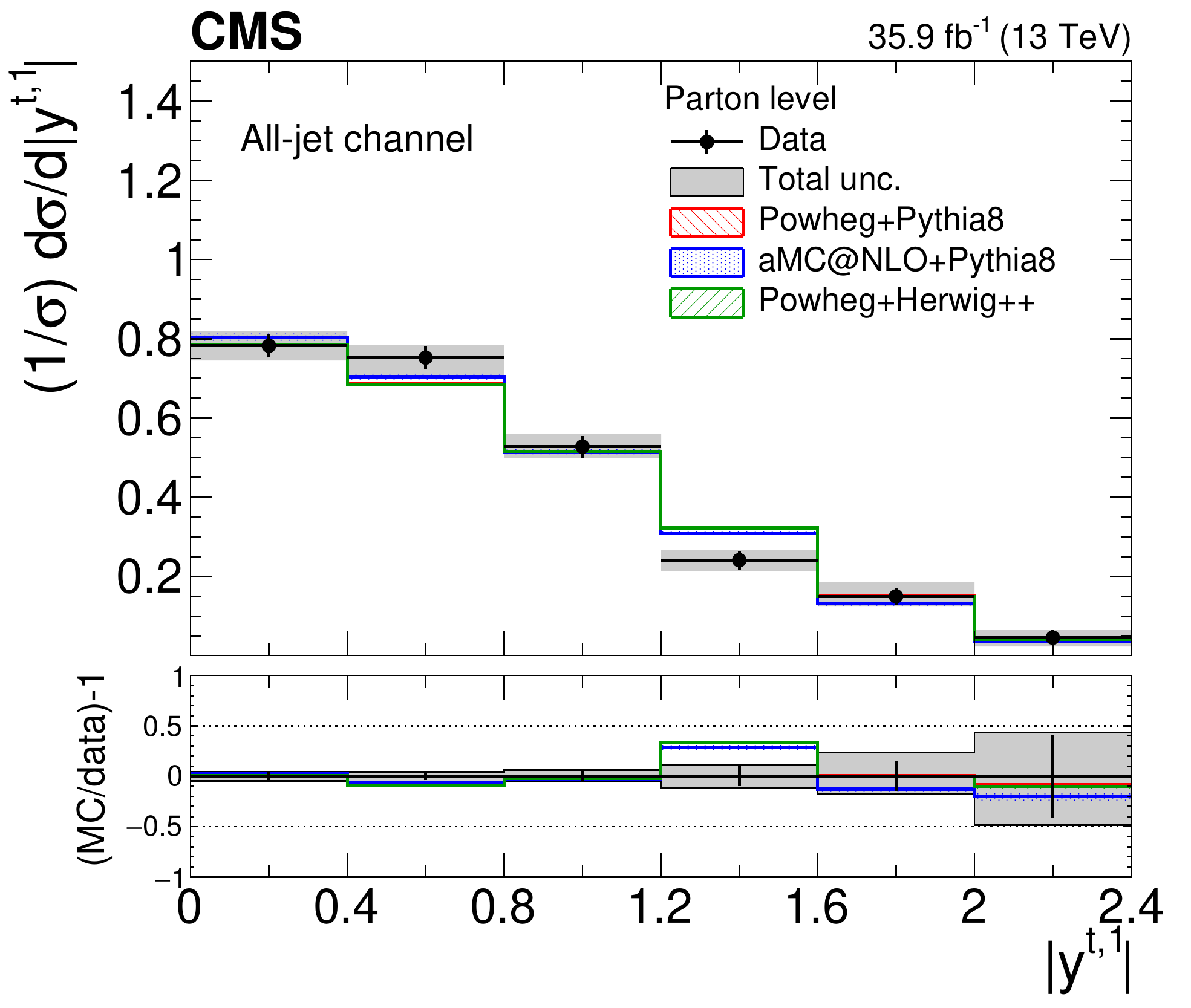} \\
    \includegraphics[width=\cmsFigWidth]{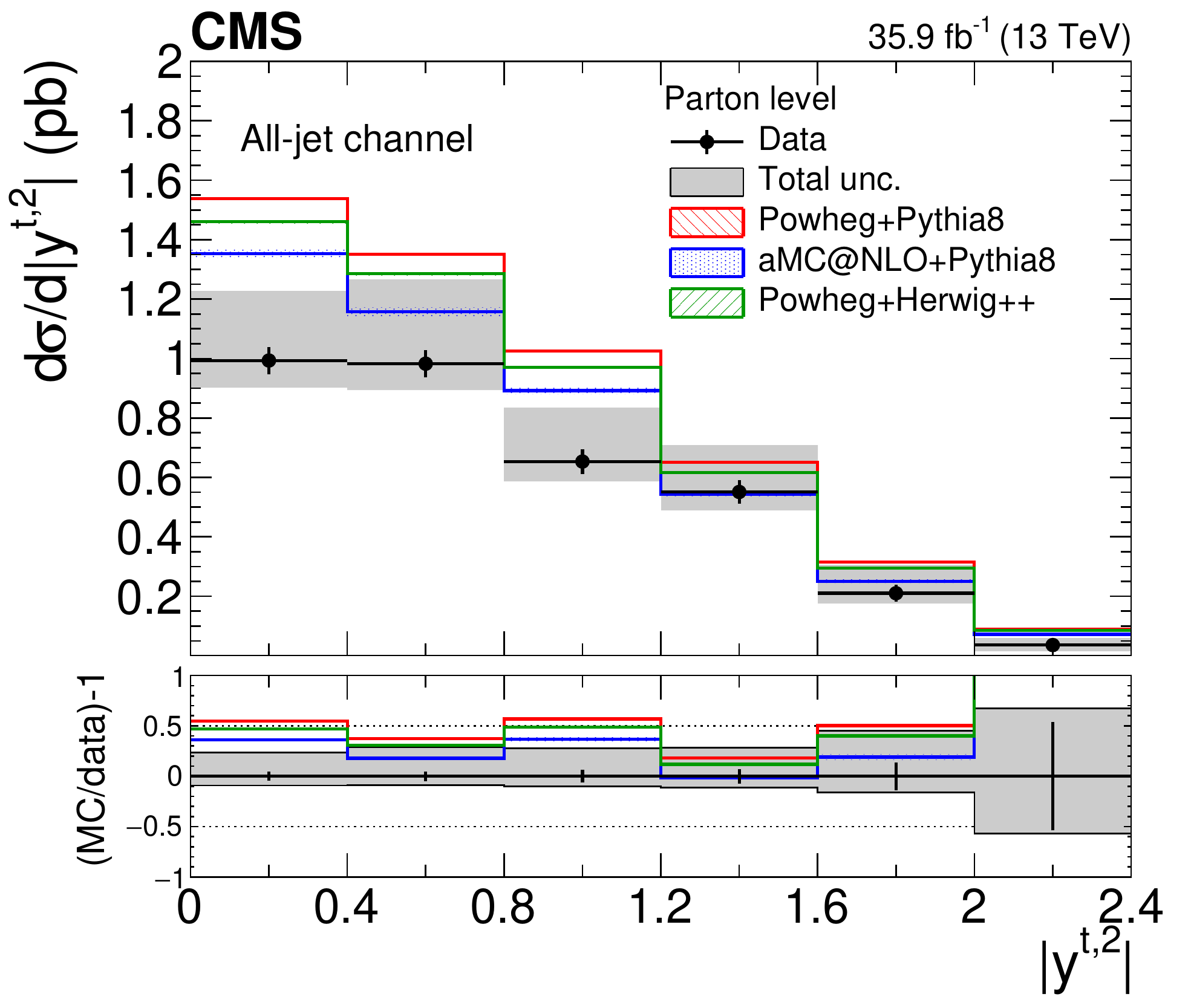}
    \includegraphics[width=\cmsFigWidth]{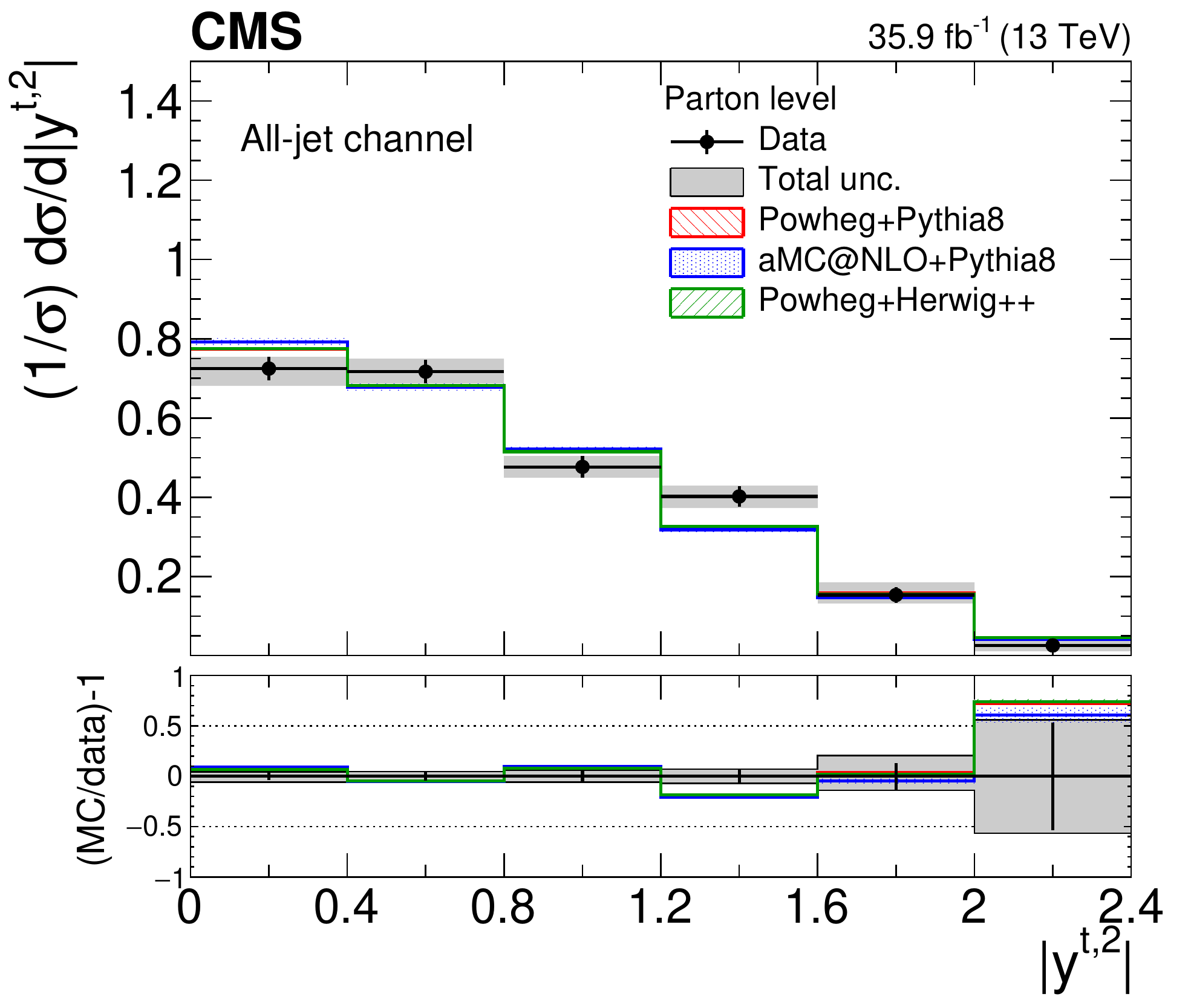}
    \caption{Differential cross section unfolded to the parton level, absolute (left) and normalized (right), as a function of the leading (upper row) and subleading (lower row) top quark $\abs{y}$ in the all-jet channel. The lower panel shows the ratio (MC/data)$-$1. The vertical bars on the data and in the ratio represent the statistical uncertainty in data, while the shaded band shows the total statistical and systematic uncertainty added in quadrature. The hatched bands show the statistical uncertainty of the MC samples.}
    \label{fig:Parton_jetY01}
\end{figure*}

\begin{figure*}[hbtp]
\centering
    \includegraphics[width=\cmsFigWidth]{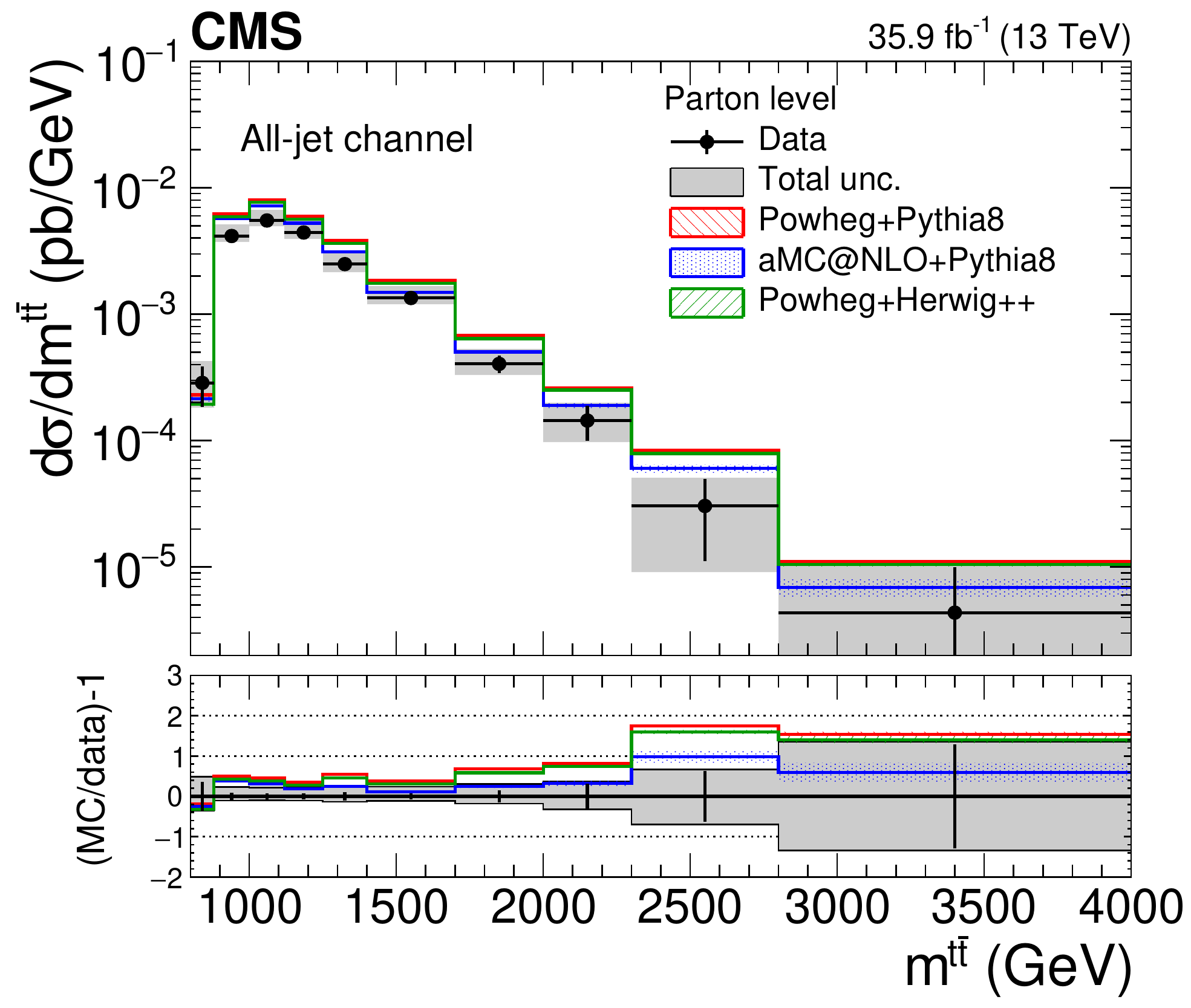}
    \includegraphics[width=\cmsFigWidth]{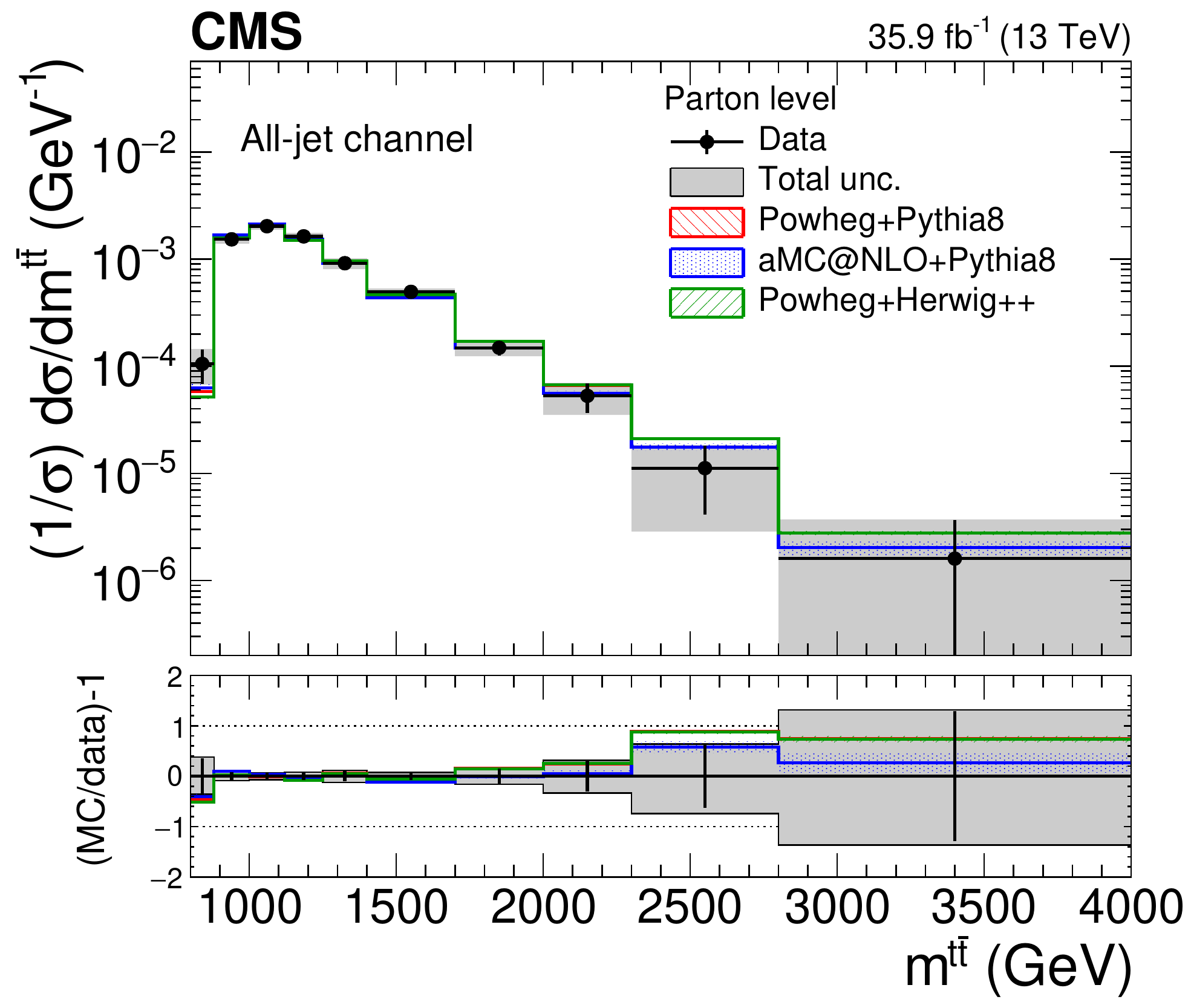} \\
    \includegraphics[width=\cmsFigWidth]{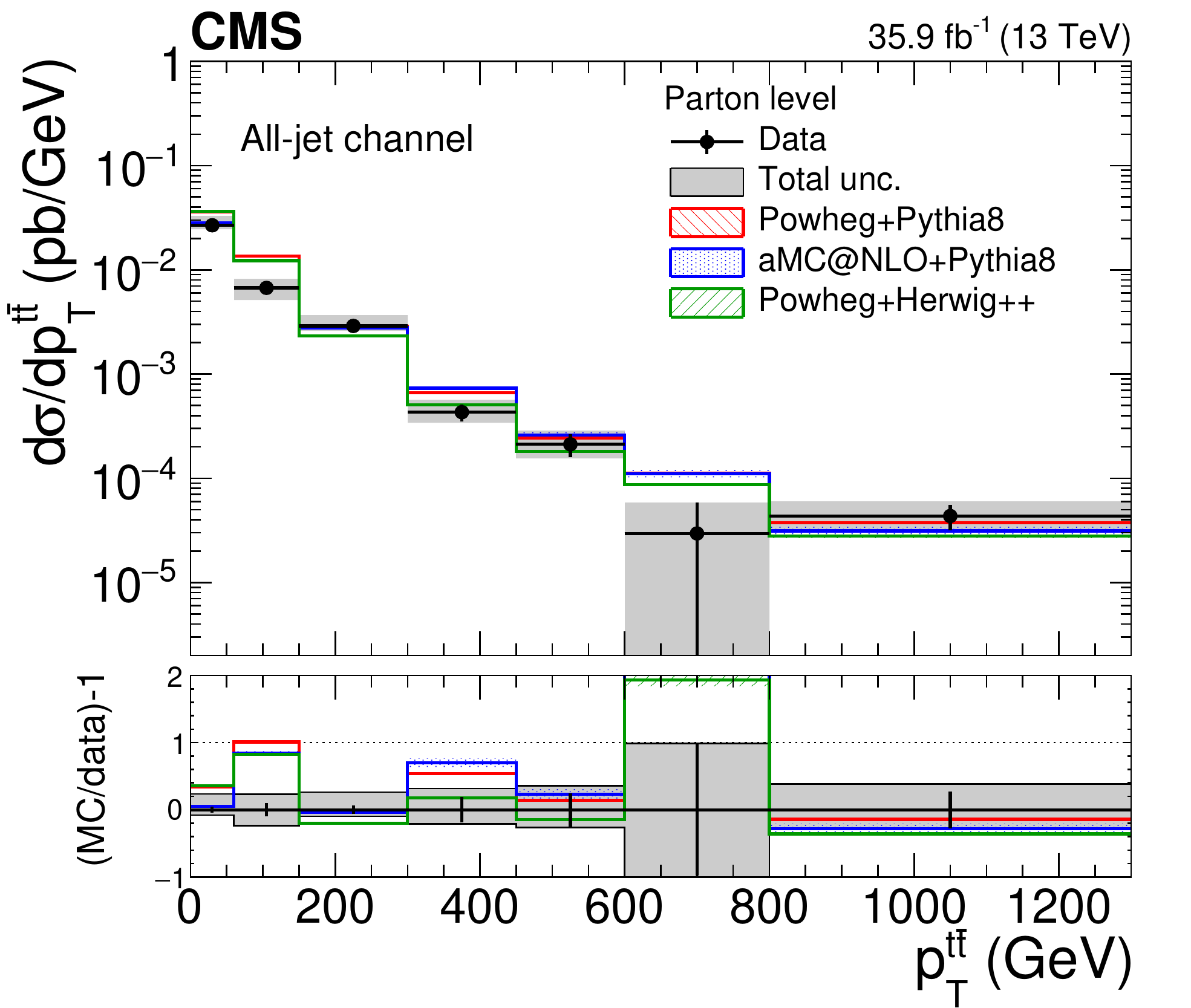}
    \includegraphics[width=\cmsFigWidth]{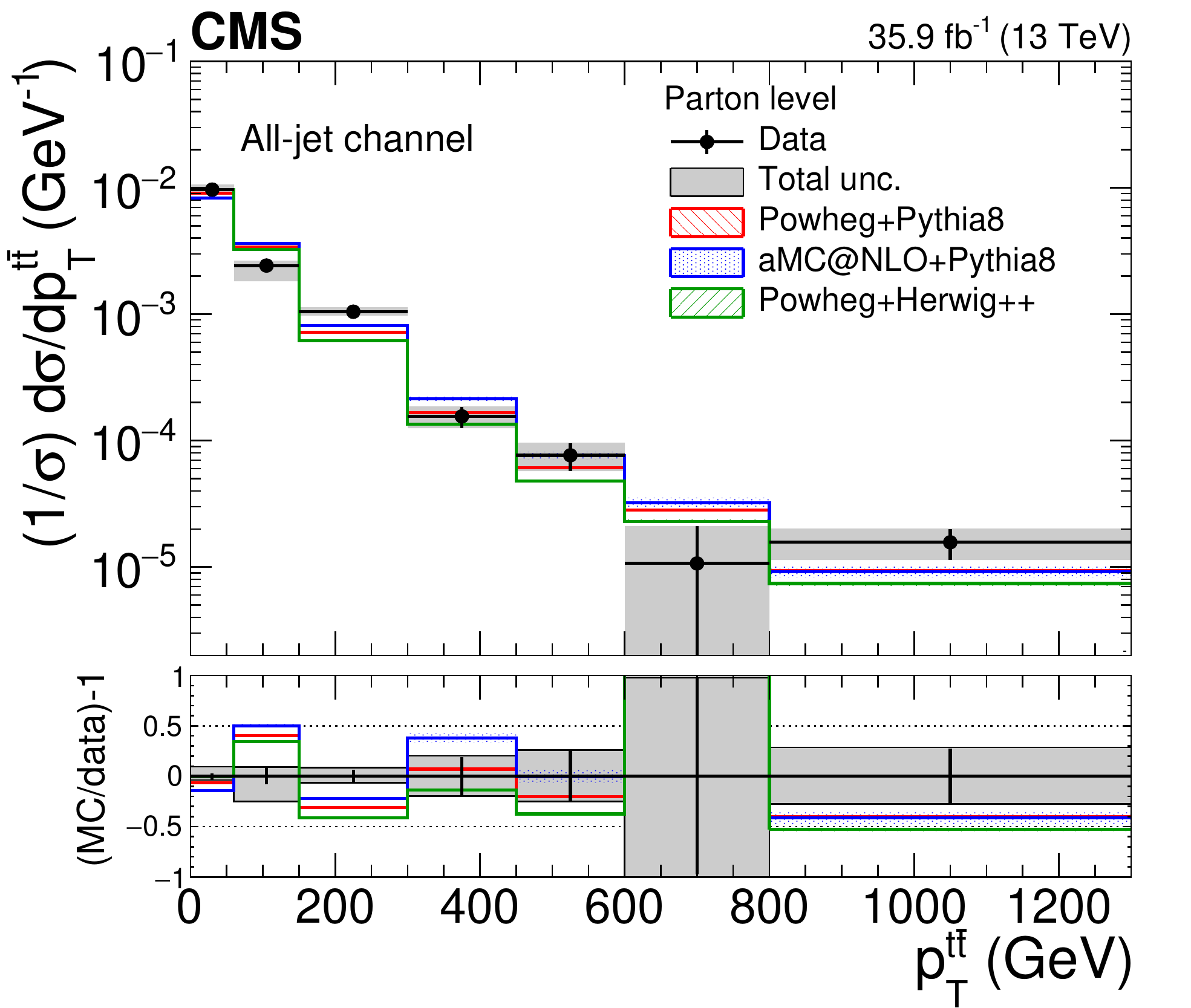} \\
    \includegraphics[width=\cmsFigWidth]{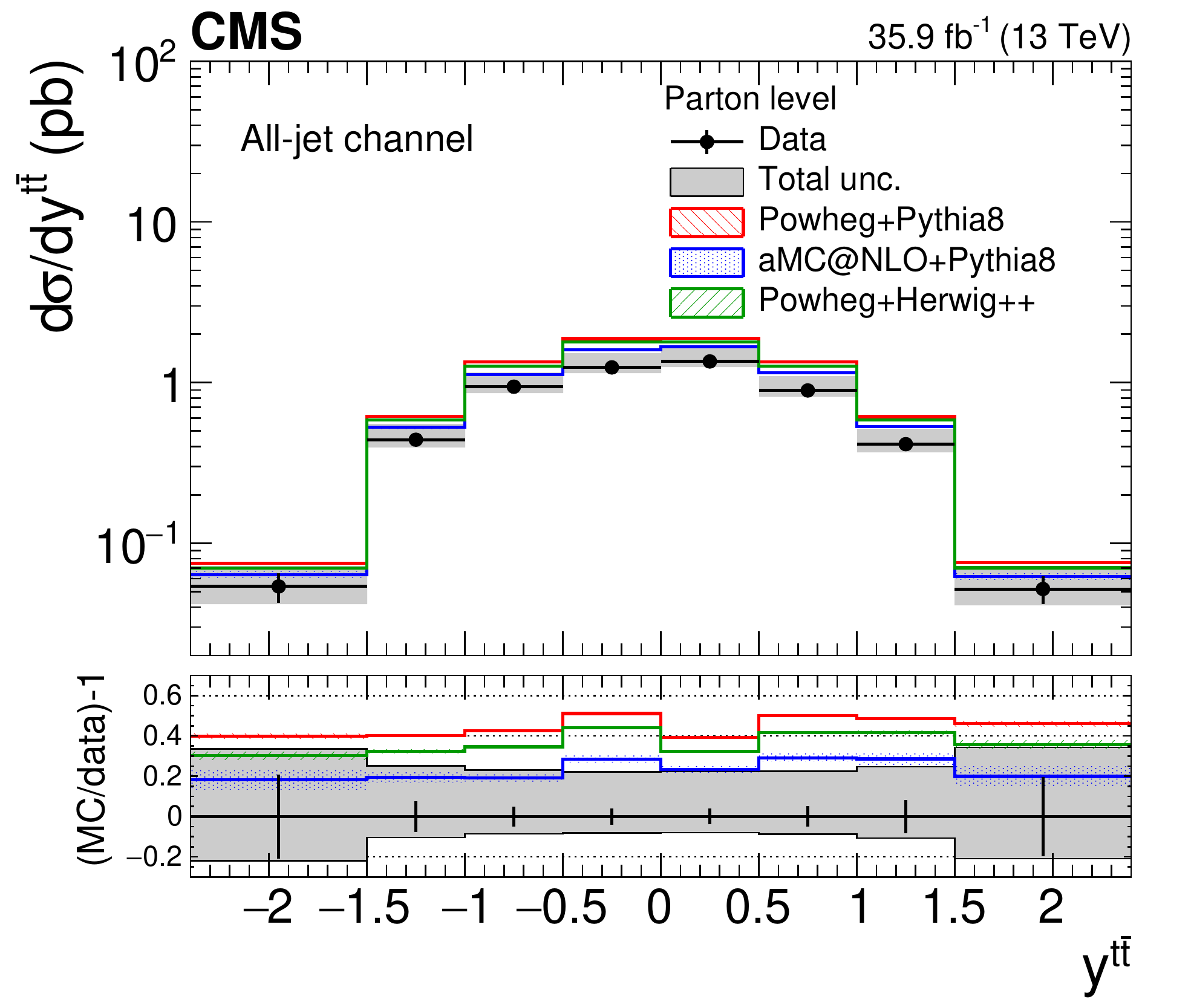}
    \includegraphics[width=\cmsFigWidth]{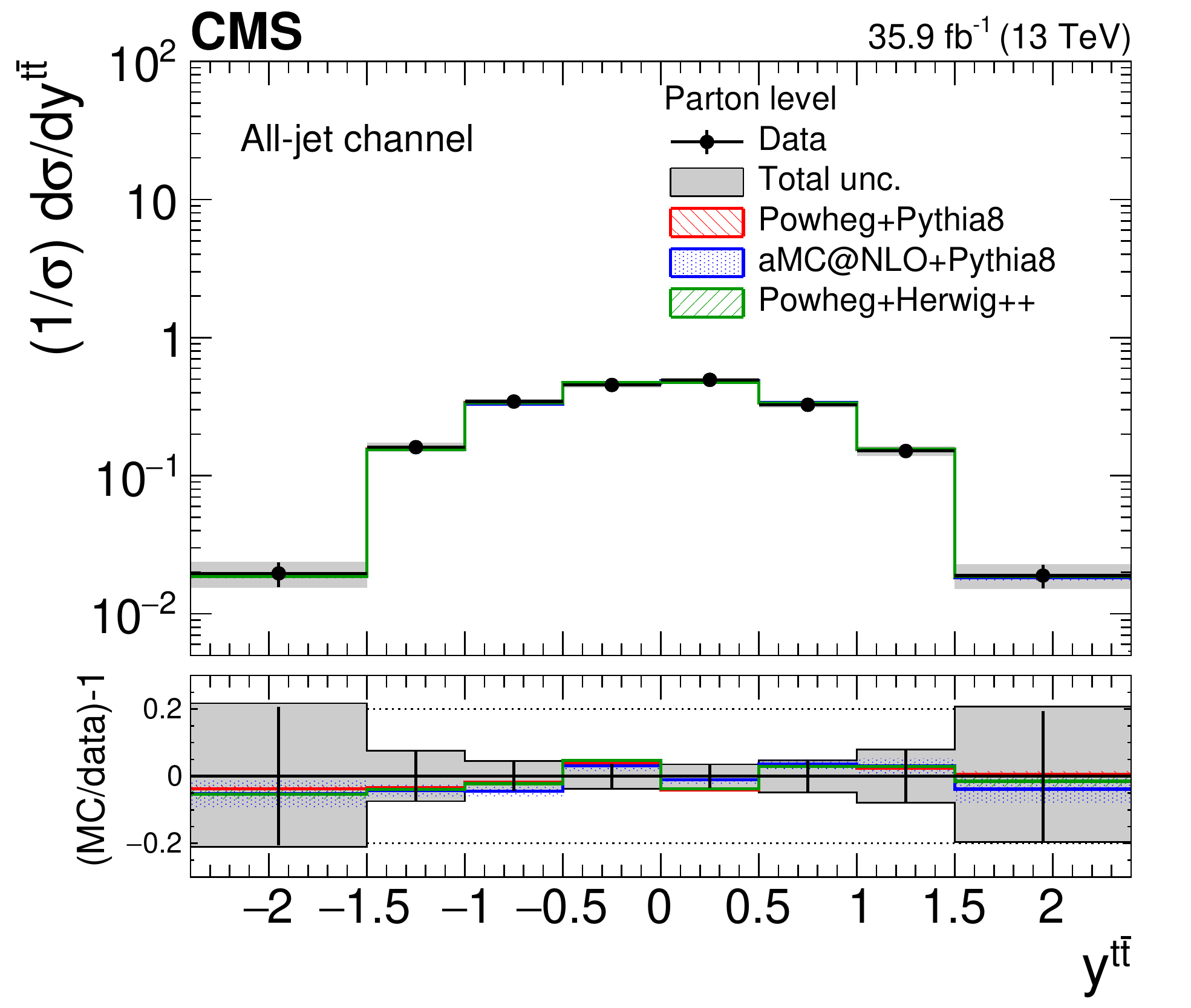}
    \caption{Differential cross section unfolded to the parton level, absolute (left) and normalized (right), as a function of $m^{\ttbar}$ (upper row), $\pt^{\ttbar}$ (middle row), and $y^{\ttbar}$ (lower row) in the all-jet channel. The lower panel shows the ratio (MC/data)$-$1. The vertical bars on the data and in the ratio represent the statistical uncertainty in data, while the shaded band shows the total statistical and systematic uncertainty added in quadrature. The hatched bands show the statistical uncertainty of the MC samples.}
    \label{fig:Parton_mJJ_ptJJ_yJJ}
\end{figure*}

\begin{figure*}[hbtp]
\centering
    \includegraphics[width=\cmsFigWidth]{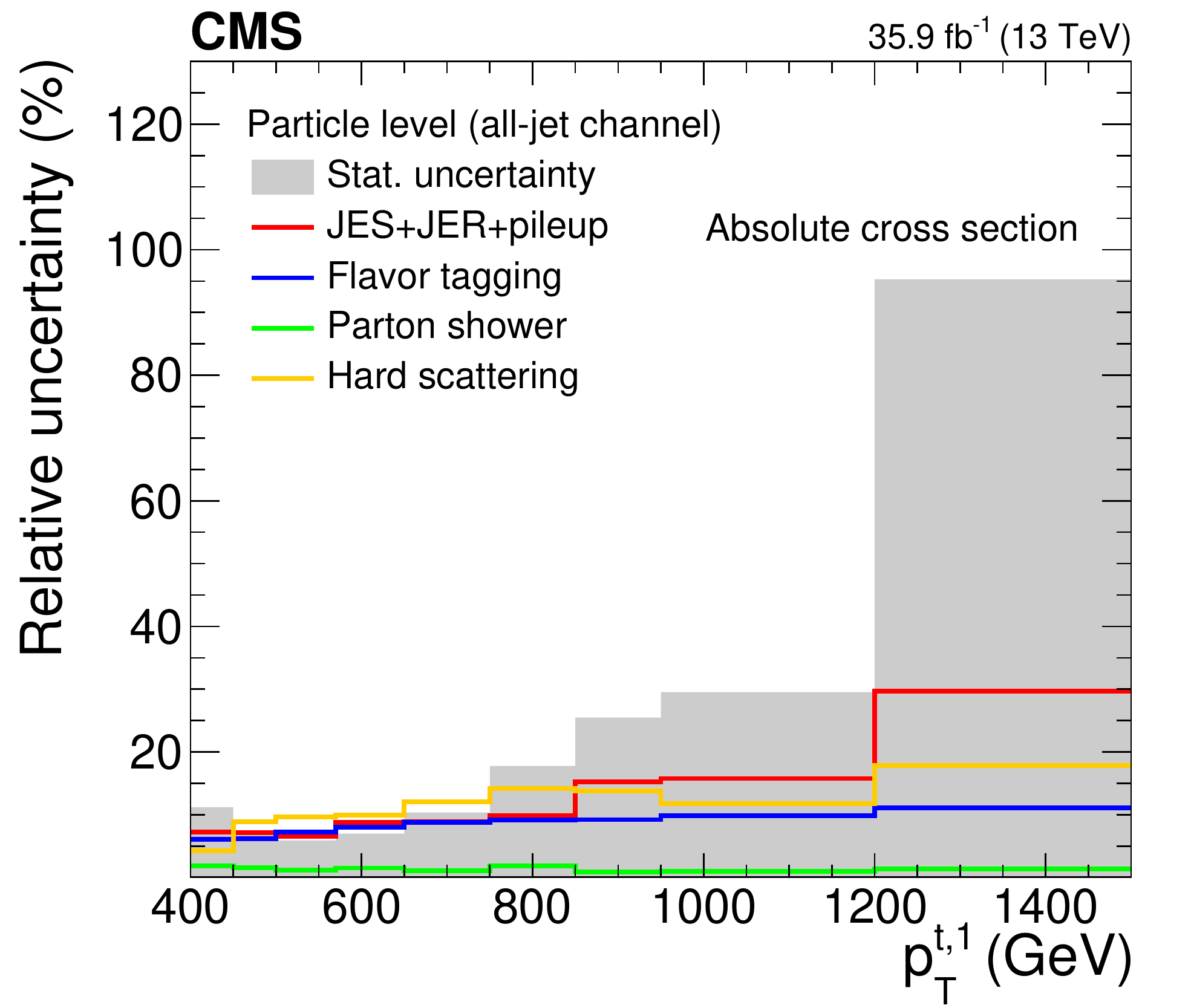}
    \includegraphics[width=\cmsFigWidth]{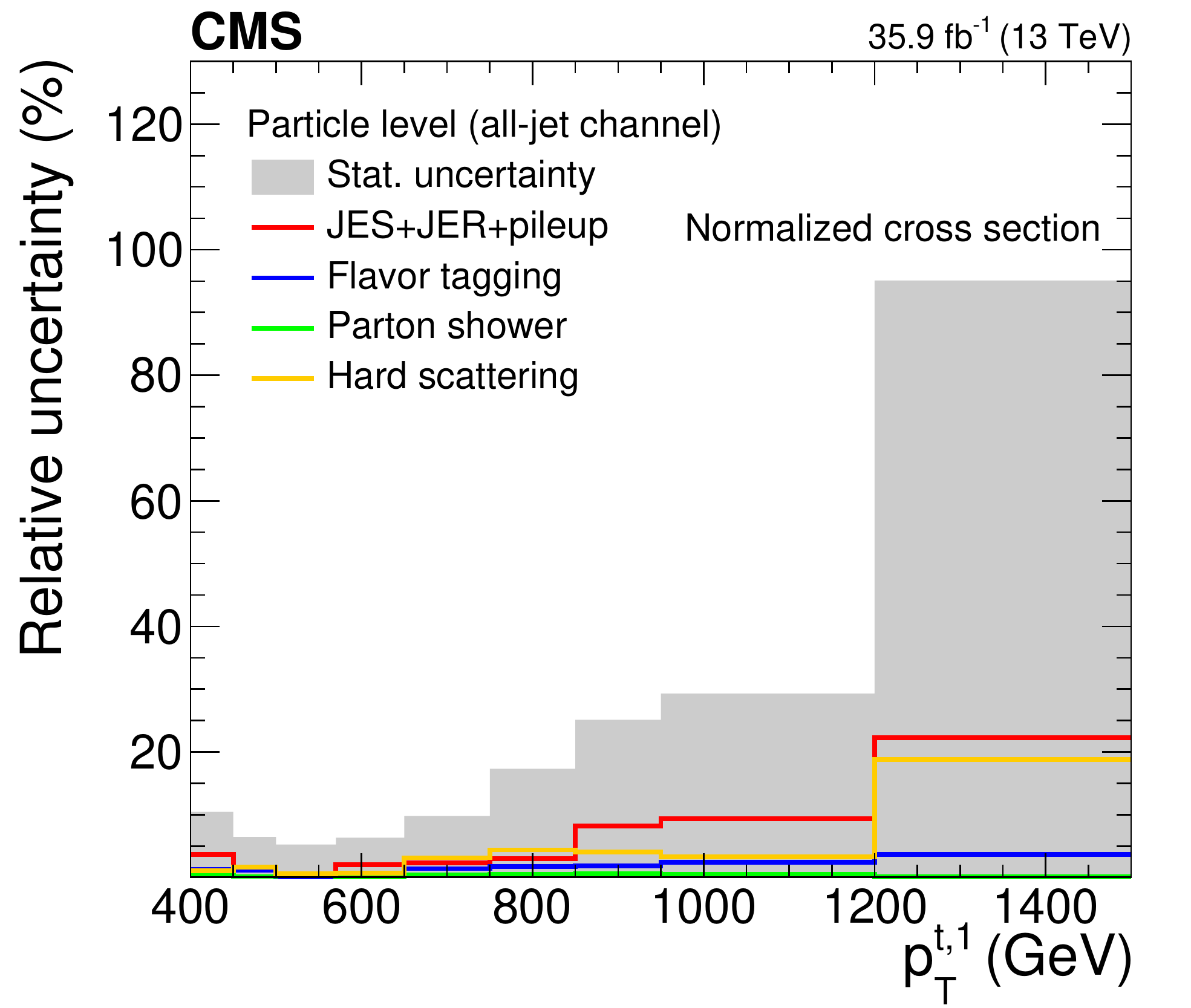}
    \includegraphics[width=\cmsFigWidth]{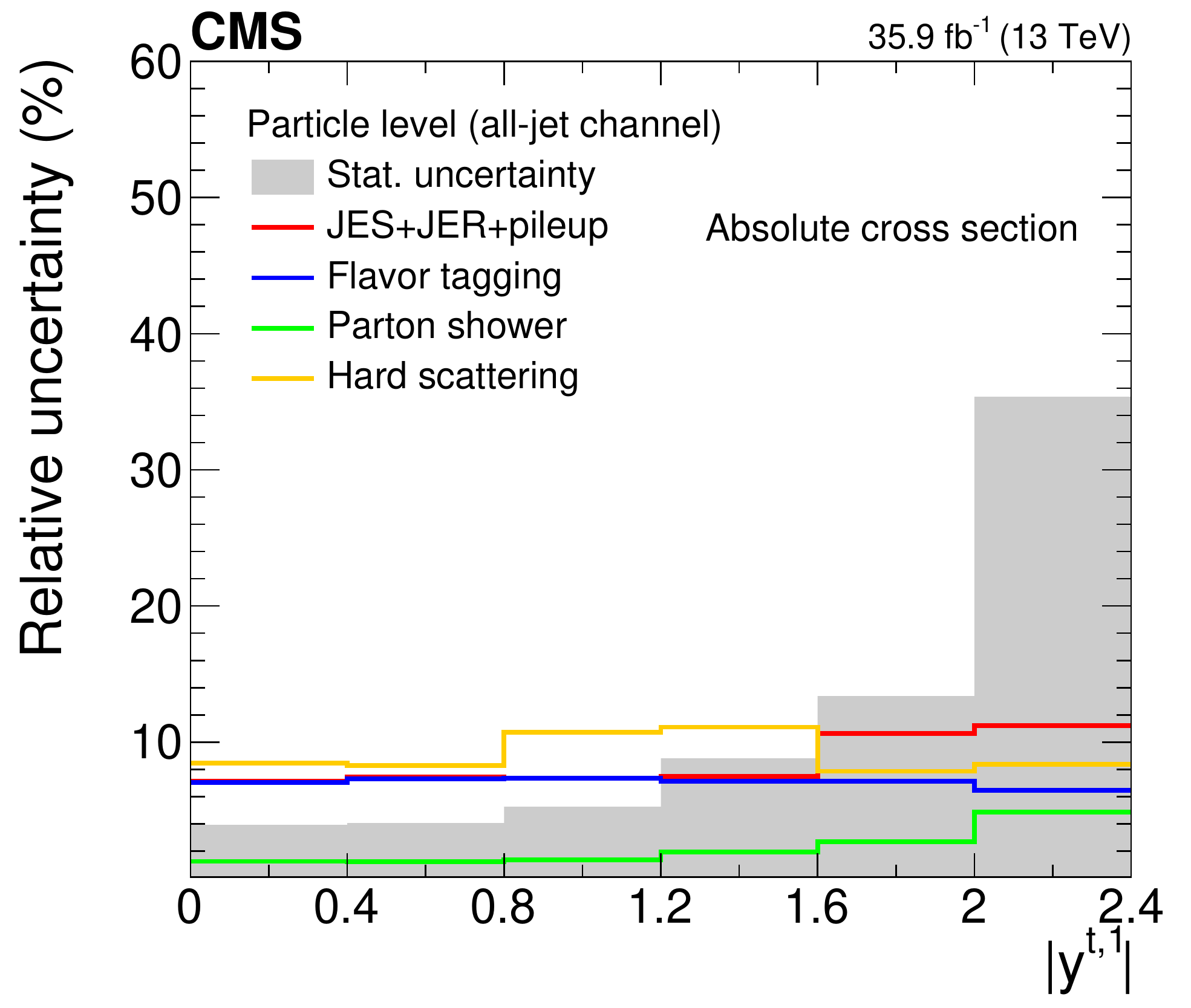}
    \includegraphics[width=\cmsFigWidth]{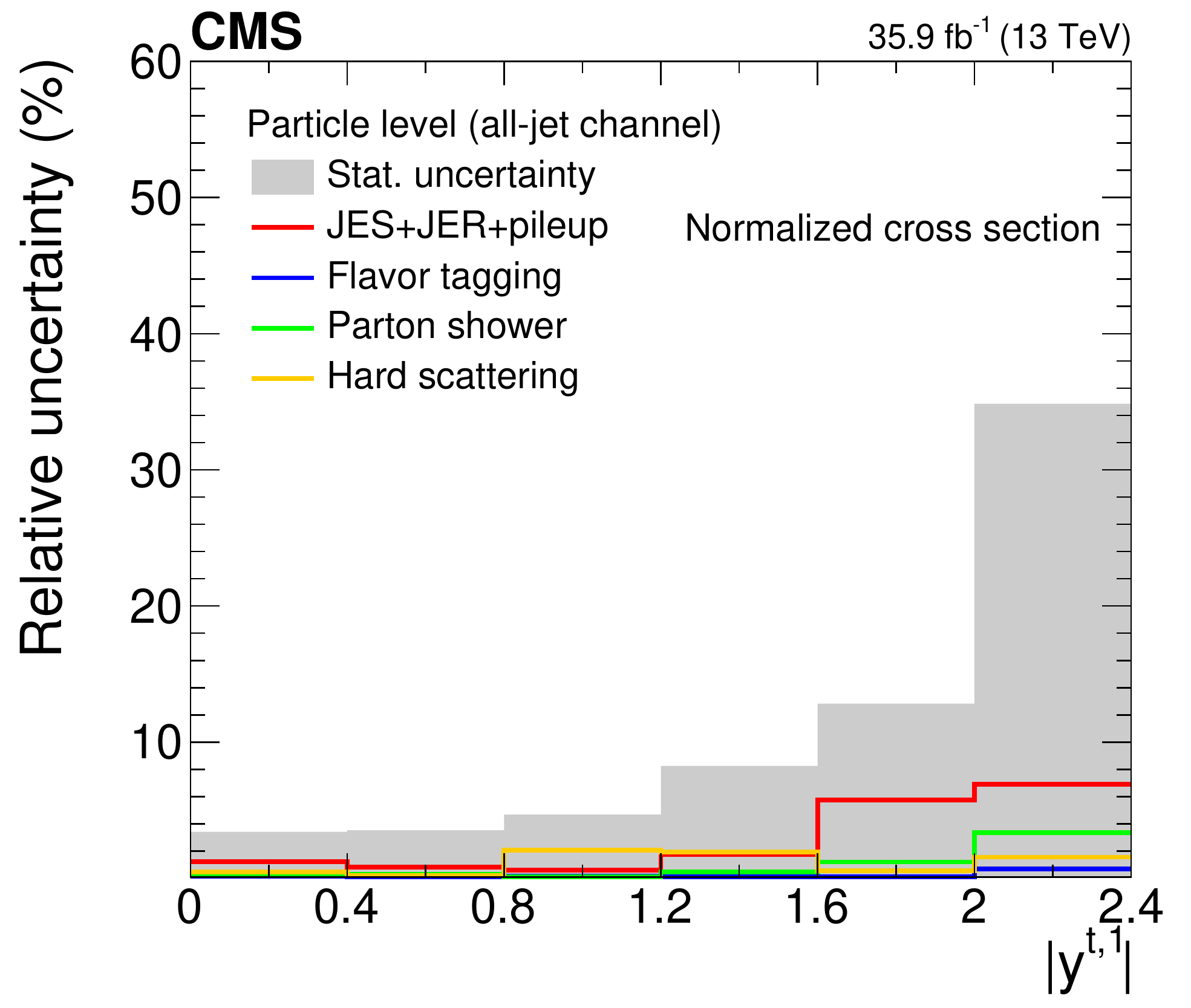}
    \caption{Breakdown of the uncertainties in the absolute (left column) and normalized (right column) measurement at the particle level, as a function of the leading top quark \pt (upper row) and $\abs{y}$ (lower row) in the all-jet channel. The shaded band shows the statistical uncertainty, while the solid lines show the systematic uncertainties grouped in four categories: a) uncertainty due to pileup and the JES and JER of the large-$R$ jets, b) uncertainty due to flavor tagging of the subjets, c) uncertainty due to the modeling of the parton shower, and d) uncertainty due to the modeling of the hard scattering.}
    \label{fig:syst_had_particle}
\end{figure*}

\begin{figure*}[hbtp]
\centering
    \includegraphics[width=\cmsFigWidth]{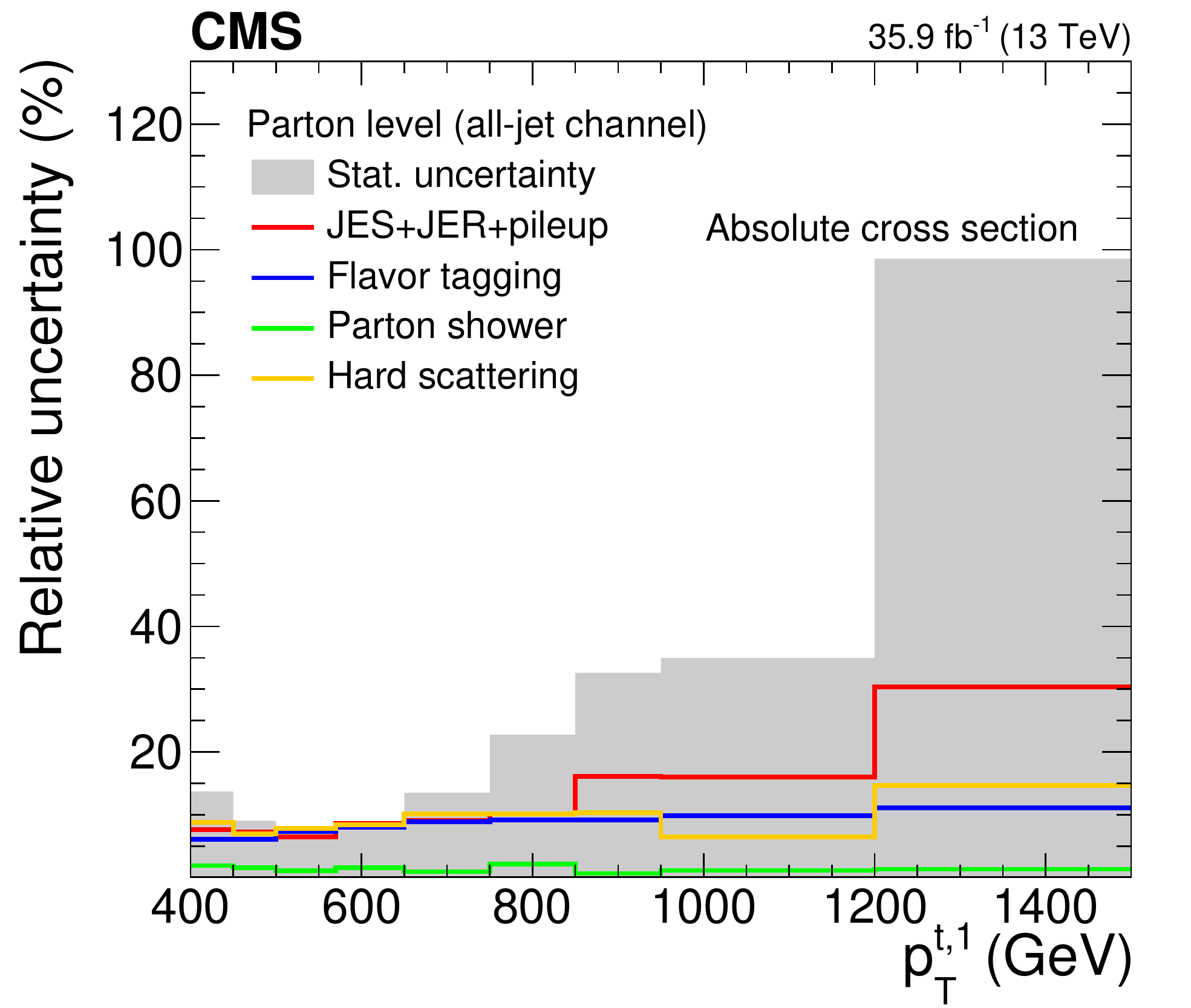}
    \includegraphics[width=\cmsFigWidth]{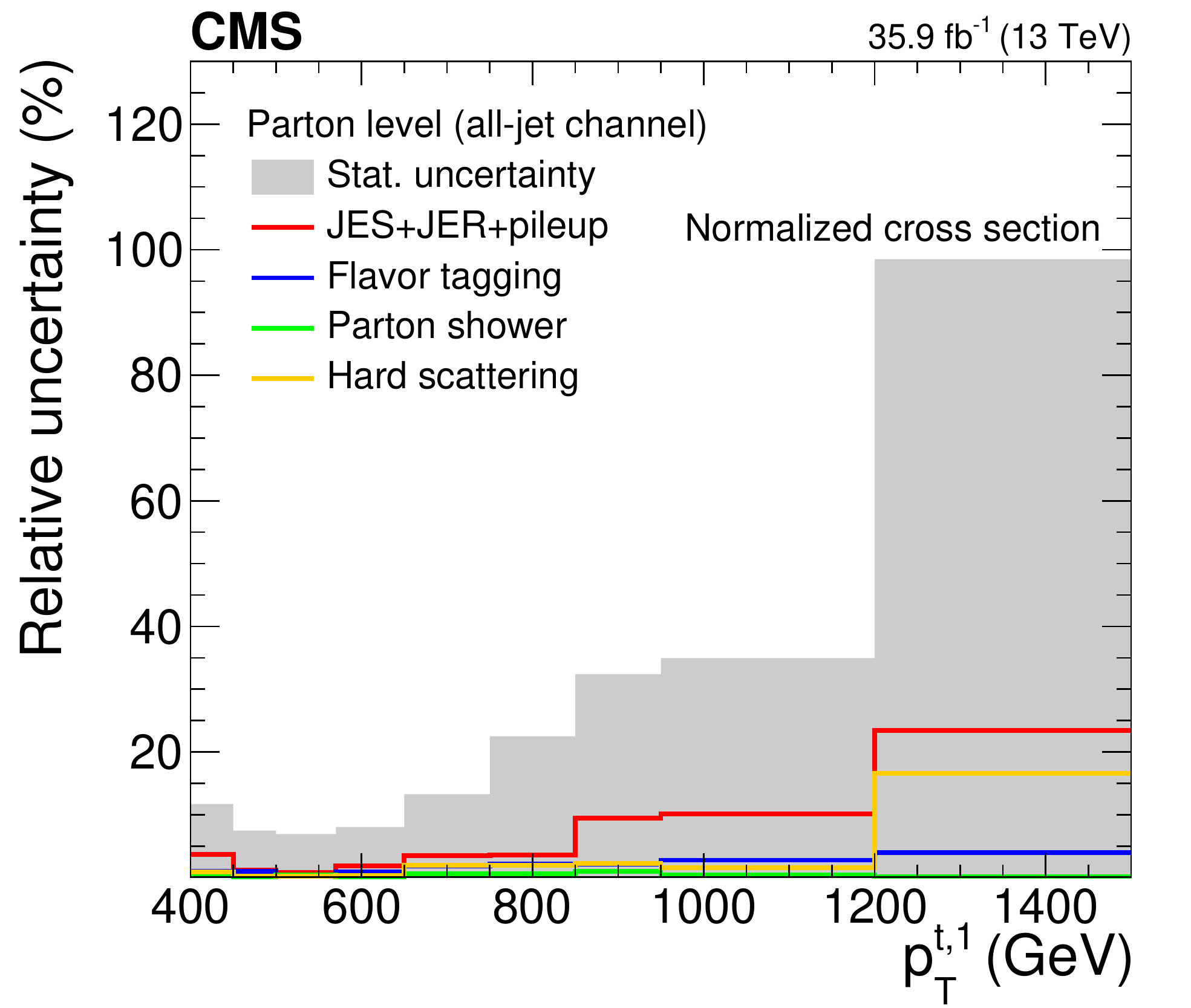}
    \includegraphics[width=\cmsFigWidth]{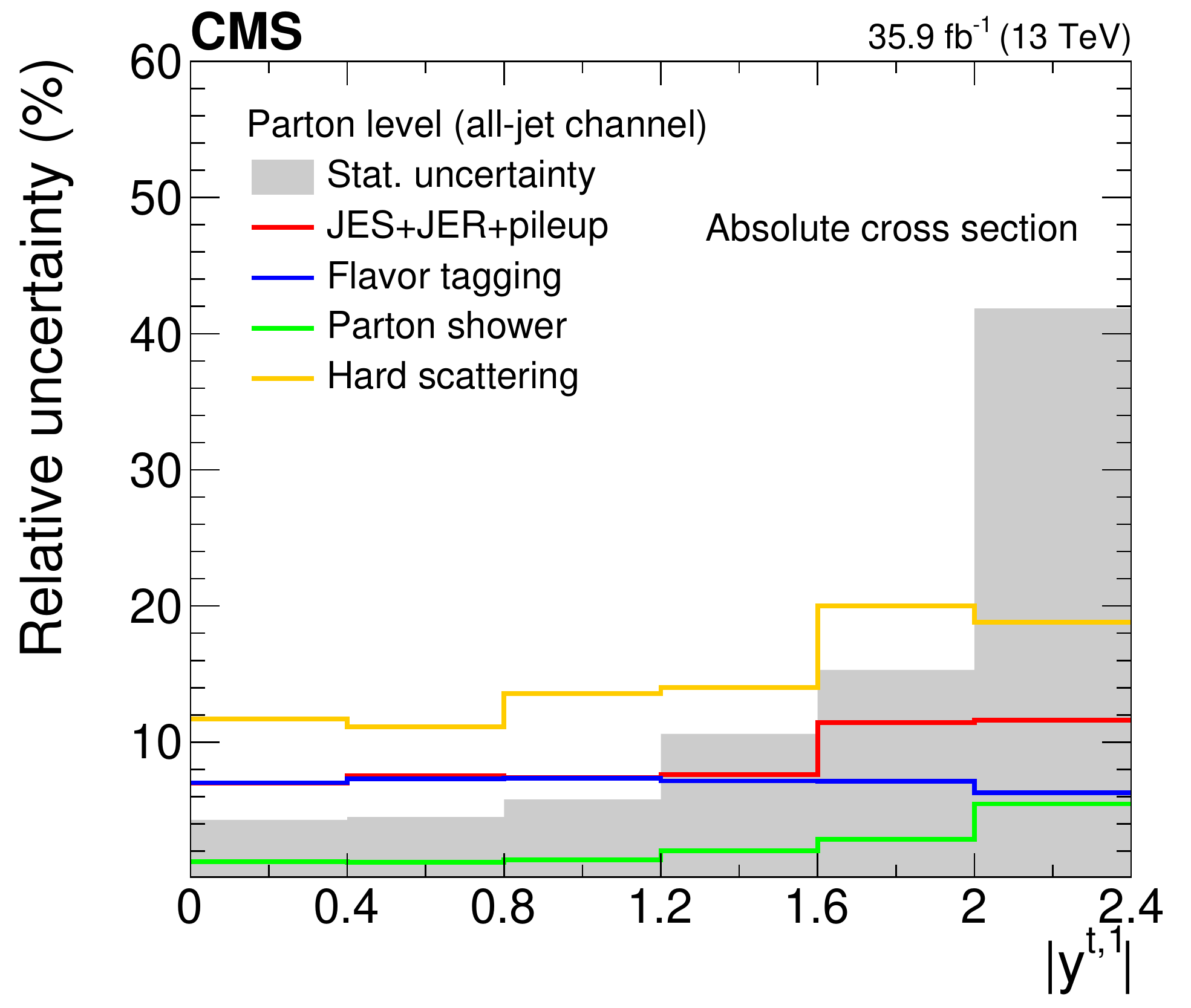}
    \includegraphics[width=\cmsFigWidth]{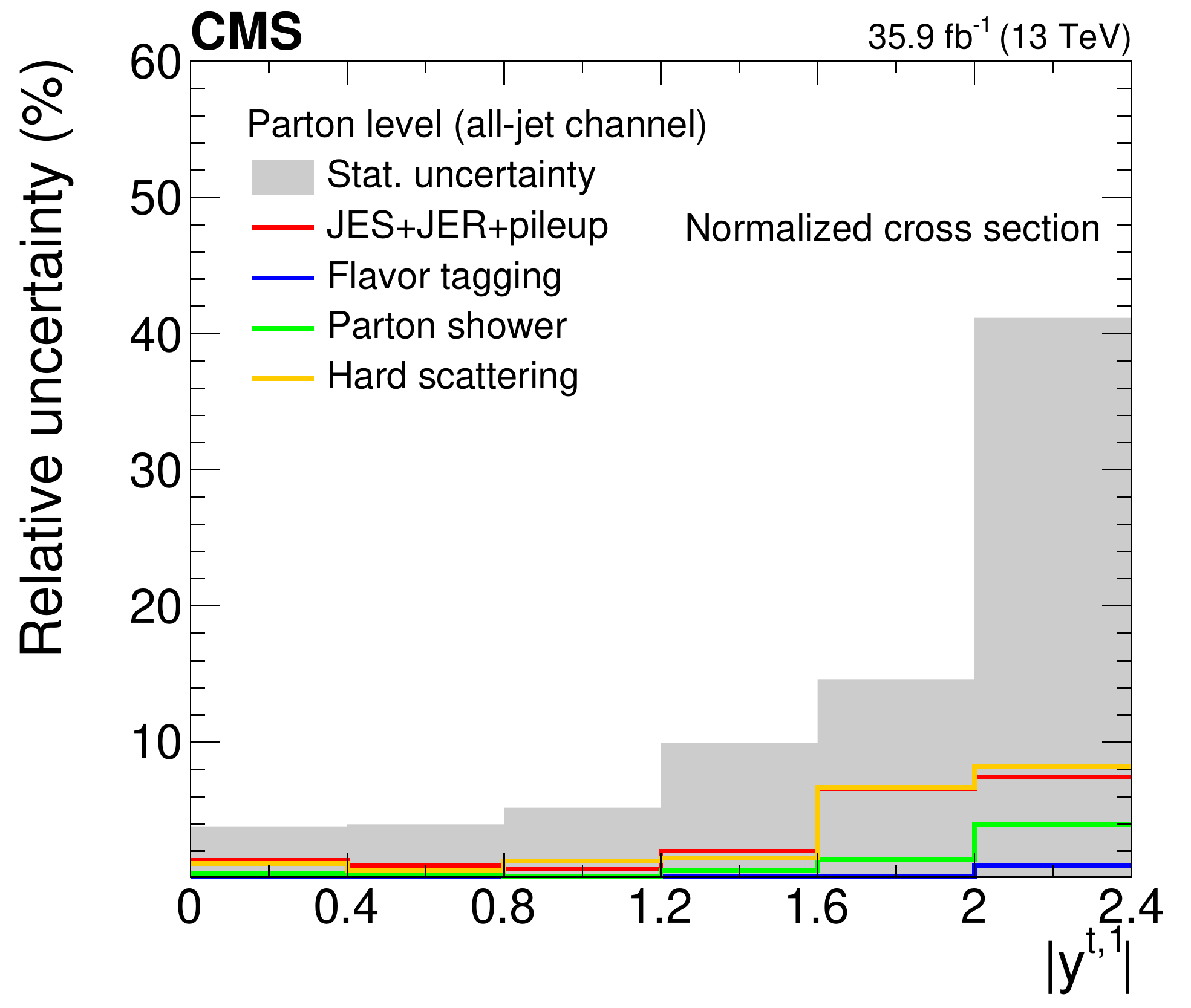}
    \caption{Breakdown of the uncertainties in the absolute (left column) and normalized (right column) measurement at the parton level, as a function of the leading top quark \pt (upper row) and $\abs{y}$ (lower row) in the all-jet channel. The shaded band shows the statistical uncertainty, while the solid lines show the systematic uncertainties grouped in four categories: a) uncertainty due to pileup and the JES and JER of the large-$R$ jets, b) uncertainty due to flavor tagging of the subjets, c) uncertainty due to the modeling of the parton shower, and d) uncertainty due to the modeling of the hard scattering.}
    \label{fig:syst_had_parton}
\end{figure*}

For the all-jet channel, the measurement of the unfolded differential cross section in bin $j$ of the variable $x$ is performed using Eq.~(\ref{eq:xs_unfolded}). To estimate the uncertainty in the measurement, the entire procedure of the signal extraction, unfolding with different response matrices, and extrapolation to the particle- or parton-level phase space is repeated for every source of uncertainty discussed in Section~\ref{sec:Sys}. The unfolded cross sections at the particle (parton) level are shown in Figs.~\ref{fig:Particle_jetPt01}--\ref{fig:Particle_mJJ_ptJJ_yJJ} (\ref{fig:Parton_jetPt01}--\ref{fig:Parton_mJJ_ptJJ_yJJ}). Figures~\ref{fig:syst_had_particle} and~\ref{fig:syst_had_parton} show a summary of the statistical and the dominant systematic uncertainties in the differential cross section, as a function of the leading top quark \pt and $\abs{y}$ at the particle and parton levels, respectively.

\subsection{\texorpdfstring{$\ell$+jets}{ell+jets} channel}

In the $\ell$+jets channel, the differential \ttbar cross section is measured as a function of the \pt and $\abs{y}$ of the top quark that decays according to $\PQt \to \PW \PQb \to \PQq \PAQq' \PQb$. The measurement at the particle level defines a region of phase space that mimics the event selection criteria as detailed in Section~\ref{sec:definitionLevels}, but at the parton level corresponds to the phase space where the non-leptonically decaying top quark has $\pt > 400\GeV$. The $\ell$+jets \ttbar events are selected at the parton level, and the properties of the non-leptonically decaying top quarks are defined to represent the true top quark \pt values. 

The differential cross section is extracted from the signal-dominated 1t1b category. The distribution in the measured signal is determined by subtracting the estimated background contributions from the distribution in data, using the posterior normalizations from the fit given in Table~\ref{tab:combFit_counts}. To account for reconstruction efficiencies and bin migrations in signal, we use unregularized unfolding as described in Section~\ref{sec:Unf}. 
The unfolding relies on response matrices that map the \pt and $\abs{y}$ distributions for the $\PQt$-tagged jet to corresponding properties for either the particle-level \PQt jet candidate or the parton-level top quark.

Systematic uncertainties in the unfolded measurement receive contributions from the experimental and theoretical sources discussed in Section~\ref{sec:Sys}. The posterior values from the likelihood fit are used for the \PQt tagging efficiency, background normalizations, and lepton efficiencies, while the a priori values are used for the remaining uncertainties. For each systematic change that affects the distribution in \pt or $\abs{y}$, we define a separate response matrix that is used to unfold the data. The resulting uncertainties are added in quadrature to obtain the total uncertainty in the unfolded distribution. 

The data in the electron and muon channels are combined before the unfolding by adding the measured distributions and their response matrices into a single channel. The background contributions are also merged into a single channel before subtracting these from the measured distributions, with the exception of the electron and muon multijet backgrounds that are treated as separate sources. 

\begin{figure*}[hbtp]
\centering
\includegraphics[width=\cmsFigWidth]{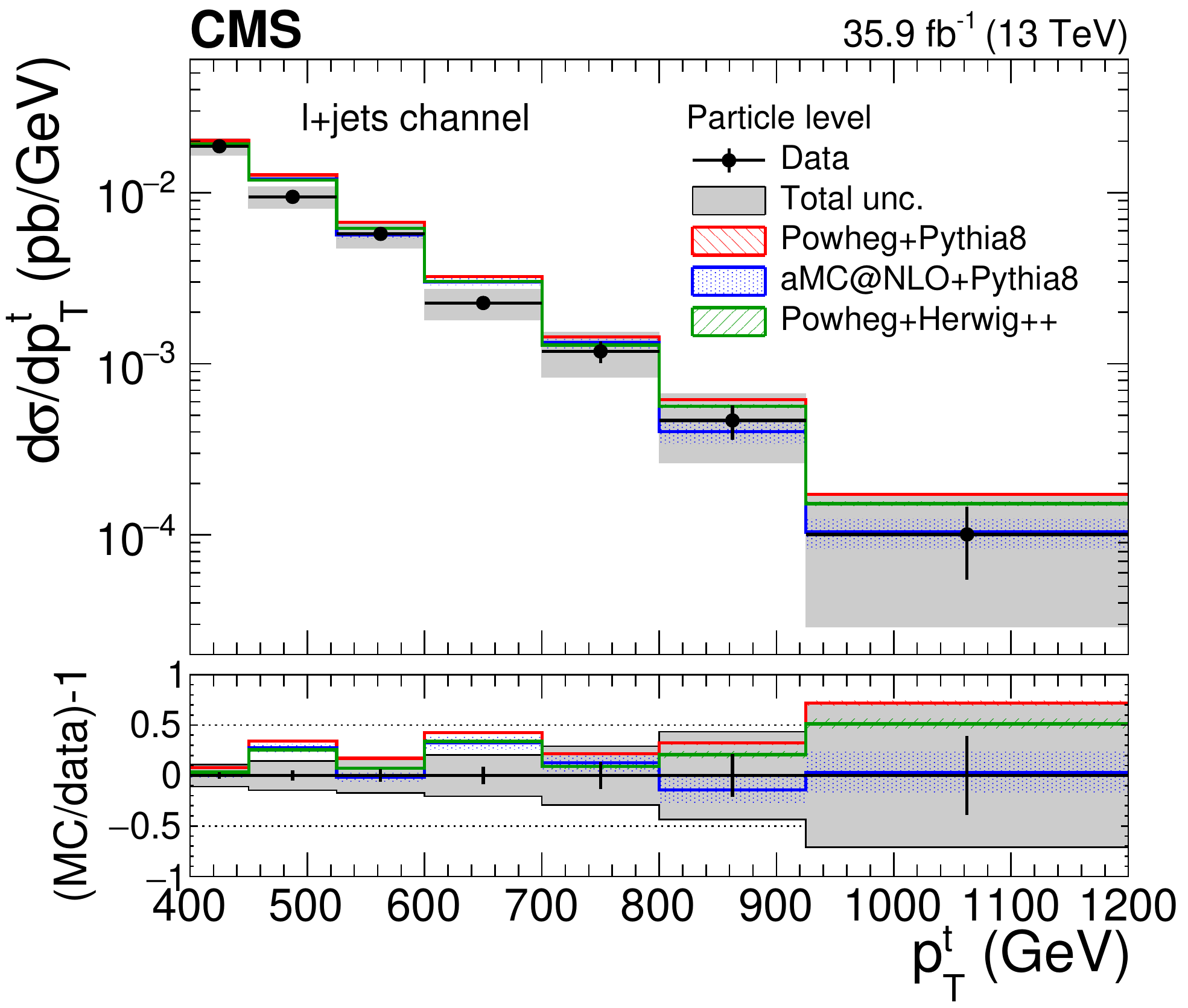}
\includegraphics[width=\cmsFigWidth]{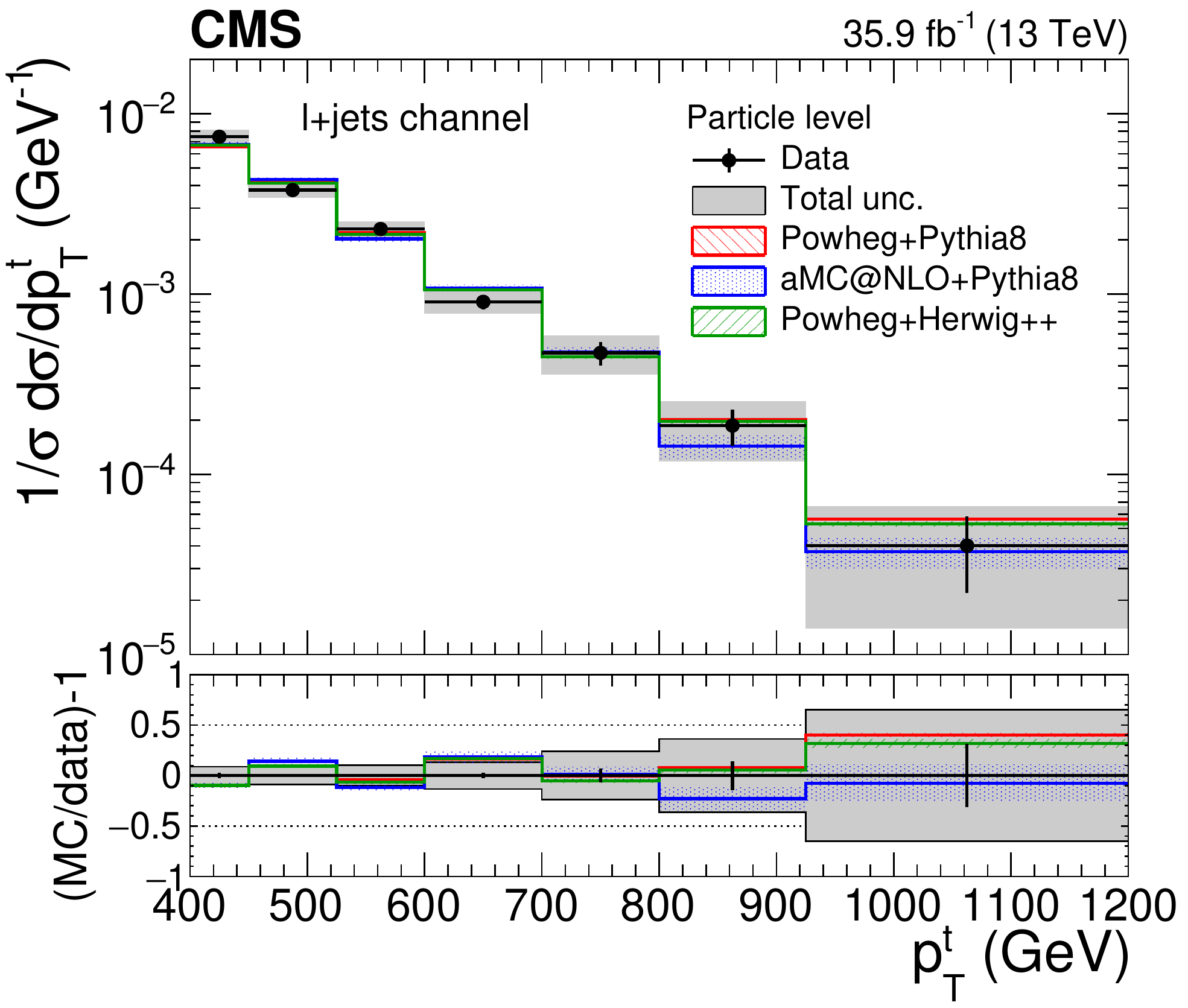} \\
\includegraphics[width=\cmsFigWidth]{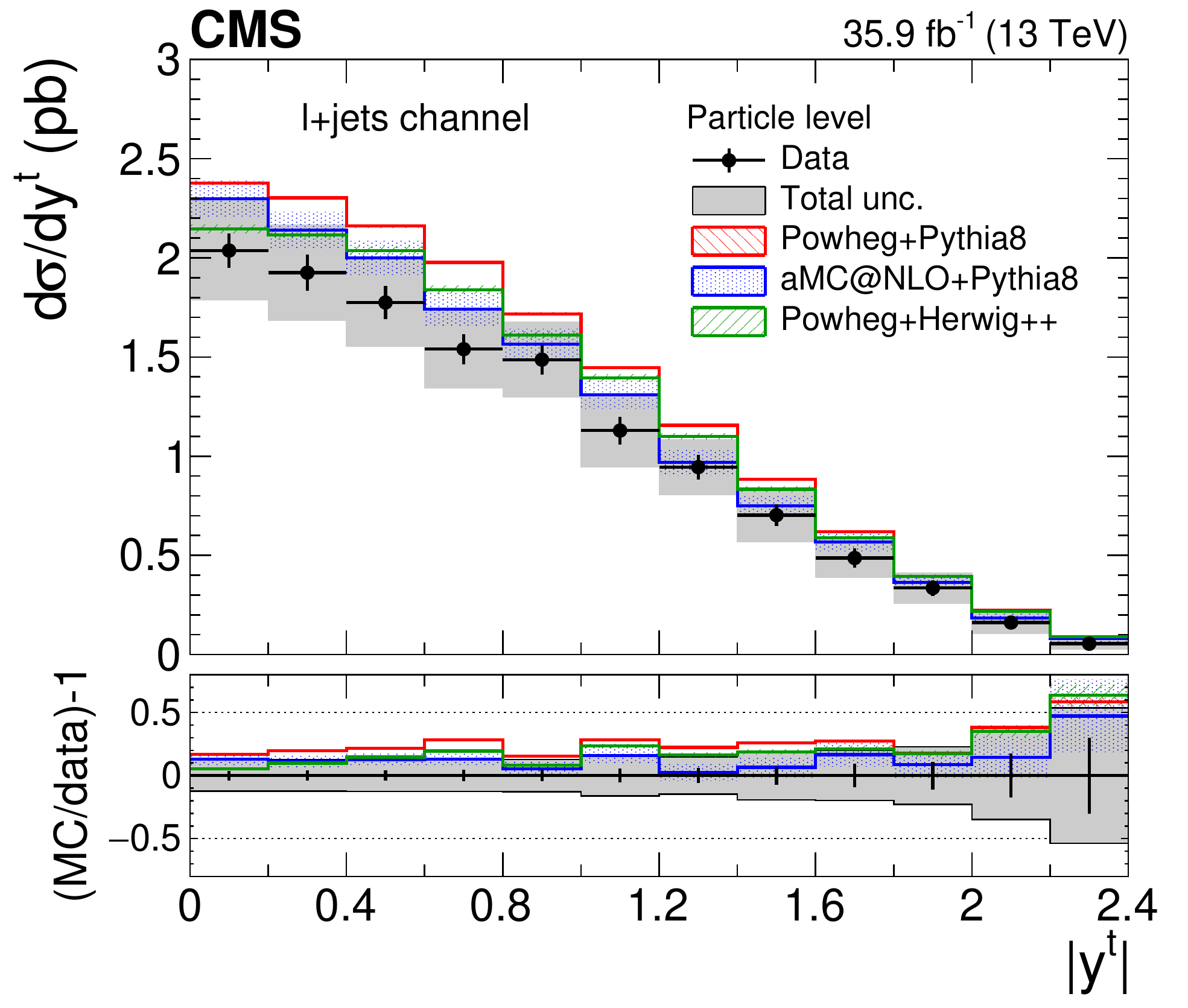}
\includegraphics[width=\cmsFigWidth]{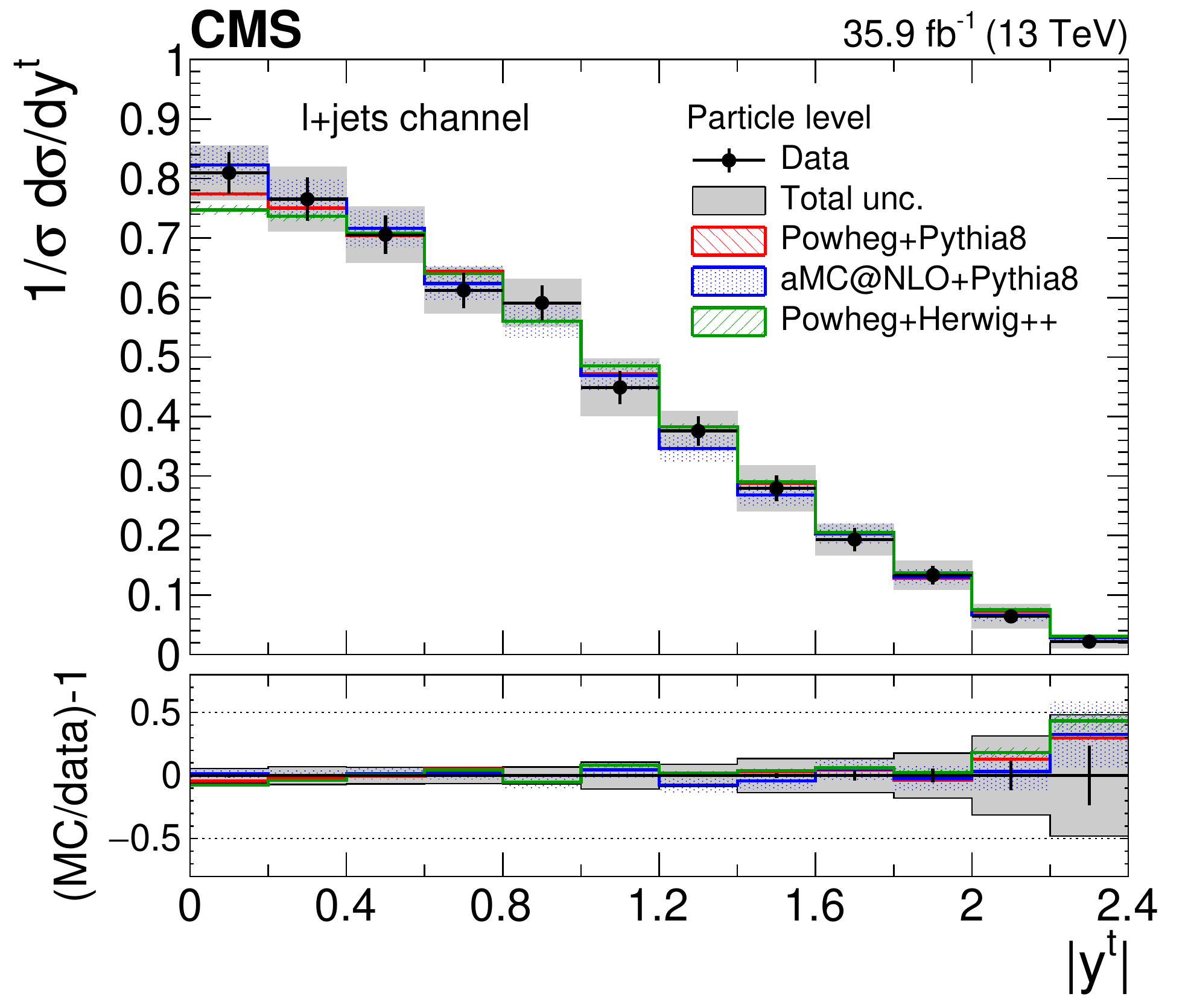}
\caption{\label{fig:lepjet_xsec_particle_pt_y} Differential cross section measurements at the particle level, as a function of the particle-level \PQt jet \pt (upper row) and $\abs{y}$ (lower row) for the $\ell$+jets channel. Both absolute (left column) and normalized (right column) cross sections are shown. The lower panel shows the ratio (MC/data)$-$1. The vertical bars on the data and in the ratio represent the statistical uncertainty in data, while the shaded band shows the total statistical and systematic uncertainty added in quadrature. The hatched bands show the statistical uncertainty of the MC samples.}
\end{figure*}

\begin{figure*}[htb]
\centering
\includegraphics[width=\cmsFigWidth]{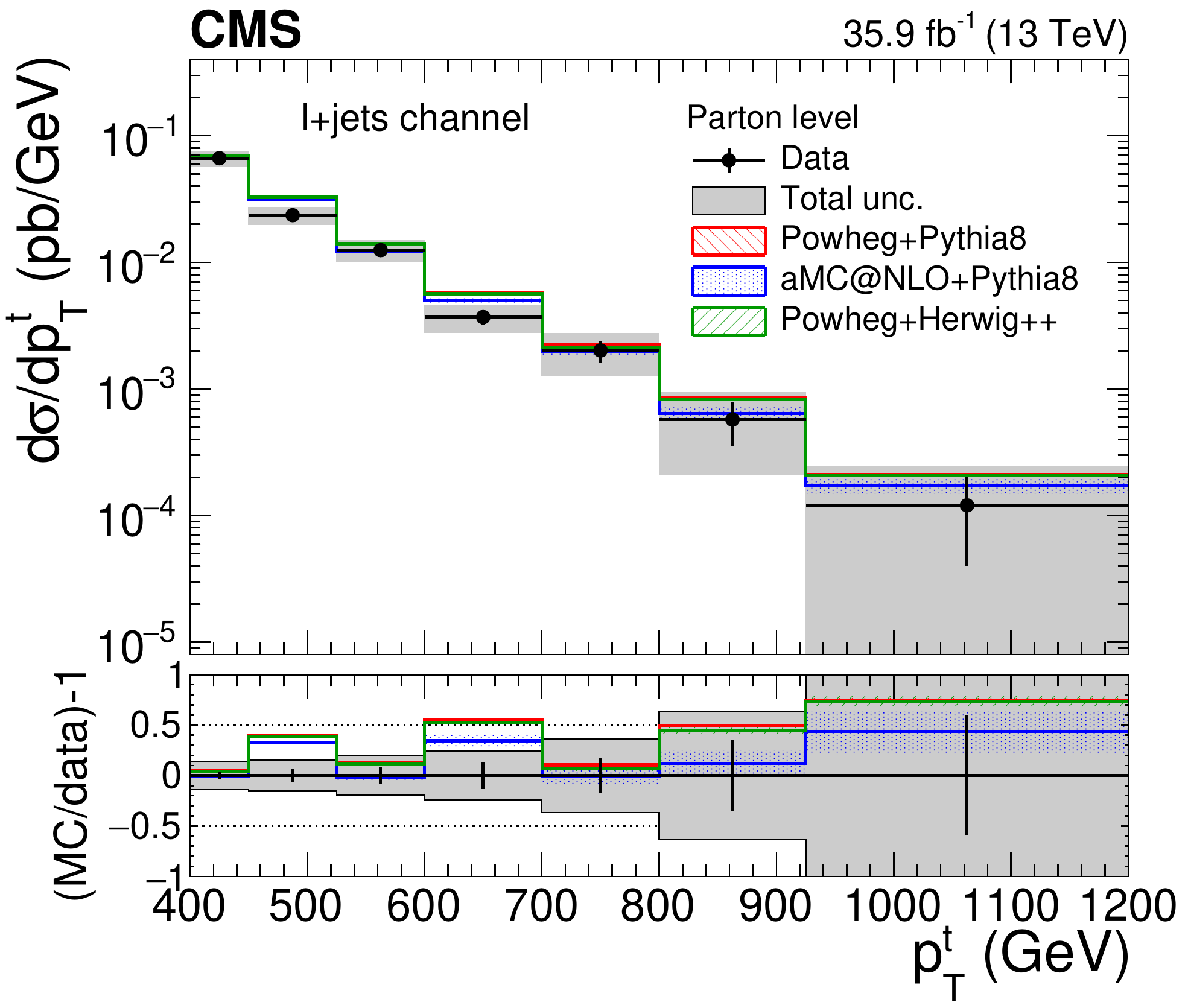}
\includegraphics[width=\cmsFigWidth]{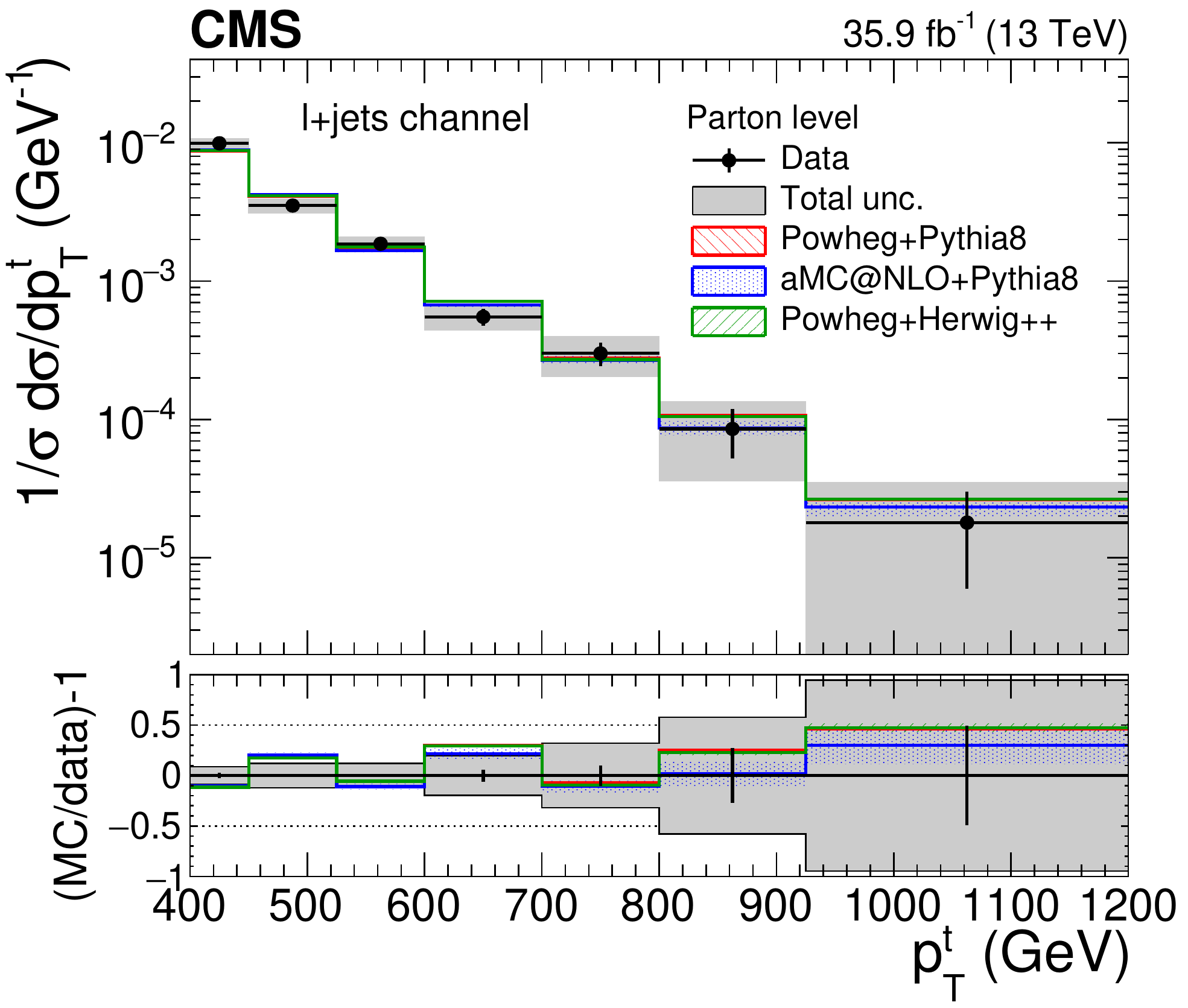} \\
\includegraphics[width=\cmsFigWidth]{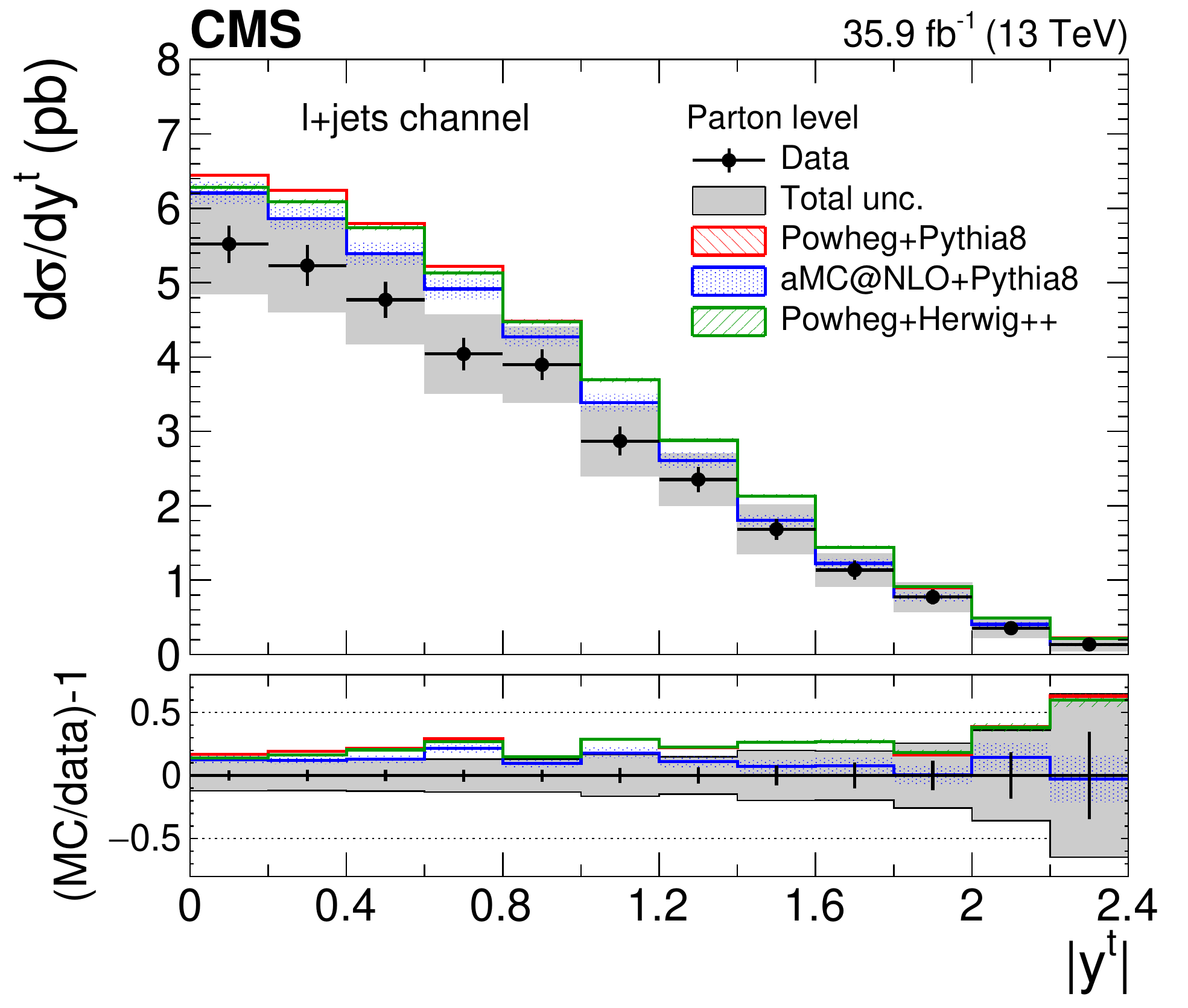}
\includegraphics[width=\cmsFigWidth]{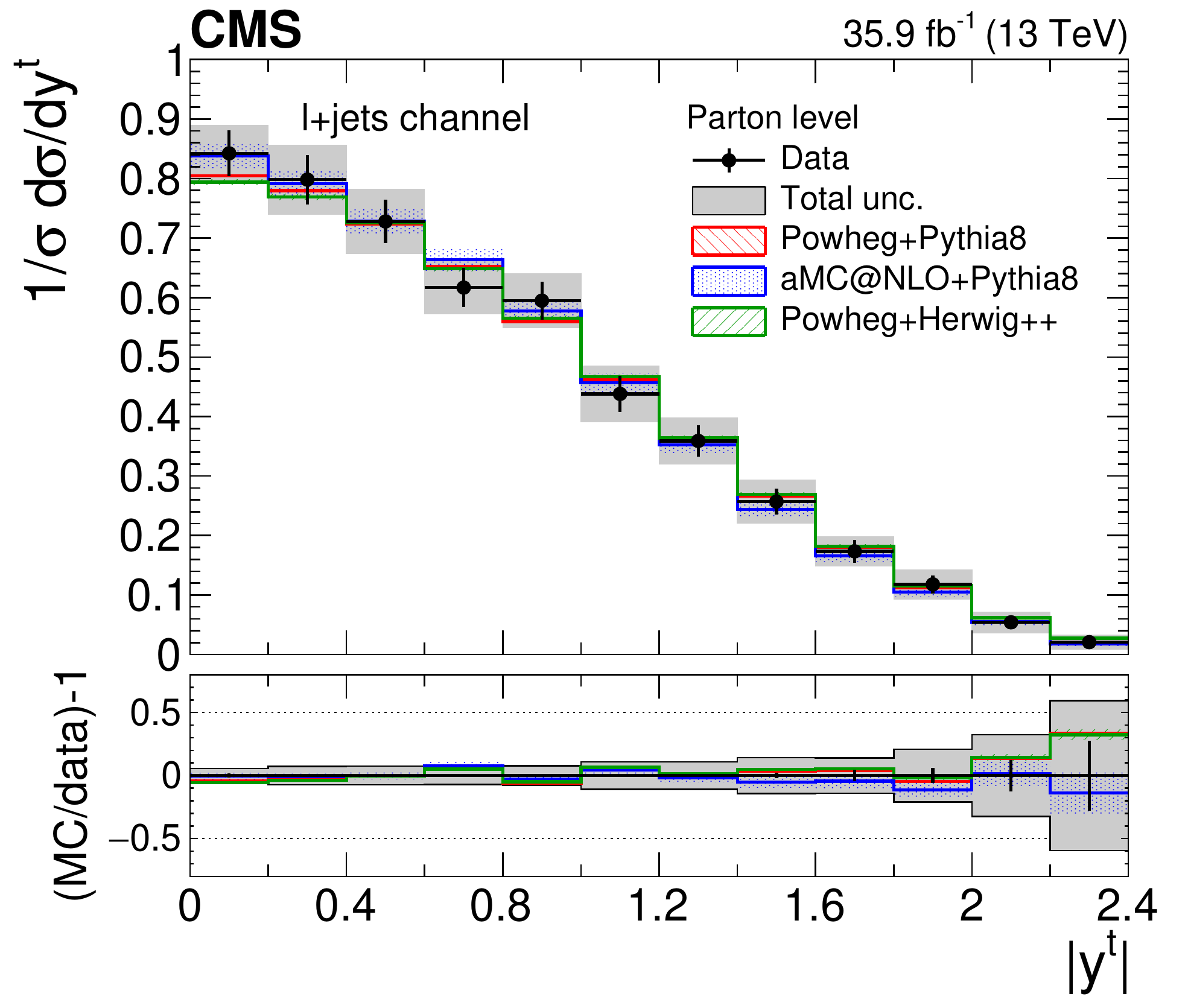}
\caption{\label{fig:lepjet_xsec_parton_pt_y} Differential cross section measurements at the parton level, as a function of the parton-level top quark \pt (upper row) and $\abs{y}$ (lower row) for the $\ell$+jets channel. Both absolute (left column) and normalized (right column) cross sections are shown. The lower panel shows the ratio (MC/data)$-$1. The vertical bars on the data and in the ratio represent the statistical uncertainty in data, while the shaded band shows the total statistical and systematic uncertainty added in quadrature. The hatched bands show the statistical uncertainty of the MC samples.}
\end{figure*}

\begin{figure*}[htb]
\centering
\includegraphics[width=\cmsFigWidth]{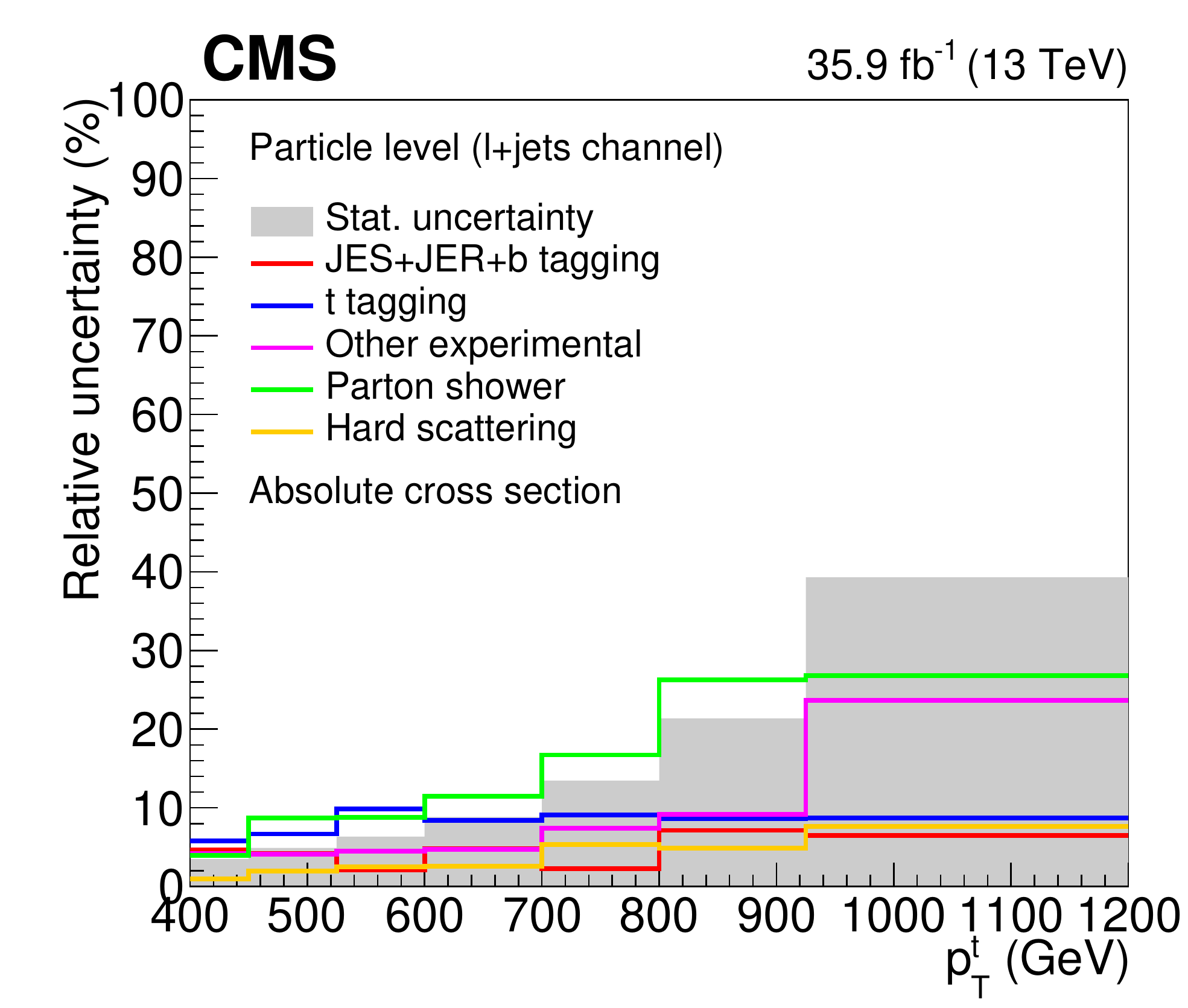}
\includegraphics[width=\cmsFigWidth]{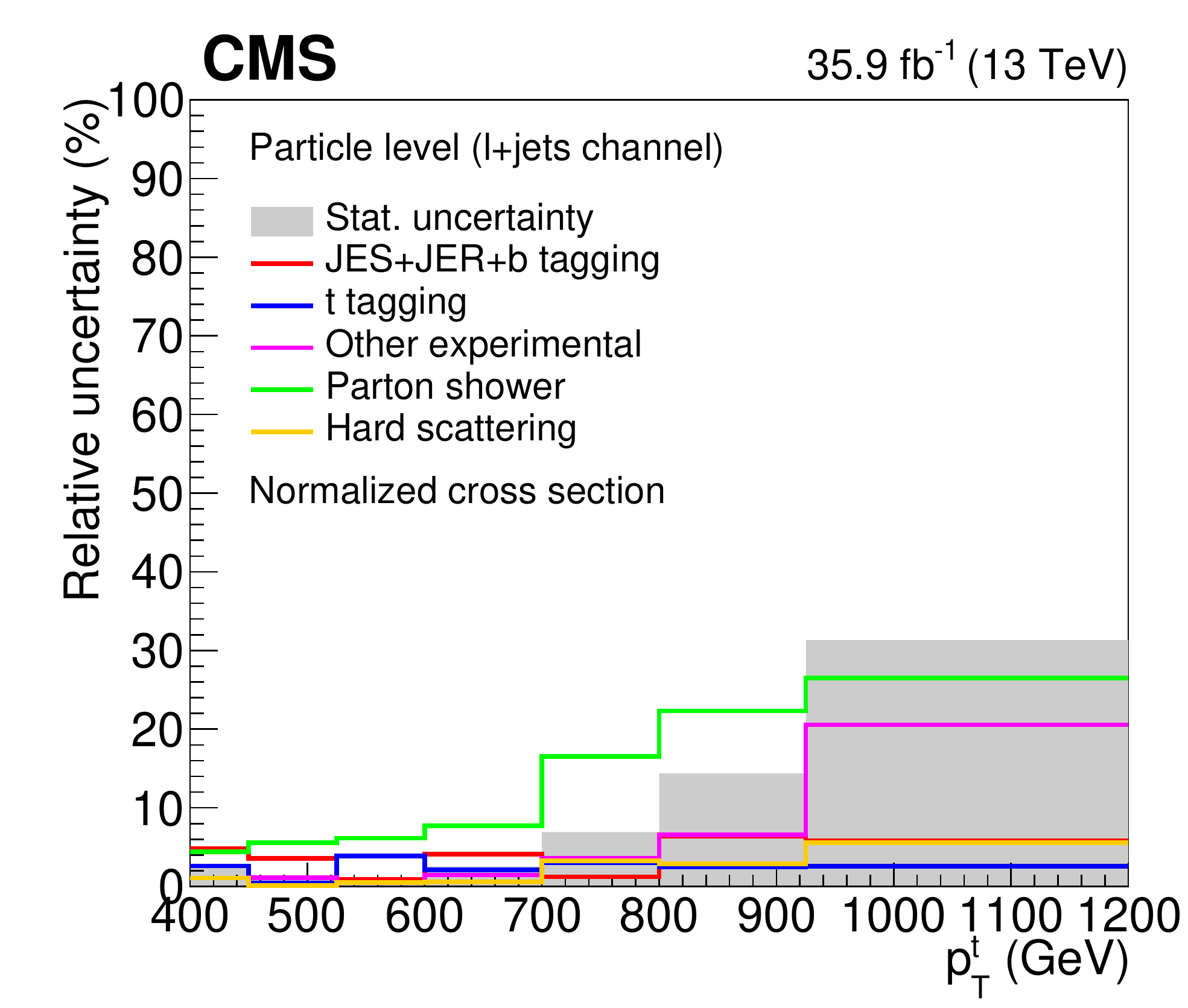}\\
\includegraphics[width=\cmsFigWidth]{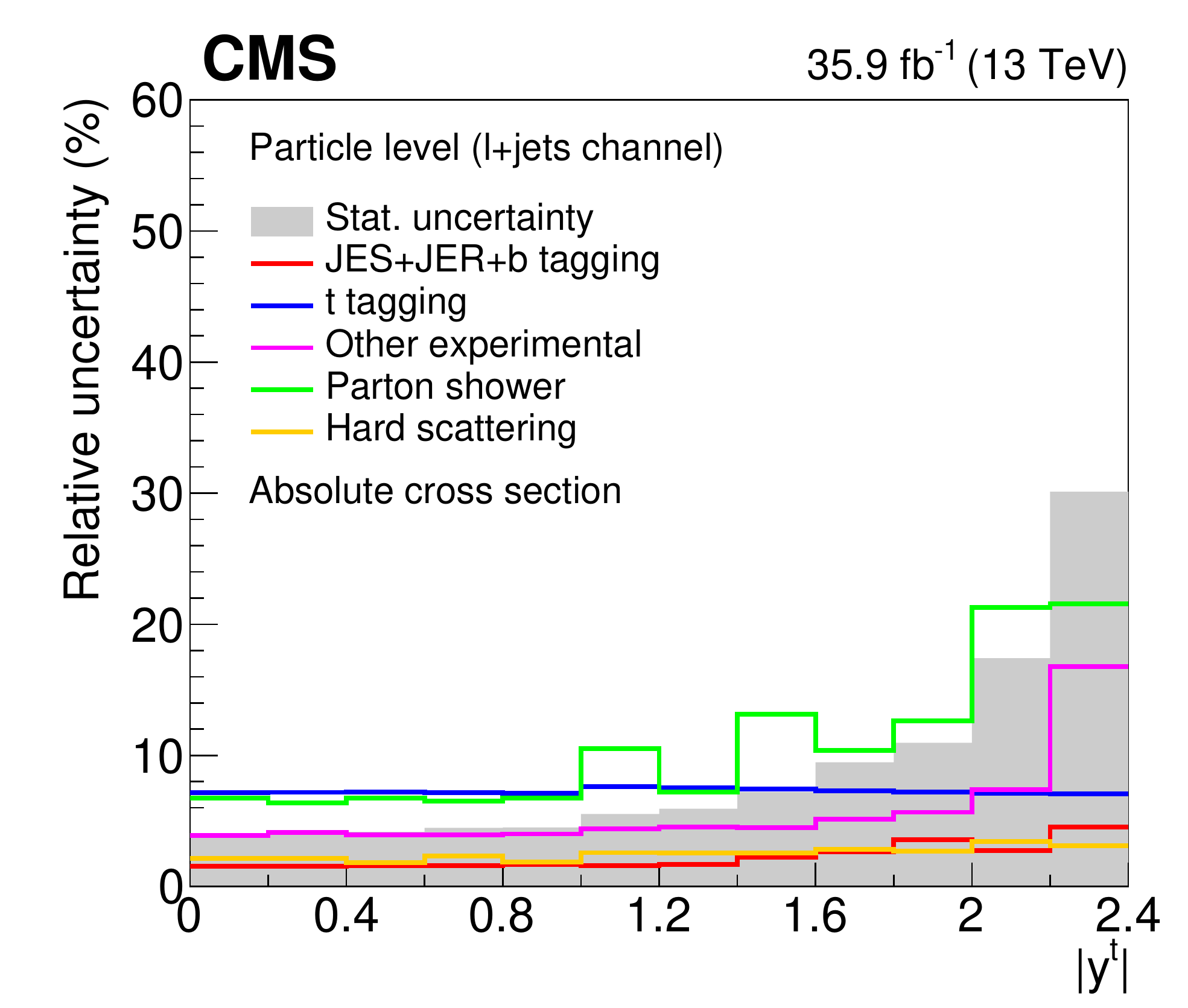} 
\includegraphics[width=\cmsFigWidth]{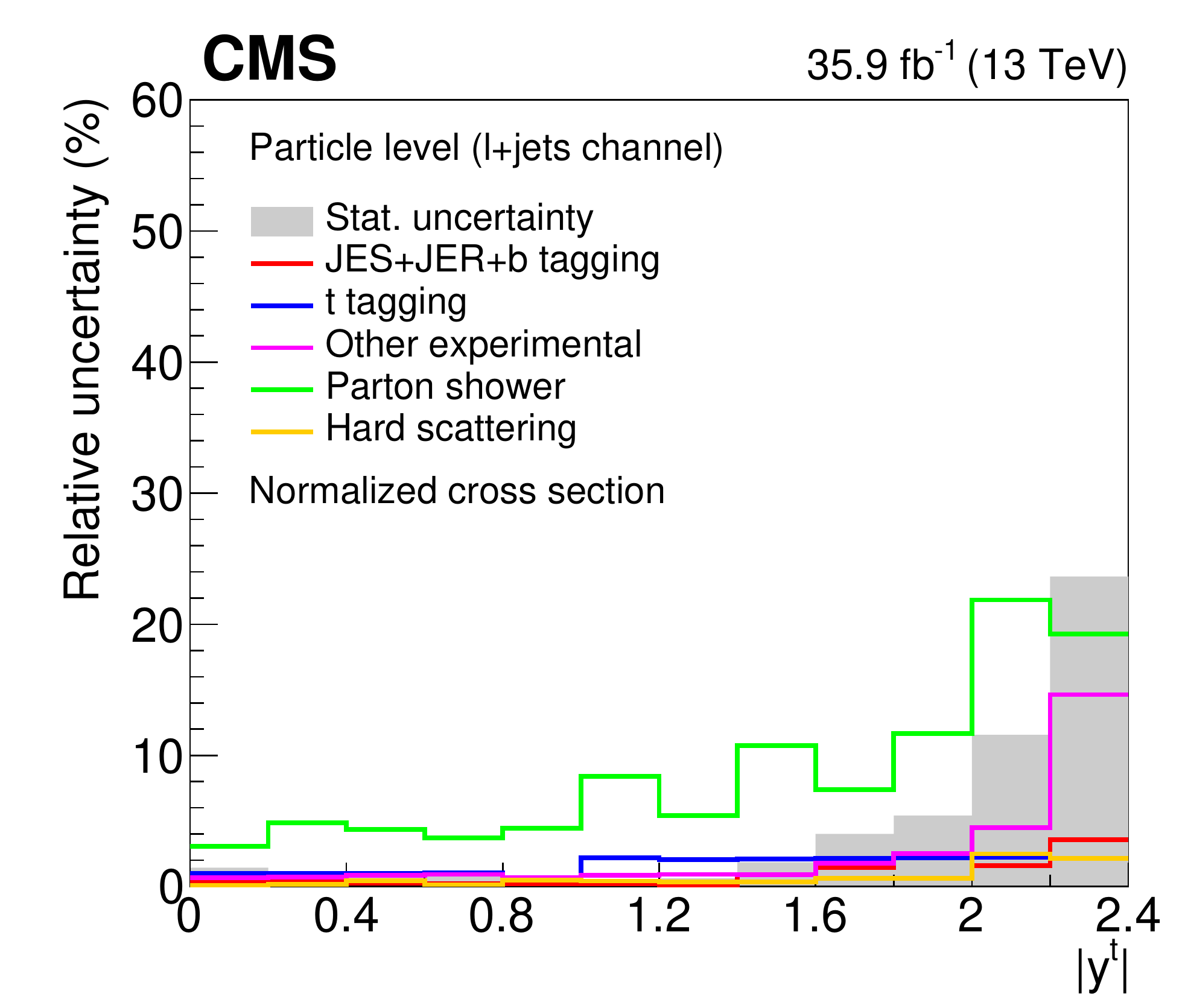}
\caption{\label{fig:lepjet_unc_particle} Breakdown of the sources of systematic uncertainty affecting the differential cross section measurements in the $\ell$+jets channel at the particle level as a function of the particle-level \PQt jet \pt (upper row) or $\abs{y}$ (lower row). Both the systematic uncertainties in the absolute (left column) and the normalized (right column) cross sections are shown. "JES+JER+$\PQb$ tagging" includes uncertainties due to the JES, JER, and small-$R$ jet \PQb tagging efficiency; "\PQt tagging" is the uncertainty associated with the large-$R$ jet \PQt tagging efficiency; "Other experimental" includes the uncertainties originating from the background estimate, pileup modeling, lepton identification and trigger efficiency, and measurement of the integrated luminosity; "Parton shower" includes contributions from ISR and FSR, underlying event tune, ME-PS matching, and color reconnection; "Hard scattering" includes the uncertainty due to PDFs, as well as renormalization and factorization scales. The grey bands shows the statistical uncertainty.}
\end{figure*}

\begin{figure*}[htb]
\centering
\includegraphics[width=\cmsFigWidth]{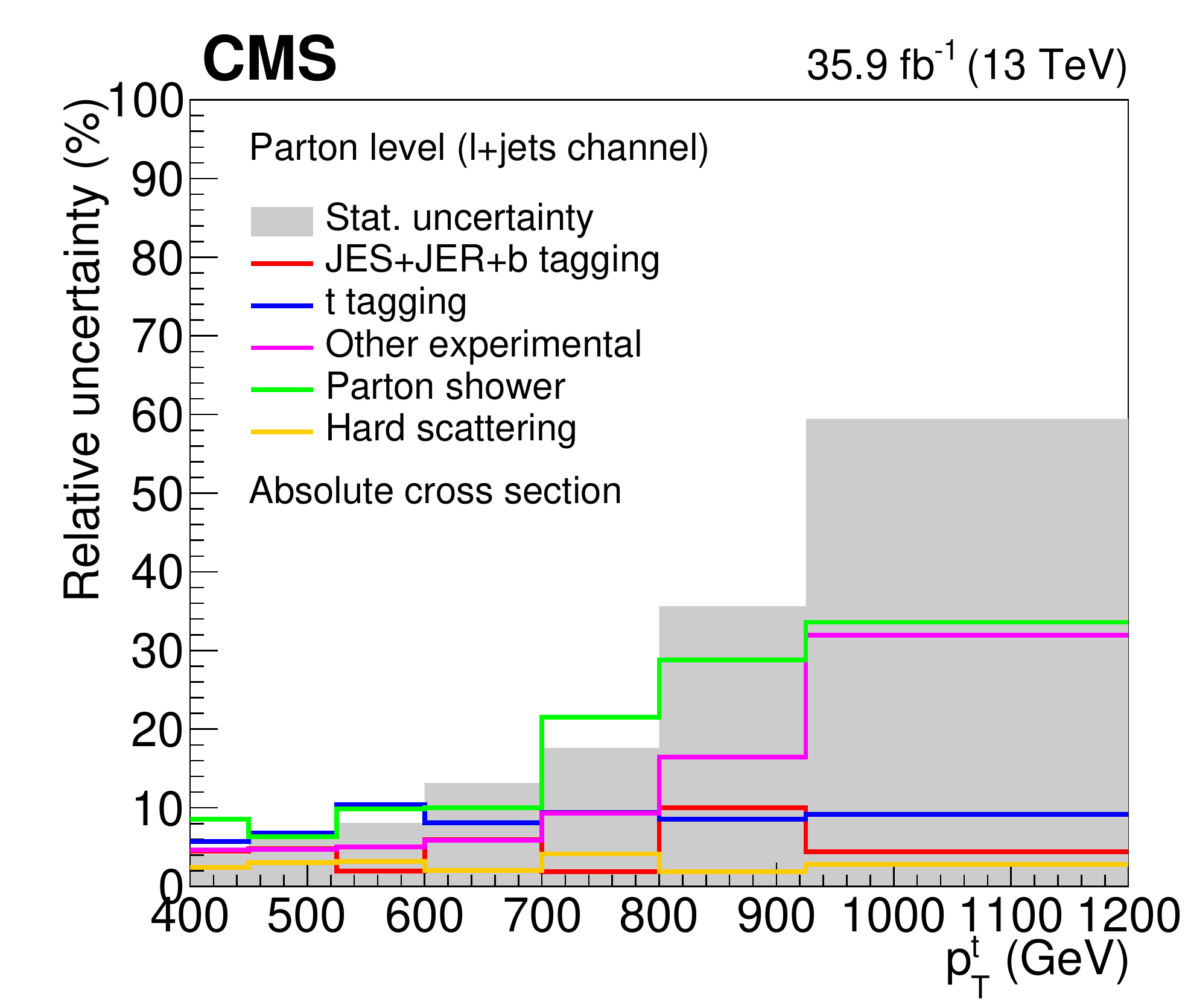}
\includegraphics[width=\cmsFigWidth]{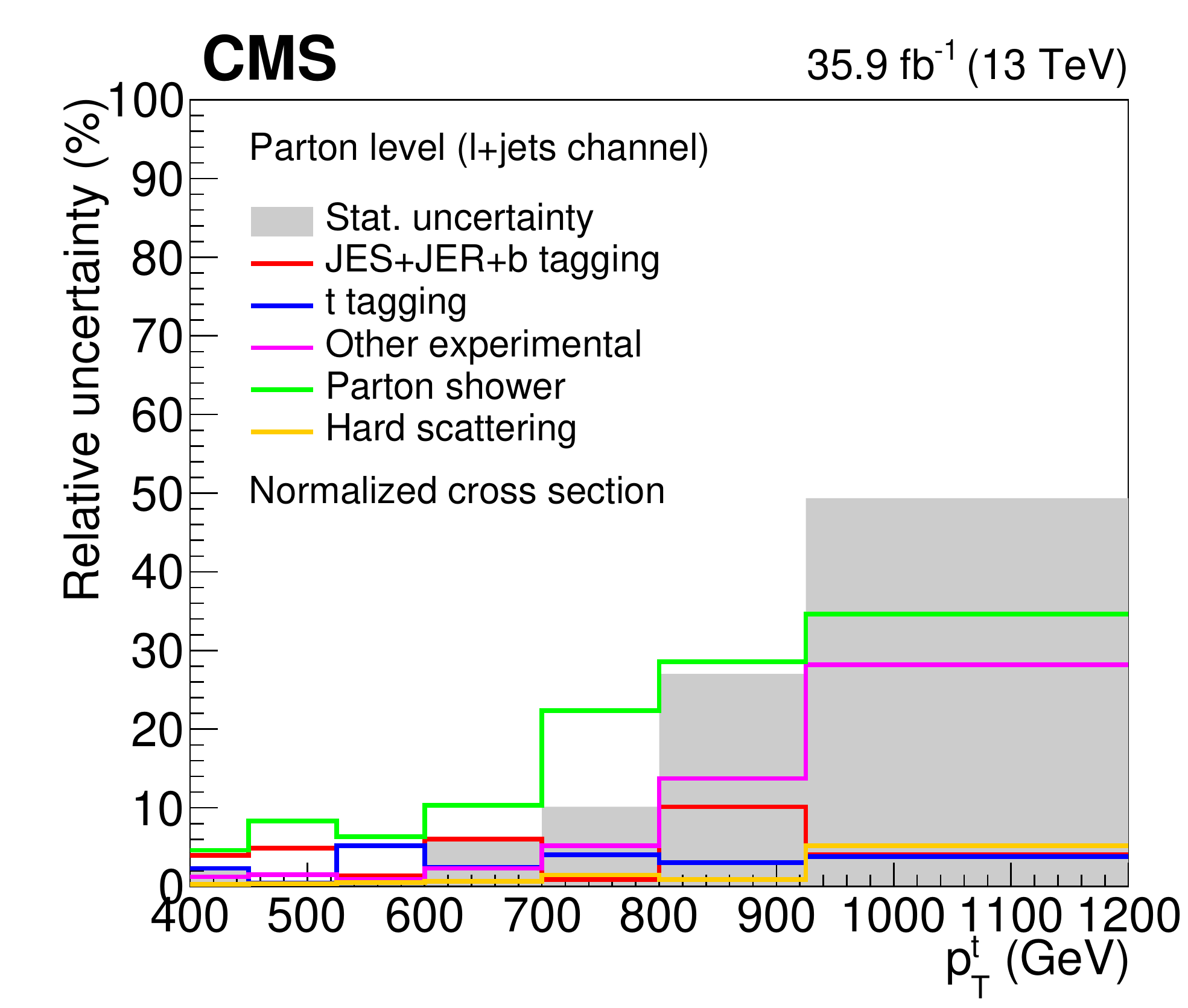}\\
\includegraphics[width=\cmsFigWidth]{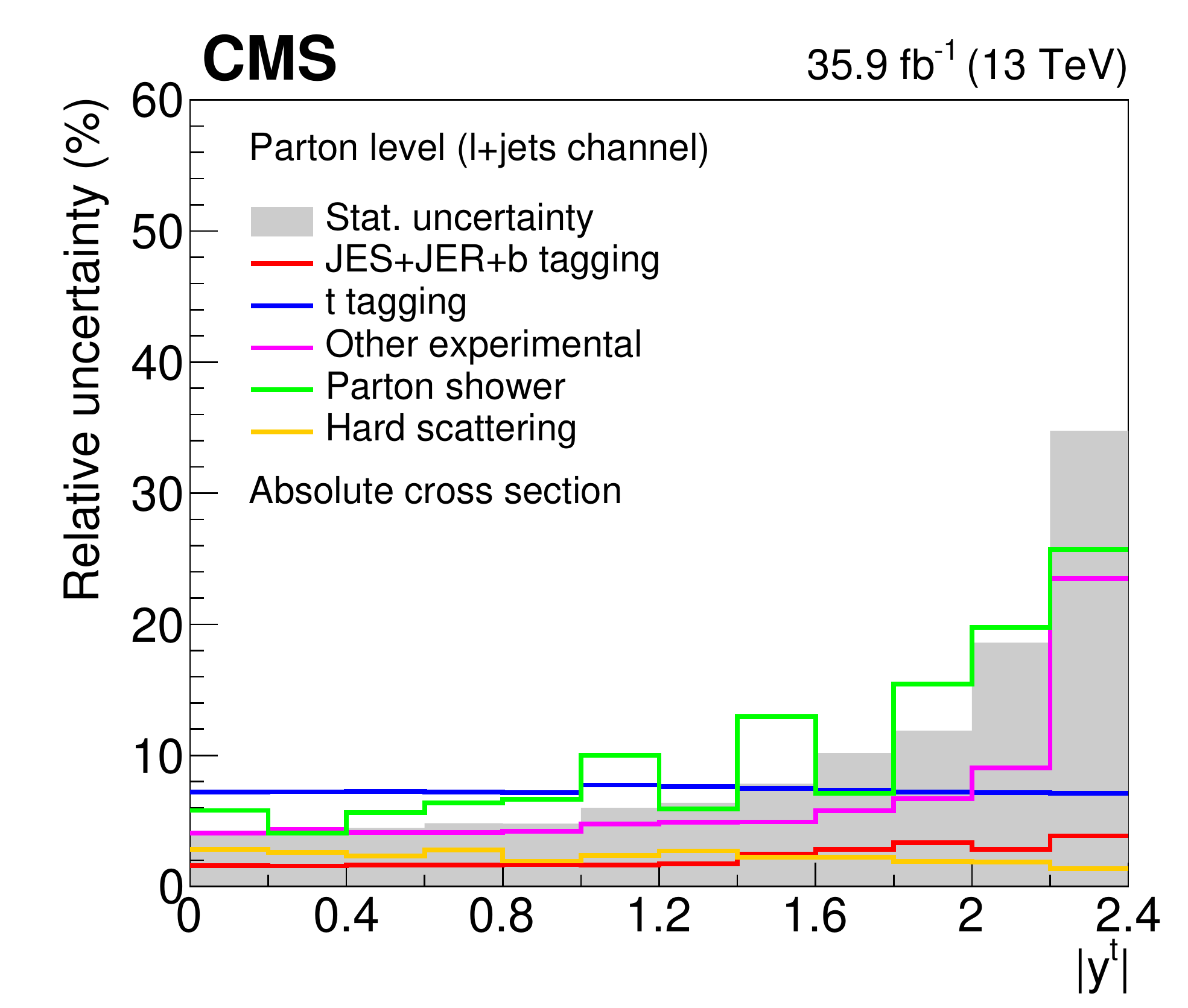}
\includegraphics[width=\cmsFigWidth]{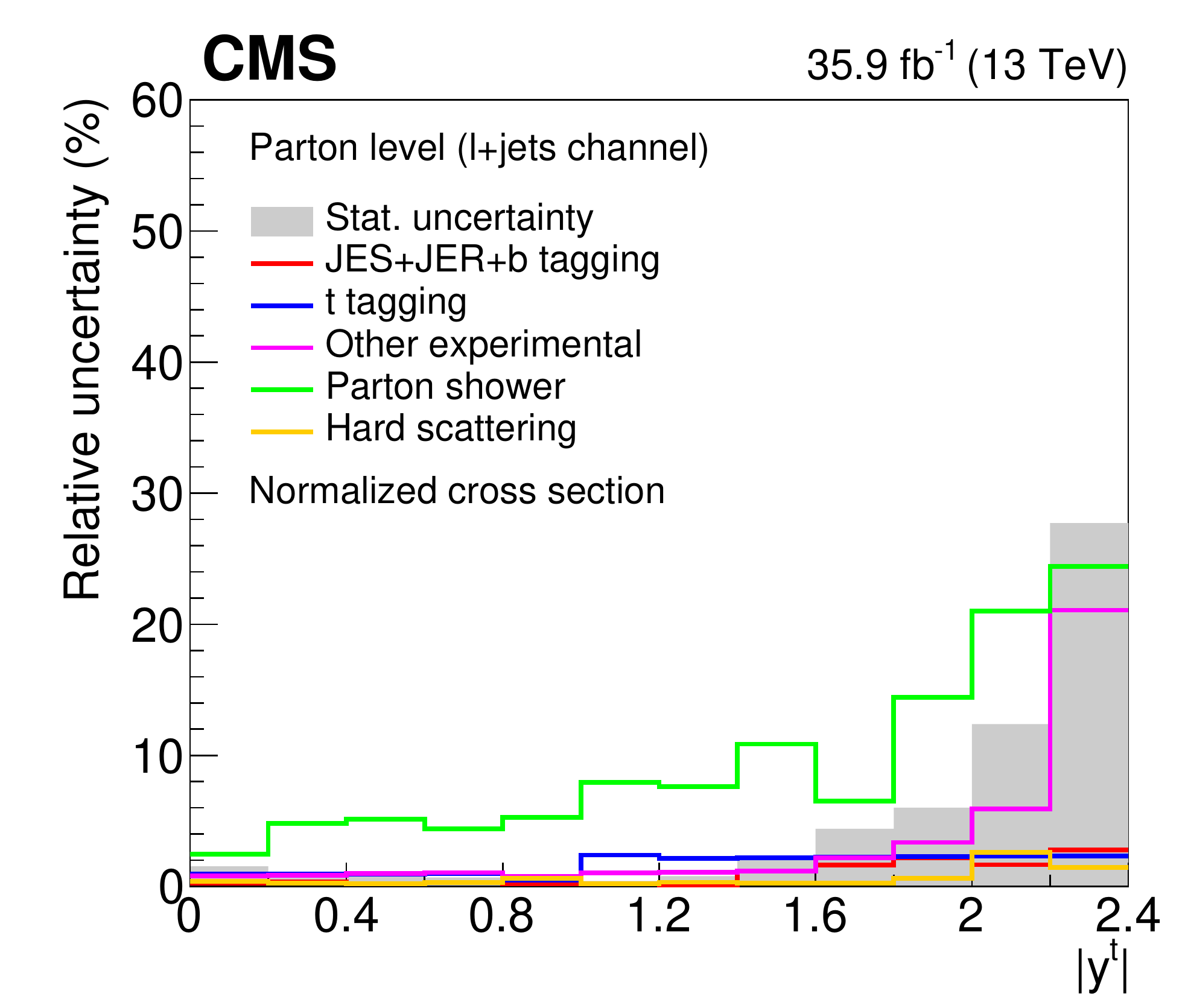}
\caption{\label{fig:lepjet_unc_parton} Breakdown of the sources of systematic uncertainty affecting the differential cross section measurements in the $\ell$+jets channel at the parton level as a function of the top quark \pt (upper row) or $\abs{y}$ (lower row). Both the systematic uncertainties in the absolute (left column) and the normalized (right column) cross sections are shown. "JES+JER+$\PQb$ tagging" includes uncertainties due to the JES, JER, and small-$R$ jet \PQb tagging efficiency; "\PQt tagging" is the uncertainty associated with the large-$R$ jet \PQt tagging efficiency; "Other experimental" includes the uncertainties originating from the background estimate, pileup modeling, lepton identification and trigger efficiency, and measurement of the integrated luminosity; "Parton shower" includes contributions from ISR and FSR, underlying event tune, ME-PS matching, and color reconnection; "Hard scattering" includes the uncertainty due to PDFs, as well as renormalization and factorization scales.}
\end{figure*}

The unfolded cross sections for top quarks are shown in Figs.~\ref{fig:lepjet_xsec_particle_pt_y}--\ref{fig:lepjet_xsec_parton_pt_y} as a function of \pt and $\abs{y}$ for the particle and parton levels, respectively, and compared to results from \POWHEG interfaced with \PYTHIA or \HERWIGpp and from \MGvATNLO interfaced with \PYTHIA. 
The breakdown of sources of systematic uncertainty are given in Figs.~\ref{fig:lepjet_unc_particle} and~\ref{fig:lepjet_unc_parton}.
The cross section at the parton level as a function of the \pt of the top quark that decays as $\PQt \to \PW \PQb \to \PQq \PAQq' \PQb$ presented in this paper can also be compared to the corresponding measurement from CMS in the resolved final state~\cite{TOP17002}. The two measurements are observed to be in agreement in the region of phase space where they overlap.

\subsection{Discussion}\label{sec:Disc}

The unfolded cross sections at the particle and parton levels reveal some important features. Theory predictions of the integrated cross sections, obtained using \POWHEG normalized as described in Section~\ref{sec:MC}, are 56 and 25\% higher than our measurement for the all-jet and $\ell$+jets channels, respectively, which agrees with previous results~\cite{Aaboud:2018eqg}. It should be noted that the two channels probe different phase spaces of the \ttbar production, due to the kinematic requirement on the subleading top quark in the all-jet channel, and therefore the integrated cross sections are not expected to be the same. 
That is, the phase space probed in the all-jet channel requires two top quarks with \pt above 400\GeV, while the $\ell$+jets channel phase space only requires one such high-\pt top quark. 
In terms of the normalized differential distributions, there is agreement between the data and theory within the uncertainties of the measurement and some qualitative observations can be made by comparing the central values of the data and theory. There is good agreement  for the leading top quark (all-jet channel) and the \pt of the top quark that decays as $\PQt \to \PW \PQb \to \PQq \PAQq' \PQb$ ($\ell$+jets channel), while the cross section as a function of the \pt of the subleading top quark in the all-jet channel appears to be softer in data than for the \POWHEG predictions, with \MGvATNLO providing the best description. The distributions in $y$ are well described by theory in both channels, with a small deviation for the subleading top quark that is related to the difference in the \pt spectrum. Finally, the measured distributions for the \ttbar system are mostly in agreement with theory, with a possible deviation in the $m^{\ttbar}$ variable, where \POWHEG tends to produce a harder spectrum, while \MGvATNLO is fully consistent with the data. Regarding systematic uncertainties, it should be noted that they are in general larger for the all-jet channel because the two leading experimental sources in JES and \PQb tagging enter twice (two large-$R$ jets). In contrast, the uncertainty in parton showering is smaller for the all-jet channel because its main contribution (FSR) is constrained through a dedicated analysis, as discussed in Section~\ref{sec:Sys}. 

\section{Summary}\label{sec:Sum}

A measurement was presented of the top quark pair (\ttbar) cross section for top quarks with high transverse momentum (\pt) produced in $\Pp\Pp$ collisions at 13\TeV. The measurement uses events in which either one or both top quarks decay to jets, and where the decay products cannot be resolved but are instead clustered in a single large-radius ($R$) jet with $\pt > 400\GeV$. The all-jet final state contains two such large-$R$ jets, while the lepton+jets final state is identified through the presence of an electron or muon, a $\PQb$-tagged jet, missing transverse momentum from the escaping neutrino, and a single $\PQt$-tagged, large-$R$ jet. The measurement utilizes a larger data set relative to previous results to explore a wider phase space of \ttbar production and to elucidate any discrepancies with theory that were reported in previous publications.
For the all-jet channel, absolute and normalized differential cross sections are measured as functions of the leading and subleading top quark \pt and absolute rapidity $\abs{y}$, and as a function of the invariant mass, \pt, and $y$ of the \ttbar system, unfolded to the particle level within a fiducial phase space and to the parton level. For the lepton+jets channel, the differential cross sections are measured as functions of the \pt and $\abs{y}$ of the top quark that decays according to $\PQt \to \PW \PQb \to \PQq \PAQq' \PQb$, both at the particle and parton levels.
The results are compared with theory using the \POWHEG matrix element generator, interfaced to either \PYTHIA or \HERWIGpp for the underlying event and parton showering, and with the \MGvATNLO matrix element generator, interfaced to \PYTHIA. 
All the models significantly exceed the absolute cross section in the phase spaces of the measurements. However, the normalized differential cross sections are consistently well described. The most notable discrepancies are observed in the invariant mass of the \ttbar system and the subleading top quark \pt in the all-jet channel, where theory predicts a higher cross section at high mass and at high \pt, respectively.
To further investigate the severity of this discrepancy, more data are needed to enhance the statistical significance of the measurement in this region of phase space. 

\clearpage

\begin{acknowledgments}

    We congratulate our colleagues in the CERN accelerator departments for the excellent performance of the LHC and thank the technical and administrative staffs at CERN and at other CMS institutes for their contributions to the success of the CMS effort. In addition, we gratefully acknowledge the computing centers and personnel of the Worldwide LHC Computing Grid for delivering so effectively the computing infrastructure essential to our analyses. Finally, we acknowledge the enduring support for the construction and operation of the LHC and the CMS detector provided by the following funding agencies: BMBWF and FWF (Austria); FNRS and FWO (Belgium); CNPq, CAPES, FAPERJ, FAPERGS, and FAPESP (Brazil); MES (Bulgaria); CERN; CAS, MoST, and NSFC (China); COLCIENCIAS (Colombia); MSES and CSF (Croatia); RIF (Cyprus); SENESCYT (Ecuador); MoER, ERC IUT, PUT and ERDF (Estonia); Academy of Finland, MEC, and HIP (Finland); CEA and CNRS/IN2P3 (France); BMBF, DFG, and HGF (Germany); GSRT (Greece); NKFIA (Hungary); DAE and DST (India); IPM (Iran); SFI (Ireland); INFN (Italy); MSIP and NRF (Republic of Korea); MES (Latvia); LAS (Lithuania); MOE and UM (Malaysia); BUAP, CINVESTAV, CONACYT, LNS, SEP, and UASLP-FAI (Mexico); MOS (Montenegro); MBIE (New Zealand); PAEC (Pakistan); MSHE and NSC (Poland); FCT (Portugal); JINR (Dubna); MON, RosAtom, RAS, RFBR, and NRC KI (Russia); MESTD (Serbia); SEIDI, CPAN, PCTI, and FEDER (Spain); MOSTR (Sri Lanka); Swiss Funding Agencies (Switzerland); MST (Taipei); ThEPCenter, IPST, STAR, and NSTDA (Thailand); TUBITAK and TAEK (Turkey); NASU (Ukraine); STFC (United Kingdom); DOE and NSF (USA).
    
    \hyphenation{Rachada-pisek} Individuals have received support from the Marie-Curie program and the European Research Council and Horizon 2020 Grant, contract Nos.\ 675440, 752730, and 765710 (European Union); the Leventis Foundation; the A.P.\ Sloan Foundation; the Alexander von Humboldt Foundation; the Belgian Federal Science Policy Office; the Fonds pour la Formation \`a la Recherche dans l'Industrie et dans l'Agriculture (FRIA-Belgium); the Agentschap voor Innovatie door Wetenschap en Technologie (IWT-Belgium); the F.R.S.-FNRS and FWO (Belgium) under the ``Excellence of Science -- EOS" -- be.h project n.\ 30820817; the Beijing Municipal Science \& Technology Commission, No. Z191100007219010; the Ministry of Education, Youth and Sports (MEYS) of the Czech Republic; the Deutsche Forschungsgemeinschaft (DFG) under Germany's Excellence Strategy -- EXC 2121 ``Quantum Universe" -- 390833306; the Lend\"ulet (``Momentum") Program and the J\'anos Bolyai Research Scholarship of the Hungarian Academy of Sciences, the New National Excellence Program \'UNKP, the NKFIA research grants 123842, 123959, 124845, 124850, 125105, 128713, 128786, and 129058 (Hungary); the Council of Science and Industrial Research, India; the HOMING PLUS program of the Foundation for Polish Science, cofinanced from European Union, Regional Development Fund, the Mobility Plus program of the Ministry of Science and Higher Education, the National Science Center (Poland), contracts Harmonia 2014/14/M/ST2/00428, Opus 2014/13/B/ST2/02543, 2014/15/B/ST2/03998, and 2015/19/B/ST2/02861, Sonata-bis 2012/07/E/ST2/01406; the National Priorities Research Program by Qatar National Research Fund; the Ministry of Science and Higher Education, project no. 02.a03.21.0005 (Russia); the Programa Estatal de Fomento de la Investigaci{\'o}n Cient{\'i}fica y T{\'e}cnica de Excelencia Mar\'{\i}a de Maeztu, grant MDM-2015-0509 and the Programa Severo Ochoa del Principado de Asturias; the Thalis and Aristeia programs cofinanced by EU-ESF and the Greek NSRF; the Rachadapisek Sompot Fund for Postdoctoral Fellowship, Chulalongkorn University and the Chulalongkorn Academic into Its 2nd Century Project Advancement Project (Thailand); the Kavli Foundation; the Nvidia Corporation; the SuperMicro Corporation; the Welch Foundation, contract C-1845; and the Weston Havens Foundation (USA).
\end{acknowledgments}

\bibliography{auto_generated}

\cleardoublepage \appendix\section{The CMS Collaboration \label{app:collab}}\begin{sloppypar}\hyphenpenalty=5000\widowpenalty=500\clubpenalty=5000\vskip\cmsinstskip
\textbf{Yerevan Physics Institute, Yerevan, Armenia}\\*[0pt]
A.M.~Sirunyan$^{\textrm{\dag}}$, A.~Tumasyan
\vskip\cmsinstskip
\textbf{Institut f\"{u}r Hochenergiephysik, Wien, Austria}\\*[0pt]
W.~Adam, F.~Ambrogi, T.~Bergauer, M.~Dragicevic, J.~Er\"{o}, A.~Escalante~Del~Valle, R.~Fr\"{u}hwirth\cmsAuthorMark{1}, M.~Jeitler\cmsAuthorMark{1}, N.~Krammer, L.~Lechner, D.~Liko, T.~Madlener, I.~Mikulec, N.~Rad, J.~Schieck\cmsAuthorMark{1}, R.~Sch\"{o}fbeck, M.~Spanring, S.~Templ, W.~Waltenberger, C.-E.~Wulz\cmsAuthorMark{1}, M.~Zarucki
% ...
\vskip\cmsinstskip
\dag: Deceased\\
1:  Also at Vienna University of Technology, Vienna, Austria\\
% ...
\end{sloppypar}
\end{document}